\documentclass[longauth]{aa}  

\usepackage{graphicx}

\usepackage{txfonts}
\usepackage{multirow}
\usepackage{array}
\usepackage[pdftex, colorlinks=true, linkcolor=blue, citecolor=blue, filecolor=blue, urlcolor=blue]{hyperref}
\newcommand{\nodata}{ ~$\cdots$~ }     
\graphicspath{{./}{figures/}}          

\newcolumntype{H}{>{\setbox0=\hbox\bgroup}c<{\egroup}@{}}  

\newcommand{\MBH}{M_\mathrm{BH}}
\newcommand{\MSeed}{M_\mathrm{seed}}
\newcommand{\MStar}{M_*}
\newcommand{\MSun}{M_\odot}
\newcommand{\LEdd}{L_\mathrm{Edd}}
\newcommand{\Lbol}{L_\mathrm{bol}}

\begin{document}

\title{Tracing the rise of supermassive black holes:}
\subtitle{A panchromatic search for faint, unobscured quasars at $z\gtrsim6$\\ with COSMOS-Web and other surveys}

\author{
	Irham T. Andika \inst{\ref{affil:tum}, \ref{affil:mpa}}
	\and Knud Jahnke \inst{\ref{affil:mpia}}
    \and Masafusa Onoue \inst{\ref{affil:ipmu}, \ref{affil:kiaa}}
	\and John D. Silverman \inst{\ref{affil:ipmu}, \ref{affil:utokyo}}
	\and Itsna K. Fitriana \inst{\ref{affil:naoj}}
	\and Angela Bongiorno \inst{\ref{affil:inaf_roma}}
	\and Malte Brinch \inst{\ref{affil:dawn}, \ref{affil:dtu}}
	\and Caitlin M. Casey \inst{\ref{affil:ut_austin}, \ref{affil:dawn}}
	\and Andreas Faisst \inst{\ref{affil:ipac}}
	\and Steven Gillman \inst{\ref{affil:dawn}, \ref{affil:dtu}}
	\and Ghassem Gozaliasl \inst{\ref{affil:aalto}, \ref{affil:helsinki}}
	\and Christopher C. Hayward \inst{\ref{affil:flatiron}}
	\and Michaela Hirschmann \inst{\ref{affil:sauverny}, \ref{affil:inaf_trieste}}
	\and Dale Kocevski \inst{\ref{affil:colby}}	
	\and Anton M. Koekemoer \inst{\ref{affil:stsci}}
	\and Vasily Kokorev \inst{\ref{affil:kapteyn}}
	\and Erini Lambrides \inst{\ref{affil:nasa}}
	\and Minju M. Lee \inst{\ref{affil:dawn}, \ref{affil:dtu}}
	\and R. Michael Rich \inst{\ref{affil:ucla}}
	\and Benny Trakhtenbrot \inst{\ref{affil:aviv}}
	\and C. Megan Urry \inst{\ref{affil:yale}}
	\and Stephen M. Wilkins \inst{\ref{affil:sussex}, \ref{affil:malta}}
	\and Aswin P. Vijayan \inst{\ref{affil:dawn}, \ref{affil:dtu}}
}

\institute{
Technical University of Munich, TUM School of Natural Sciences, Department of Physics, James-Franck-Str. 1, D-85748 Garching, Germany \\ \email{irham.andika@tum.de} \label{affil:tum}
\and Max-Planck-Institut f\"{u}r Astrophysik, Karl-Schwarzschild-Str. 1, D-85748 Garching, Germany \label{affil:mpa}
\and Max-Planck-Institut f\"{u}r Astronomie, K\"{o}nigstuhl 17, D-69117 Heidelberg, Germany \label{affil:mpia}
\and Kavli Institute for the Physics and Mathematics of the Universe (Kavli IPMU, WPI), The University of Tokyo, 5-1-5 Kashiwanoha, Kashiwa, Chiba 277-8583, Japan \label{affil:ipmu}
\and Kavli Institute for Astronomy and Astrophysics, Peking University, Beijing 100871, China \label{affil:kiaa}
\and Department of Astronomy, School of Science, The University of Tokyo, 7-3-1 Hongo, Bunkyo, Tokyo 113-0033, Japan \label{affil:utokyo}
\and National Astronomical Observatory of Japan, 2-21-1, Osawa, Mitaka, Tokyo 181-8588, Japan \label{affil:naoj} 
\and INAF, Osservatorio Astronomico di Roma, Via di Frascati 33, 00078 Monte Porzio Catone, Italy \label{affil:inaf_roma}
\and Cosmic Dawn Center (DAWN), Denmark \label{affil:dawn} 
\and DTU-Space, Technical University of Denmark, Elektrovej 327, DK2800 Kgs. Lyngby, Denmark \label{affil:dtu}
\and Department of Astronomy, The University of Texas at Austin, 2515 Speedway Blvd Stop C1400, Austin, TX 78712, USA \label{affil:ut_austin}
\and Caltech/IPAC, MS 314-6, 1200 E. California Blvd. Pasadena, CA 91125, USA \label{affil:ipac}
\and Department of Computer Science, Aalto University, PO Box 15400, Espoo, FI-00 076, Finland \label{affil:aalto} 
\and Department of Physics, Faculty of Science, University of Helsinki, 00014-Helsinki, Finland \label{affil:helsinki}
\and Center for Computational Astrophysics, Flatiron Institute, 162 Fifth Avenue, New York, NY 10010, USA \label{affil:flatiron}
\and Institute for Physics, Laboratory for Galaxy Evolution and Spectral Modelling, EPFL, Observatoire de Sauverny, Chemin Pegasi 51, 1290 Versoix, Switzerland \label{affil:sauverny} 
\and INAF, Osservatorio Astronomico di Trieste, Via Tiepolo 11, 34131 Trieste, Italy \label{affil:inaf_trieste} 
\and Department of Physics and Astronomy, Colby College, Waterville, ME 04901, USA \label{affil:colby}  
\and Space Telescope Science Institute, 3700 San Martin Dr., Baltimore, MD 21218, USA \label{affil:stsci}
\and Kapteyn Astronomical Institute, University of Groningen, P.O. Box 800, 9700AV Groningen, The Netherlands \label{affil:kapteyn} 
\and NASA-Goddard Space Flight Center, Code 662, Greenbelt, MD, 20771, USA \label{affil:nasa}
\and Department of Physics and Astronomy, University of California, Los Angeles, LA, CA 90095-1547 \label{affil:ucla} 
\and School of Physics and Astronomy, Tel Aviv University, Tel Aviv 69978, Israel \label{affil:aviv}
\and Department of Physics, Yale University, P.O. Box 208120, New Haven, CT 06520-8120, USA \label{affil:yale} 
\and Astronomy Centre, University of Sussex, Falmer, Brighton BN1 9QH, UK \label{affil:sussex} 
\and Institute of Space Sciences and Astronomy, University of Malta, Msida MSD 2080, Malta \label{affil:malta} 
}


\date{}

 
\abstract{
We report the identification of 64 new candidates of compact galaxies, potentially hosting faint quasars with bolometric luminosities of $L_\mathrm{bol} = 10^{43}$--10$^{46}$~erg~s$^{-1}$, residing in the reionization epoch within the redshift range of $6 \lesssim z \lesssim 8$.
These candidates were selected by harnessing the rich multiband datasets provided by the emerging JWST-driven extragalactic surveys, focusing on COSMOS-Web, as well as JADES, UNCOVER, CEERS, and PRIMER.
Our search strategy includes two stages: applying stringent photometric cuts to catalog-level data and detailed spectral energy distribution fitting.
These techniques effectively isolate the quasar candidates while mitigating contamination from low-redshift interlopers, such as brown dwarfs and nearby galaxies.
The selected candidates indicate physical traits compatible with low-luminosity active galactic nuclei, likely hosting $\approx10^5$--$10^7~M_\odot$ supermassive black holes (SMBHs) living in galaxies with stellar masses of $\approx10^8$--$10^{10}~M_\odot$.
The SMBHs selected in this study, on average, exhibit an elevated mass compared to their hosts, with the mass ratio distribution slightly higher than those of galaxies in the local Universe. 
As with other high-$z$ studies, this is at least in part due to the selection method for these quasars.
An extensive Monte Carlo analysis provides compelling evidence that heavy black hole seeds from the direct collapse scenario appear to be the preferred pathway to mature this specific subset of SMBHs by $z\approx7$.
Notably, most of the selected candidates might have emerged from seeds with masses of $\sim10^5~M_\odot$, assuming a thin disk accretion with an average Eddington ratio of $f_\mathrm{Edd}=0.6\pm0.3$ and a radiative efficiency of $\epsilon = 0.2\pm0.1$.
This work underscores the significance of further spectroscopic observations, as the quasar candidates presented here offer exceptional opportunities to delve into the nature of the earliest galaxies and SMBHs that formed during cosmic infancy.
}


\titlerunning{Tracing the rise of supermassive black holes}
\authorrunning{Andika et al.} 

\maketitle

%

\nolinenumbers 

\section{Introduction} \label{sec:intro}

Powered by gas and dust accretion onto supermassive black holes (SMBHs), quasars are among the brightest entities in the Universe with the corresponding active galactic nucleus (AGN) bolometric luminosities reaching $L_\mathrm{bol}\gtrsim10^{46}$~erg~s$^{-1}$. 
Thanks to various wide-field sky surveys, to date, more than 200 quasars hosting $\gtrsim 10^9~\MSun$ black holes have been discovered at $z\gtrsim6$, with a select number of them already shining brightly when the cosmos was just less than 800~Myr old \citep[see, e.g.,][for a recent review]{2023ARA&A..61..373F}. 
Assuming that such SMBHs originate from less massive seeds (i.e., $\approx10^2$--$10^6~\MSun$), assembling those enormous amounts of mass is challenging, requiring highly efficient matter accretions with additions of black hole mergers \citep{2019PASA...36...27W,2020ApJ...895...95P}.
Hence, these high-$z$ quasars, with their extreme characteristics compared to inactive galaxies, are ideal targets for examining the assembly of the earliest galaxies and SMBHs during cosmic infancy \citep{2022MNRAS.509.1885P}.

Several studies have proposed explanations for constructing the black hole seeds, although the comprehensive solution to this problem is still open-ended.
These theories include the idea that the first generation of low-mass black holes are presumably produced at the same time when the first-generation stars (hereafter Population III stars) are populating the Universe at $z\sim20$--30, or around 200 Myr since the Big Bang \citep{2021NatRP...3..732V}.
In line with that, black hole seeds are often separated into two classes, depending on their initial mass: (i) heavy seeds with a mass range of $10^4$--$10^6~\MSun$ and (ii) light seeds with masses of 10--100~$\MSun$ \citep[see, e.g.,][and references therein]{2020ARA&A..58...27I}.

One challenge of growing light seeds to form $10^9~\MSun$ SMBHs by $z\approx7$ is there is simply not enough time unless episodes of super- or even hyper-Eddington accretion can be sustained \citep[e.g.,][]{2013Natur.493..187M,2014ApJ...784L..38M,2014MNRAS.440.1590D,2016MNRAS.457.3356V,2016MNRAS.458.3047P,2017ApJ...850L..42P,2021MNRAS.501.1413N}.
While the super-Eddington accretion rate is just slightly above the Eddington-limit rate but still around the same order of magnitude, hyper-Eddington events can have values that are hundreds of times higher owing to photon trapping mechanisms reducing the radiation pressure effect on the infalling matter \citep{2017MNRAS.464.1102B}.
However, since most of the quasars discovered today are observed as having instantaneous accretion rates below or around the Eddington limit \citep{2017ApJ...836L...1T,2023arXiv230814986F}, the theory on heavy seeds is thus being explored further to ease the time-limited SMBH growth issue and possibly jump-start the formation of high-$z$ quasars \citep[e.g.,][]{2004ApJ...614L..25Y,2010A&ARv..18..279V,2019RPPh...82a6901M}.
As the first possibility, heavy seeds could form by collapsing primeval gas residing in the atomic-cooling halo, potentially producing short-lived supermassive stars (or quasi-stars) as by-products with a mass range of $10^5$--$10^6~\MSun$ \citep{2003ApJ...596...34B,2006MNRAS.371.1813L,2013ApJ...778..178H,2019ConPh..60..111S}.
The second possibility of heavy seed formation is that runaway collisions and mergers of either black holes or Population III stars within a gas-dense environment -- namely, a dense star cluster -- could produce seeds with masses of $10^3$--$10^4~\MSun$ \citep{2014Sci...345.1330A,2016MNRAS.456.2993L,2018MNRAS.476..366B,2021MNRAS.508.1756L,2023A&A...670A.180M,2023MNRAS.519.4753T}.
Heavy seeds might reduce the discrepancy between the theoretical model of SMBH growth and the observed quasar properties. However, such objects have yet to be detected \citep{2023arXiv230807260N,2023arXiv230802654N}.

Discovering more quasars in the reionization era is one obvious pathway for understanding early SMBH formation.
In particular, finding less massive black holes ($\approx10^6$--$10^8~\MSun$) at higher redshifts might give more information on whether heavy seeds are the dominant channel to explain the majority of the $z\gtrsim7$ quasar population.
Only the most luminous quasars, and hence, the largest, rarest SMBHs, could be discovered before the launch of the James Webb Space Telescope \citep[JWST;][]{2011Natur.474..616M,2019ApJ...881L..23B,2019ApJ...883..183M,2020ApJ...904..130V,2021ApJ...907L...1W,2021ApJ...923..262Y,2021ApJ...914...36I,2022AJ....163..251A}. 
Today, JWST is allowing for high-$z$ lower-luminosity AGNs ($L_\mathrm{bol}\approx10^{43}$--$10^{45}$~erg~s$^{-1}$) to be hunted where the stellar light might dominate the total emission or where the central emission from the accretion process is obscured \citep[e.g.,][]{2023arXiv230607320L,2023arXiv230801230M,2023arXiv230512492M,2023ApJ...953L..29L,2023arXiv230811609F,2023ApJ...955L..24G,2023arXiv230805735F,2023arXiv230811610K,2023arXiv230905714G,2023arXiv231107483W,2024arXiv240109981K,2024arXiv240108782P}.
About 30 lower mass SMBHs have been reported so far, and these objects might be the missing connection between the earliest bright quasars and black hole seeds.

Given the necessity of understanding how the first SMBHs and galaxies evolve, we present 64 new compact sources, potentially harboring quasars with $L_\mathrm{bol}\lesssim10^{46}$~erg~s$^{-1}$ and $\lesssim 10^8 \MSun$ SMBHs at $6 \lesssim z \lesssim 8$, selected utilizing various ground- and space-based imaging data.
Specifically, we exploit publicly available archival datasets covering the COSMOS, GOODS-S/N, Abell 2744, EGS HST legacy, and PRIMER extragalactic fields.
If spectroscopically confirmed, our candidates will double the number of quasars in the mass, luminosity, and redshift ranges mentioned earlier.
Furthermore, combining our samples with other published quasars in the literature will allow us to perform statistical analysis on this intriguing population and check their black hole and host galaxy characteristics.

The outline of this paper is as follows.
We start with the details on data acquisition and main database construction in Section~\ref{sec:data}.
Then, the method for identifying quasar candidates via photometric and spectral energy distribution (SED) modeling will be presented in Section~\ref{sec:sedfit}.
After that, we deliver the results and discuss the properties of the new candidates in Section~\ref{sec:result}.
Finally, we end this paper with a summary and conclusions in Section~\ref{sec:conclusion}.
For simplification and ease of reference within this paper, we subsequently define ``quasar'' as an interchangeable term for quasi-stellar object (QSO) and active galactic nucleus (AGN).
On several occasions, low-luminosity AGNs with $L_\mathrm{bol}\lesssim10^{46}$~erg~s$^{-1}$, whose emission could be overwhelmed by the host galaxy's light but the AGN contribution is still detectable are also considered as quasars.
The magnitudes written in this paper are reported using the AB system.
We further adopt the flat $\Lambda$CDM cosmological framework, where we assume $\Omega_\Lambda=0.7$, $\Omega_\mathrm{m}=0.3$, and $H_0 = 70~\rm km~s^{-1}~Mpc^{-1}$.
Consequently, at $z = 7$, the Universe's age is 0.748~Gyr, and the angular scale of $\theta = 1\arcsec$ corresponds to a linear scale of 5.3~kpc.

\section{Multi-survey datasets} \label{sec:data}

\begin{table*}[htb!]
	\caption{Overview of the employed selection that we used to detect the high-$z$ quasars.}
	\label{tab:preselection}
	\centering
	\small
	\begin{tabular}{clccccccc}
		\hline\hline
		Step & Selection & COSMOS-Web & JADES/ & GOODS-N & UNCOVER  & CEERS & PRIMER- & PRIMER-\\
		     &           &            & GOODS-S &        &          &       & COSMOS  & UDS\\
		\hline
		1 & All sources          & 342,435 & 70,899 & 37,890 & 61,648 & 76,300 & 118,794     & 143,552\\
		2 & SED modeling         & 247     & 383    & 61     & 105    & 237    & 172         & 185 \\
		3 & Visual inspection    & 30      & 58     & 16     & 32     & 54     & 69          & 91 \\
		\multirow{2}{*}{4} & $\MBH> 10^5~\MSun$ & \multirow{2}{*}{18} & \multirow{2}{*}{11} & \multirow{2}{*}{6} & \multirow{2}{*}{3} & \multirow{2}{*}{6} & \multirow{2}{*}{13} & \multirow{2}{*}{7} \\
		& and $f_\mathrm{AGN}\geq0.2$ & & & & & & & \\
		\hline\hline
		$\bullet$ & Sky coverage (arcmin$^2$)        & 1,008 & 57 & 55 & 49 & 91 & 164 & 212 \\
		$\bullet$ & Faintest magnitude        & 26.1 & 29.4 & 27.4 & 28.1 & 27.7 & 28.0 & 27.8 \\
		\hline
	\end{tabular}
	\tablefoot{
		At the end of our search, we found 350 compact sources, including 64 showing attributes consistent with low-luminosity AGNs. 
		We also report the sky area covered by each dataset and the faintest F444W magnitude of the candidates.
	}
\end{table*}

This section outlines the multiband photometric datasets used for the high-$z$ quasar selection in several major JWST extragalactic fields: COSMOS, GOODS-S/N, Abell 2744, EGS HST legacy, and PRIMER.
Some details on each of these surveys, data processing, and catalog construction will also be discussed here.
The unified database is then utilized to perform preselection and SED modeling to find promising candidates.

\subsection{The COSMOS-Web survey}

The first dataset is based on the COSMOS-Web program (GO \#1727, PIs Kartaltepe \& Casey), a deep imaging program covering 0.54~$\deg^2$ with 255 hours total integration time.
COSMOS-Web uses four JWST/NIRCam bands (F115W, F150W, F277W, and F444W) and one MIRI filter (F770W) in parallel.
More details on the survey description and observing strategy are presented by \cite{2023ApJ...954...31C}.
Our work utilizes the first two epochs of COSMOS-Web data obtained in January and April 2023.
The current available NIRcam mosaics cover approximately 0.28~$\deg^2$; on the other hand, MIRI data contains 0.07~$\deg^2$ of the COSMOS-Web field.

Data reduction for the NIRCam images is carried out utilizing the standard JWST Calibration Pipeline \citep{2022zndo...7325378B}.
In addition to that, custom processing steps are implemented to improve the image quality.
This includes 1/f noise and low-level background subtraction \citep[e.g.,][]{2022arXiv220512980B} and astrometric correction bootstrapped from the Hubble Space Telescope (HST) imaging in the F814W filter \citep{2007ApJS..172..196K} and the COSMOS2020 catalogs \citep{2022ApJS..258...11W}, anchored to the Gaia-EDR3 data \citep{2023A&A...674A...1G}.
The resulting multiband image mosaics with 0\farcs03/pixel have an astrometric normalized median absolute deviation below 12~mas.
Accordingly, MIRI data are reduced using a similar process to produce 0\farcs06/pixel mosaics.
While we only give a short overview here, two forthcoming papers will discuss details of the reduction process (Franco et al.; Harish et al., in prep.).

We complement the JWST data with multiwavelength information from various surveys performed on the Cosmic Evolution Survey (COSMOS) field.
This includes photometric datasets from HST/F814W \citep{2007ApJS..172....1S,2007ApJS..172..196K}, Spitzer/IRAC \citep{2022A&A...658A.126E}, Subaru/HSC PDR3 \citep{2022PASJ...74..247A}, and UltraVISTA DR5 \citep{2012A&A...544A.156M}.
A detailed summary of how these data are compiled and reprocessed is provided by \cite{2022ApJS..258...11W}.
Furthermore, we add submillimeter measurements from the A3COSMOS catalog \citep{2019ApJS..244...40L} when available.

The COSMOS-Web photometric catalog is produced using the SourceXtractor++ code \citep[SE++;][]{2020ASPC..527..461B,2022ascl.soft12018B}.
To create a detection image for reference, we first stack all four NIRCam bands via a chi-square ($\chi^2$) combination \citep{1999AJ....117...68S}.
Flux measurements are then performed on each band using model-based photometry, including the ancillary data from HST and other ground-based observations.
We note that model-based photometry enables flux extraction on images with diverse point spread functions (PSFs) without degrading their quality \citep[see also][]{2023arXiv231007757W}.
Specifically, this approach allows us to include constraints from ground-based data without sacrificing space-based data's resolution and, consequently, photometric accuracy.
In total, 342,435 sources are obtained from this catalog.

It should be noted that the flux errors of faint or undetected targets are often underestimated due to the flexibility given to the SE++ catalog construction.
To handle this issue, we set a noise floor in each band equivalent to the shot noise calculated using circular apertures placed randomly with sizes of 0\farcs3 and 1\arcsec\ for space-based and ground-based data, respectively.
Furthermore, to compute the source detection's significance, parameterized with the signal-to-noise ratio (S/N), we also consider the flux-to-error ratio extracted using an aperture of 1\farcs5 diameter.
This measurement is more robust than the model-based photometry S/N, and the aperture size is large enough to capture the whole source light, given the different PSF sizes between image filters.
The details on the photometric catalog creation will be described in a separate work (Shuntov et al., in prep.).

\subsection{The JADES project} \label{sec:jades}

Multiband data of the Great Observatories Origins Deep Survey South (GOODS-S) sky field is taken from the first public release of the JWST Advanced Deep Extragalactic Survey \citep[JADES\footnote{\url{https://jades-survey.github.io}};][]{2023arXiv230602465E} observations.
This dataset covers the ``deep'' portion of the images with exposure time per filter of 3.9--16.7 hours obtained in September/October 2022, resulting in a sky area of 25~arcmin$^2$ with a nominal 5$\sigma$ depth of around 29.9 mag.
The JWST/NIRCam filters utilized by the JADES project include F090W, F115W, F150W, F200W, F277W, F335W, F356W, F410M, F444W -- that is, spanning the wavelengths of 0.8--$5.0~\mu$m.
Photometry for 47,181 unique targets is provided in the catalog, where the source extractions and measurements are explained in detail by \cite{2023arXiv230602468H}.

The JADES catalog also makes use of the JWST Extragalactic Medium-band Survey \citep[JEMS;][]{2023arXiv230109780W} data, adding F182M, F210M, F430M, F460M, and F480W filters.
Moreover, observations from the First Reionization Epoch Spectroscopic COmplete survey \citep[FRESCO;][]{2023MNRAS.525.2864O} are also included, complementing the JADES catalog with F182M, F210M, and F444W filters when available.
As for the bluer wavebands, JADES utilized the existing deep HST/ACS and WFC3 mosaics from the Cosmic Assembly Near-infrared Deep Extragalactic Legacy Survey \citep[CANDELS;][]{2011ApJS..197...35G, 2011ApJS..197...36K} and the Hubble Legacy Field dataset \citep{2019ApJS..244...16W} containing F435W, F606W, F775W, F814W, and F850LP images.
Finally, it is worth mentioning that all fluxes we use for the SED fitting later are based on the ones measured within a circular aperture with a radius of 0\farcs15, corrected for flux losses, in the ``CIRC\_CONV'' table of the JADES catalog.

\subsection{The UNCOVER program}
The search on the lensing cluster Abel 2744 region will be conducted using the data provided by the Ultradeep NIRSpec and NIRCam ObserVations before the Epoch of Reionization \citep[UNCOVER\footnote{\url{https://jwst-uncover.github.io}};][]{2022arXiv221204026B} Cycle 1 JWST Treasury program.
The second version of the photometric catalog released by this program is constructed based on the 49~arcmin$^2$ image mosaics of seven NIRCam filters -- that is, F115W, F150W, F200W, F277W, F356W, F410M, and F444W -- together with numerous HST/ACS and WFC3 ancillary data.

For photometry purposes, all UNCOVER mosaics are PSF-matched to the F444W band, and the detection images for source extractions are created by combining the F277W, F356W, and F444W mosaics exploiting the noise-equalized technique.
Specifically, the so-called ``SUPER'' catalog that we will use here, where the photometry is calculated from optimally selected color apertures in the range of 0\farcs32 -- 1\farcs4 diameter on 0.04\arcsec/pixel mosaics, reaches a nominal 5$\sigma$ magnitude limit of around 30~mag.
We note that the fluxes in that catalog are corrected to total values using the Kron Radius measured in the detection image, with an additional correction of approximately 5-10\% applied to account for missing light beyond a 1\arcsec\ radius, guided by the F444W curve of growth \citep{2023arXiv230102671W}.
By default, fluxes for 61,648 unique sources in the UNCOVER catalog are reported in the unit of 10~nJy or correspond to the AB magnitude zero point of 28.9.
This catalog is further enriched with submillimeter measurements from the Deep UNCOVER-ALMA Legacy High-Z (DUALZ) Survey, featuring ALMA band 6 observations with a 30-GHz wide frequency band down to a sensitivity of 32.7~$\mu$Jy~beam$^{-1}$ \citep{2023arXiv230907834F}.

\subsection{Additional archival data}

In addition to the previously mentioned datasets, we also make use of the JWST data targeting some public extragalactic fields, which were processed with grizli \citep{2022zndo...6672538B} and msaexp \citep{2022zndo...7299500B} by the Cosmic Dawn Center (DAWN), stored in the DAWN JWST Archive \citep[DJA\footnote{\url{https://dawn-cph.github.io/dja}};][]{2023ApJ...947...20V}.
Specifically, we first mined the Cosmic Evolution Early Release Science Survey \citep[CEERS\footnote{\url{https://ceers.github.io}};][]{2023ApJ...946L..13F} data provided by DJA to expand our candidates list.
We refer the reader to \cite{2023ApJ...946L..12B} for a complete description of the official CEERS data products.
In short, the dataset consists of NIRCam imaging in F115W, F150W, F200W, F277W, F356W, F410M, and F444W bands targeting the Extended Groth Strip (EGS) HST legacy field with the current area coverage of 91~arcmin$^2$ and a 5$\sigma$ depth of 28.3--28.8 mag.

Along with that, we also exploited the DJA's version of the Public Release IMaging for Extragalactic Research \citep[PRIMER\footnote{\url{https://primer-jwst.github.io}};][]{2021jwst.prop.1837D} dataset.
The PRIMER survey was performed utilizing the NIRCam and MIRI imaging on two contiguous equatorial regions, namely, the Ultra-Deep Survey \citep[UDS;][]{2007MNRAS.379.1599L} and COSMOS \citep{2007ApJS..172....1S} fields.
While the used filters are similar to CEERS, PRIMER further enriches the covered wavelengths by adding F090W, F770W, and F1800W bands.
In total, the areas covered by the PRIMER-UDS and PRIMER-COSMOS reach about 212~arcmin$^2$ and 164~arcmin$^2$, respectively, with a 5$\sigma$ limiting magnitude of 27.4--27.9~mag.
As further information, complementary to the CEERS and PRIMER's JWST data, DJA also provides photometric measurements based on the existing HST archival images \citep[i.e.,][]{2011ApJS..197...35G,2011ApJS..197...36K,2022ApJS..263...38K}.

At the time of writing, the photometric catalog of the GOODS-N field, along with some GOODS-S regions from the ``non-deep'' portion of the JADES programs, has yet to be released by the official JADES collaboration \citep{2023arXiv230602465E}.
Fortunately, a subset of their NIRCam mosaics is publicly available and processed by DJA, covering the area of about 55~arcmin$^2$ and 57~arcmin$^2$ for the northern and southern datasets, respectively.
These images include additional data from the observing programs of JADES Medium, 1210/1286 Parallel, and northwest and southeast pointings.
Similar to the dataset introduced in Section~\ref{sec:jades}, DJA complemented the JADES GOODS-S/N data with the NIR imaging from the JEMS \citep{2023arXiv230109780W} and FRESCO \citep{2023MNRAS.525.2864O} projects, as well as the optical photometry from the Hubble Legacy Fields program \citep{2019ApJS..244...16W}.
With all data in hand, we ultimately consider the aperture-based photometry, corrected for flux losses, calculated with a diameter of 0\farcs36 for CEERS, PRIMER-UDS, PRIMER-COSMOS, GOODS-N, GOODS-S catalogs produced by DJA, each containing 76,300, 143,552, 118,794, 37,890, and 52,427 objects, respectively.
It is important to note that the GOODS-S dataset constructed here and the one obtained in Section~\ref{sec:jades} are then merged to remove duplicated sources by crossmatching these two catalogs using a 1~\arcsec\ radius.
This combined catalog is hereafter called ``JADES'' to differentiate them from the GOODS-N data.

At this point, we then compile the catalogs from the COSMOS-Web, JADES GOODS-S/N, UNCOVER, CEERS, and PRIMER projects.
All fluxes are converted to nJy unit, corresponding to AB zero point of 31.4~mag.
To further ensure that bright flux values do not excessively influence the SED fitting process later and to accommodate potential uncertainties in photometric calibration, we designate a lower limit of 5\% as the error floor for the photometric measurements \citep[e.g.,][]{2023arXiv230602468H}.
The effect of Galactic extinction is then corrected using the dust map of \cite{1998ApJ...500..525S} and reddening correction of \cite{1999PASP..111...63F}, applied using the software from \cite{2018JOSS....3..695G}.

This resulting parent catalog comprises 851,518 unique sources (see Table~\ref{tab:preselection} for the breakdown), which includes a mix of galaxies and quasars at all redshifts, stars, substellar objects, and artifacts. 
The following sections will describe various steps to extract the actual quasar content and resulting AGN properties, with the steps already listed in Table~\ref{tab:preselection}.
First, the SEDs of all objects are modeled with composite SED templates representative of galaxies with and without AGN, as well as stars and substellar objects, including dust reddening. 
This SED modeling will robustly remove all non-galaxies from the catalog, low-redshift galaxies, and AGN with a photometric $z<5.5$. 
The resulting much smaller sample of high-$z$ candidates is then visually inspected to remove objects with SEDs impacted by cosmic ray hits, hot pixels, stray light residuals, etc. 
This approach will provide a high-probability set of high-$z$ candidate galaxies and AGN we already discussed.
Then, in the final step, detailed independent SED fitting is used to extract relative galaxy and AGN flux contributions in these high-probability candidates. 
We demonstrate the robustness and limits of these estimates and then use the resulting AGN flux to infer AGN properties for the sample.


\section{Quasar search via SED fitting} \label{sec:sedfit}

\subsection{Photometric redshift estimation and initial selection}

We implement the first SED modeling step -- to find the quasar candidates and separate them from other contaminants, such as low-$z$ galaxies, brown dwarfs, detector artifacts, etc. \citep[e.g.,][]{2020ApJ...903...34A,2023ApJ...943..150A} -- using \texttt{eazy-py}\footnote{\url{https://github.com/gbrammer/eazy-py}}, a Python-based photometric redshift estimator \citep{2008ApJ...686.1503B}.
By iterating through a user-defined grid of spectral templates and redshifts, \texttt{eazy-py} tries to find the best model that matches the observed photometry.

Here, the templates for quasar SEDs are derived empirically from the observational data of XMM-COSMOS AGNs and galaxies, provided and discussed in detail by \cite{2017ApJ...850...66A}.
Although the original template list includes a wide variety of galaxy types, we exclusively use the spectra of bright quasars showing broad emission lines and a blue rest-frame ultraviolet (UV) continuum for our purposes \citep[e.g.,][]{2023A&A...678A.103A}.
As done by \cite{2021A&A...648A...4D}, we further append the effect of dust extinction using attenuation levels ($A_V$) ranging from 0 to 2 with a step of 0.2, following the model from \cite{2000ApJ...533..682C}.

Also, the built-in templates for inactive galaxies provided by \texttt{eazy-py} are constructed based on the Flexible Stellar Population Synthesis code \citep[FSPS;][]{2009ApJ...699..486C,2010ApJ...708...58C,2010ApJ...712..833C} and one high-equivalent-width galaxy from \cite{2010ApJ...719.1168E}.
These SEDs contain a mixture of stellar, nebular, and dust-reprocessed emission components.
It is important to note that, since young, high-$z$ galaxies could show very blue UV continuum slopes due to their high star formation rate (SFR), lower metallicity, and less dusty nature, we put to use additional bluer templates from \cite{2022arXiv221110035L} complementing the available SED models.
That is, we use the ``reduced Ly$\alpha$'' sets in \cite{2022arXiv221110035L}, optimized to fit galaxies at $4 \leq z \leq 7$.

\clearpage
\begin{figure*}[htb!]
	\centering
	\raisebox{0.03\height}{\includegraphics[width=0.5\textwidth]{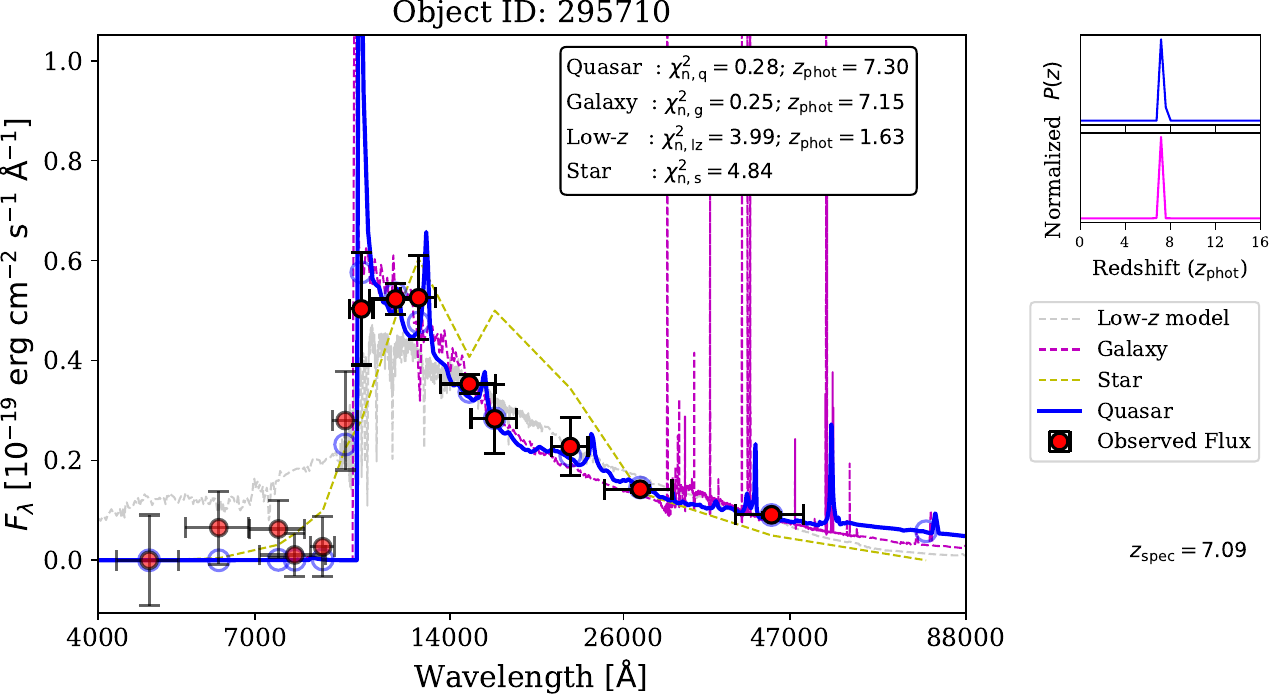}}
	\hspace{10pt}
	\raisebox{0.128\height}{\includegraphics[width=0.4\textwidth]{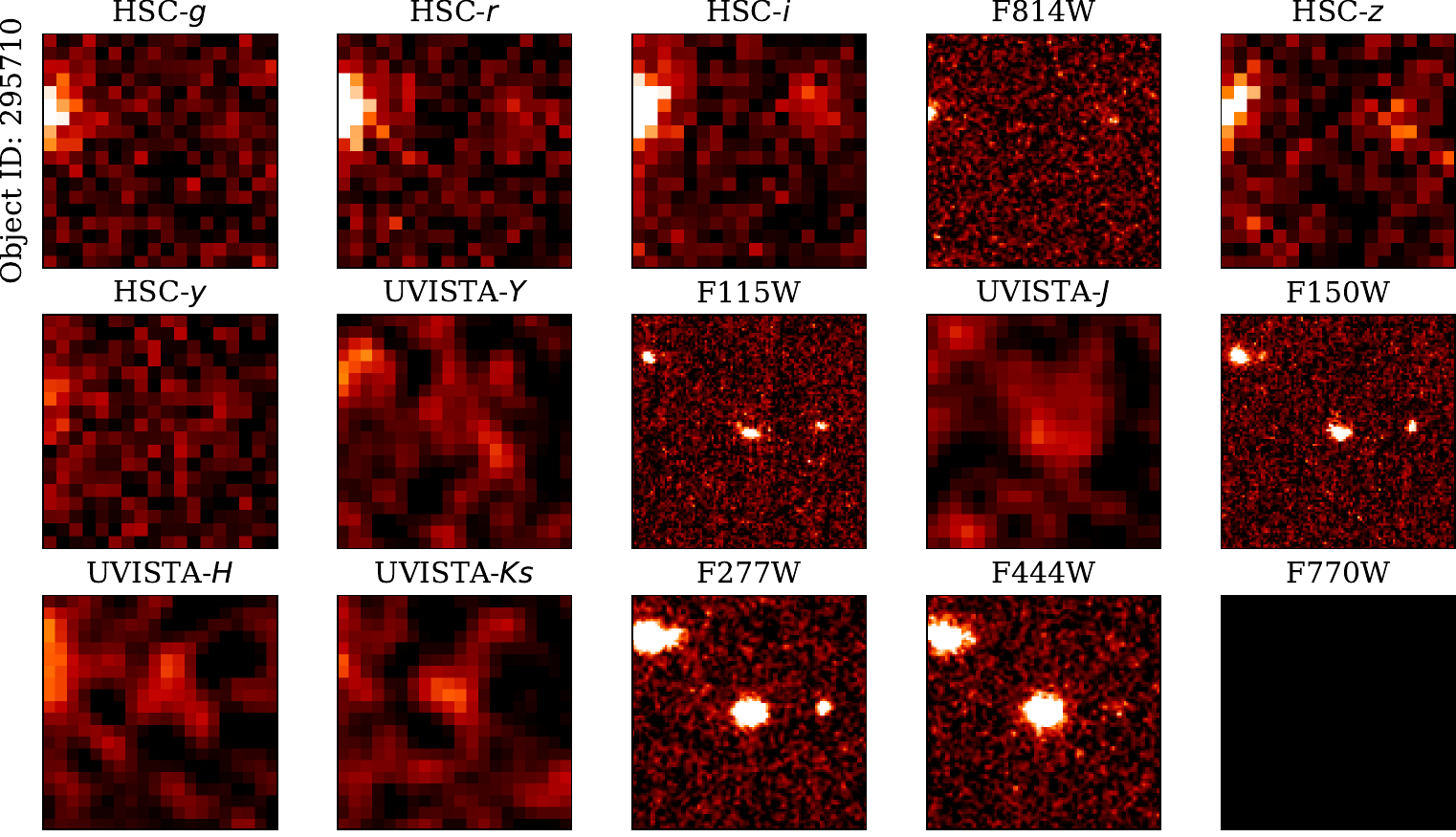}}
	\raisebox{0.03\height}{\includegraphics[width=0.5\textwidth]{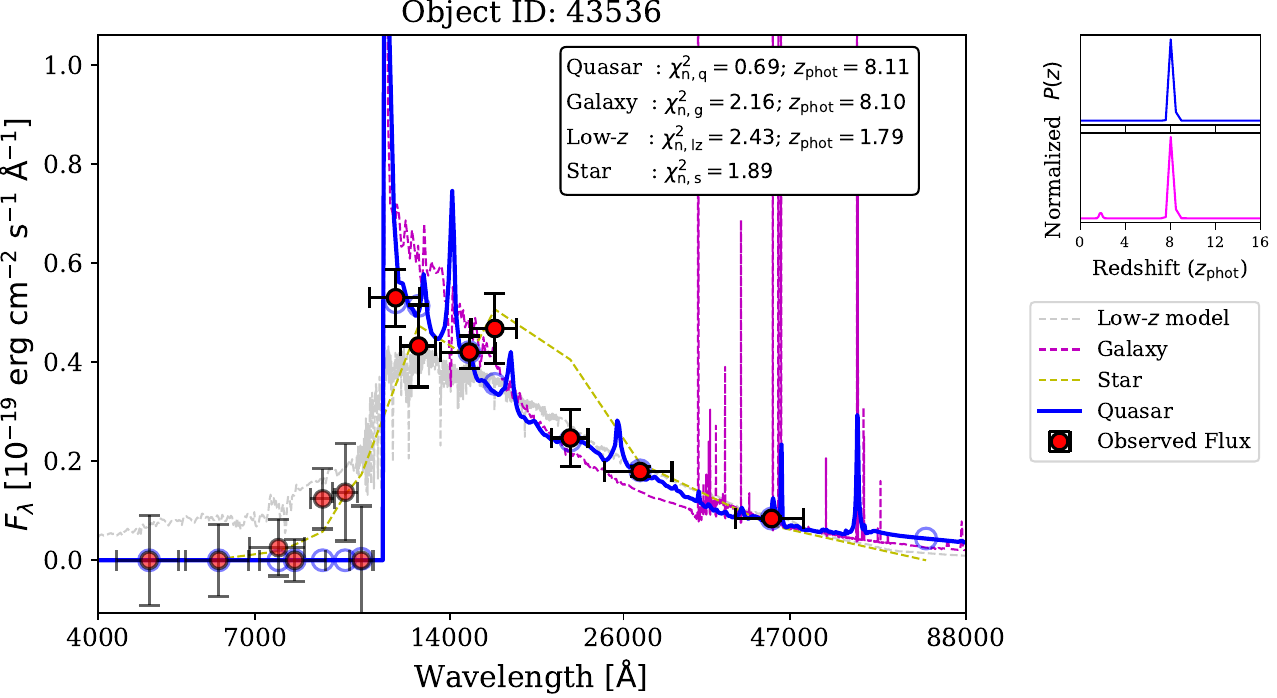}}
	\hspace{10pt}
	\raisebox{0.128\height}{\includegraphics[width=0.4\textwidth]{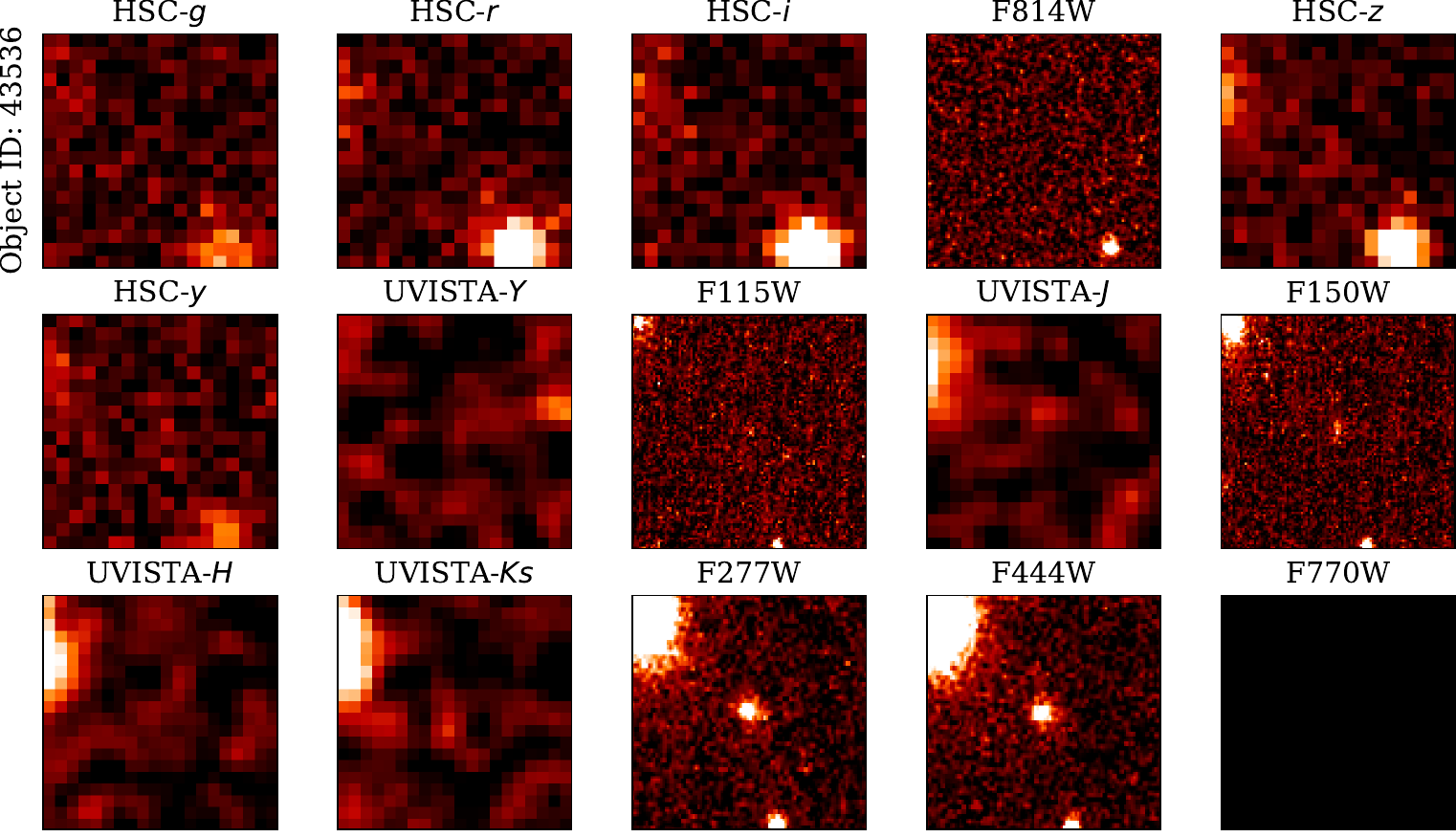}}
	\raisebox{0.03\height}{\includegraphics[width=0.5\textwidth]{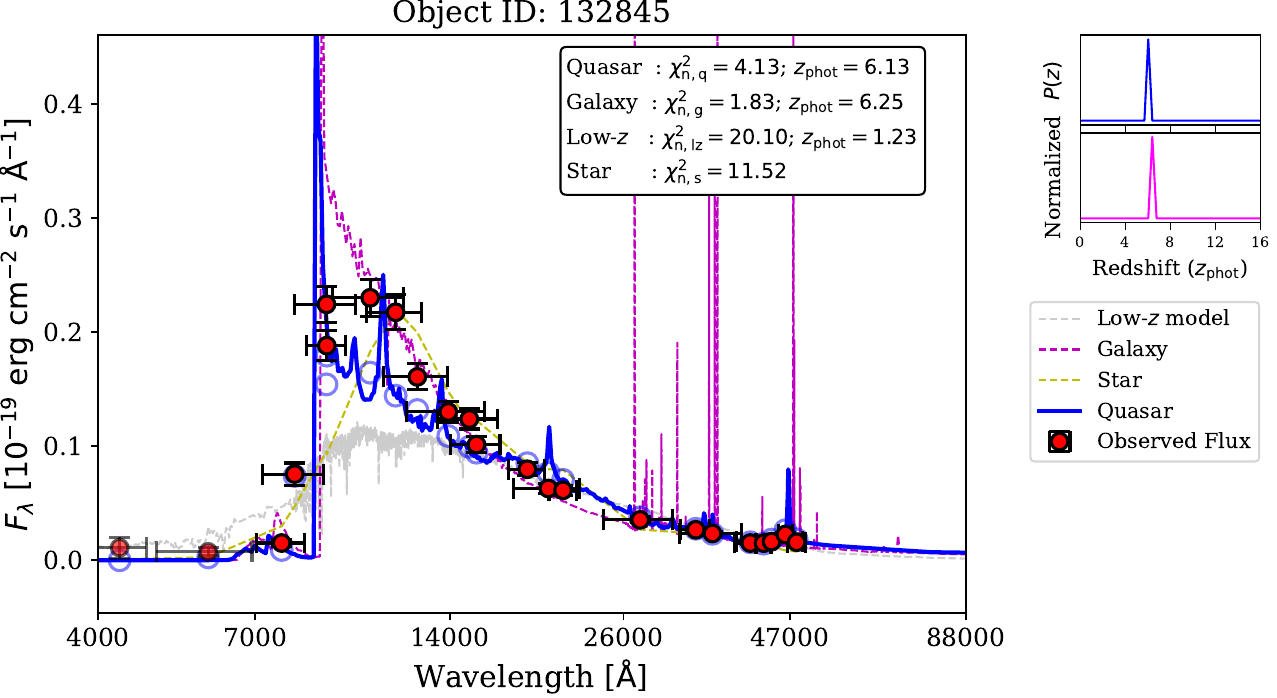}}
	\hspace{10pt}
	\raisebox{0.128\height}{\includegraphics[width=0.4\textwidth]{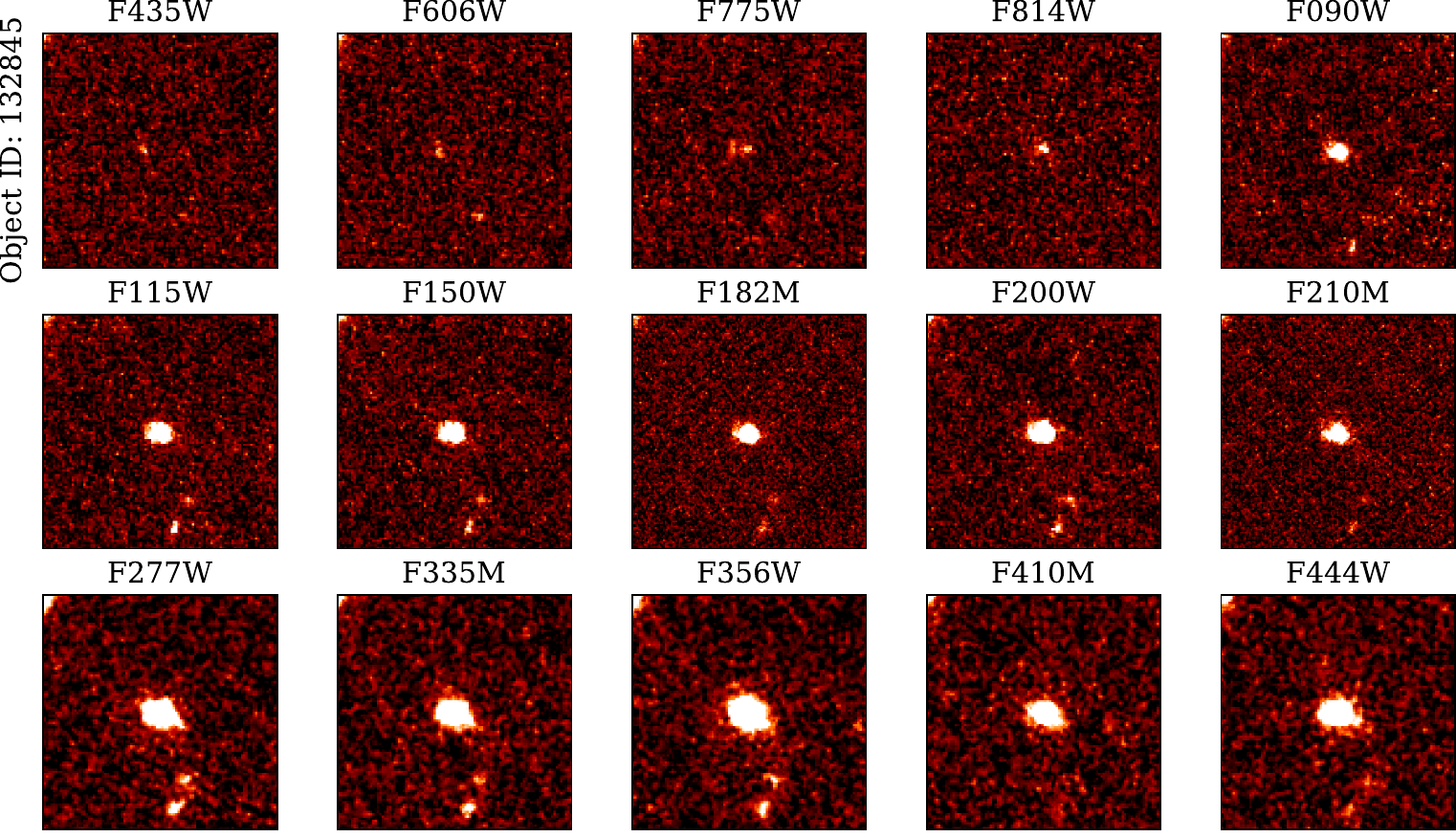}}
	\raisebox{0.03\height}{\includegraphics[width=0.5\textwidth]{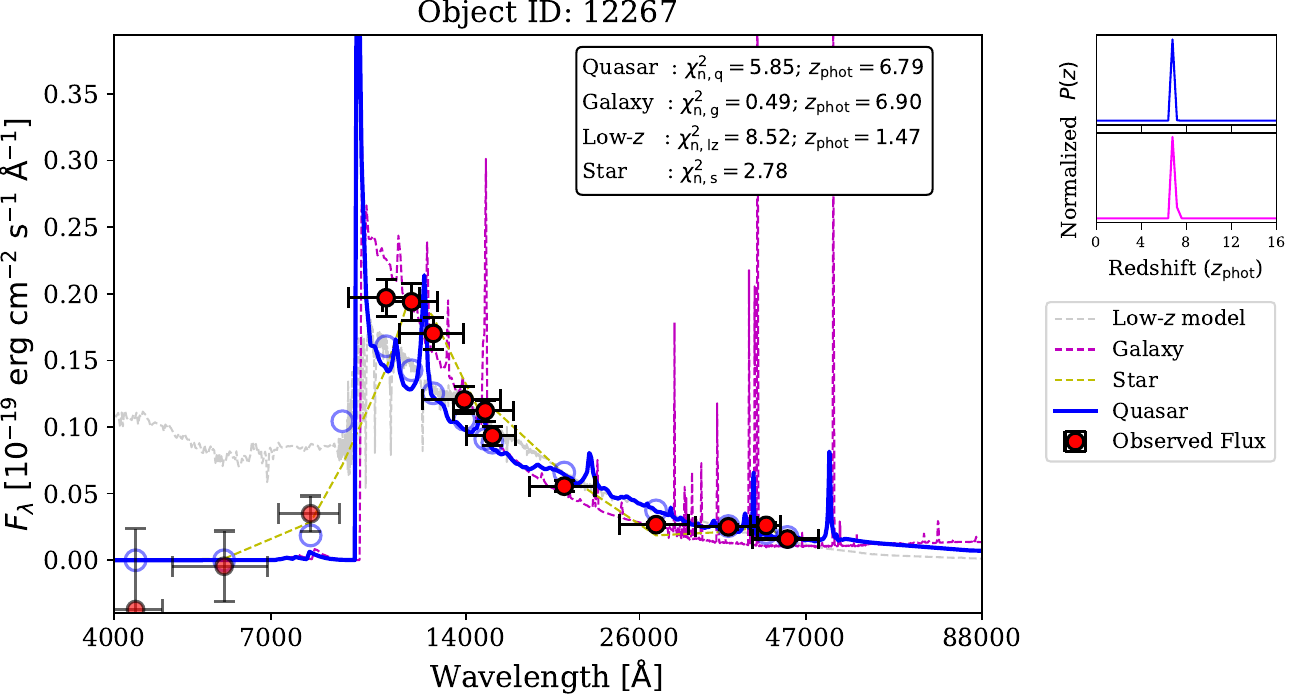}}
	\hspace{10pt}
	\raisebox{0.128\height}{\includegraphics[width=0.4\textwidth]{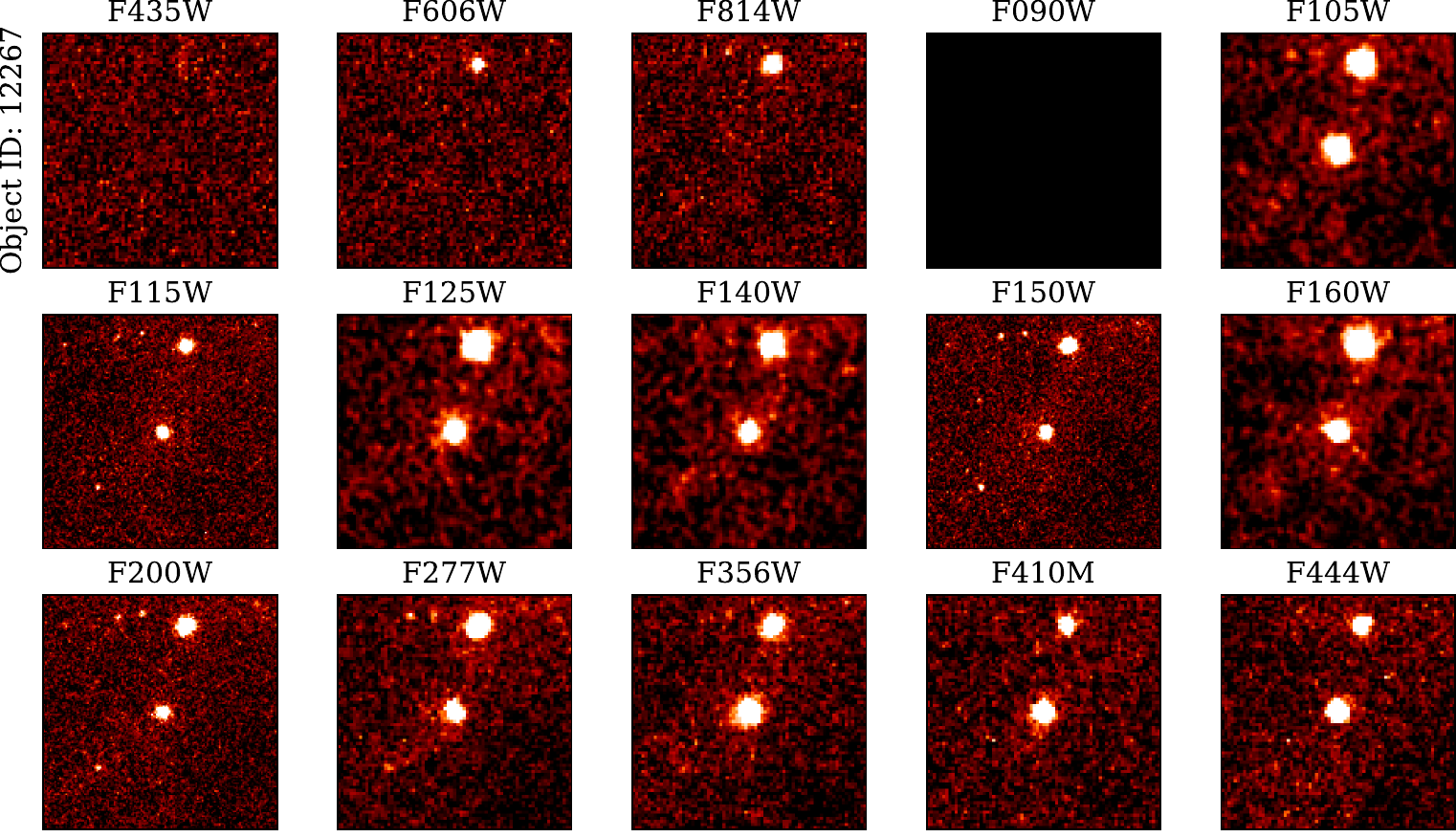}}
	\caption{
		Photometric SEDs of some quasar candidates.
		In the left part of each panel, we model the observed photometric data points with four types of spectral templates.
		Fluxes with S/N > 3 are marked with red circles, while those with lower values are shown with a bit transparent color.
		The best-fit quasar template and its associated synthetic photometry are presented with blue lines and circles.
		Models based on the galaxy and star/brown dwarf spectra are shown with magenta and yellow lines, with an additional fitted model of low-$z$ sources displayed in gray color.
		The goodness-of-fit of each model and the estimated redshift are also reported (see main text for a detailed definition).
		We further indicate the calculated redshift probability distribution function, $P(z)$ for quasar and galaxy models.
		Finally, the right part of each panel shows the multiband images of the $z\gtrsim6$ quasar candidate, each trimmed to 6\arcsec\ size on a side.
	}
	\label{fig:sedfit}
\end{figure*}
\clearpage

The set of main sequence stellar SEDs is taken directly from the PHOENIX stellar library, encompassing a wide range of spectral types, luminosities, and effective temperatures \citep{2013A&A...553A...6H}.
In line with that, brown dwarf spectra are obtained from the Sonora models, covering diverse properties of self-luminous extrasolar planets along with type L, T, and Y brown dwarfs \citep{2021ApJ...920...85M}.
After compiling the required SED models, we define a redshift grid of $0.05 \leq z \leq 20$ with $\Delta z = 0.05$ step and distribute the quasar and galaxy templates accordingly.
Furthermore, we create one additional grid for the galaxy template, forcing the redshifts to be $z \leq 5.5$ to ensure that our candidates are distinct from the low-$z$ sources.
After that, depending on the $z$ values, intergalactic medium (IGM) attenuation is applied following the analytical equation proposed by \cite{2014MNRAS.442.1805I}.
In contrast, we set the redshifts to be close to zero for the star and brown dwarf models.

Each quasar candidate will be modeled with four classes of templates: quasar, galaxy, star, and low-$z$ source.
The likelihood of the source being a quasar is subsequently determined by comparing its $\chi^2$ divided by the number of bands employed in the SED fitting (hereafter $\chi^2_\mathrm{n}$).
To be exact, we define the goodness-of-fit for quasar, galaxy, star, and low-z source model as $\chi^2_\mathrm{n,\,q}$, $\chi^2_\mathrm{n,\,g}$, $\chi^2_\mathrm{n,\,s}$, and $\chi^2_\mathrm{n,\,lz}$, respectively, and compare their values.
Examples of the resulting SED fit and the image cutouts are displayed in Figure~\ref{fig:sedfit}.
A preliminary quasar selection is then performed utilizing the criteria as follows:
\begin{enumerate}
	\item Detections in four NIRCam bands (F115W, F150W, F277W, and F444W) with more than 5$\sigma$. These bands cover the region redward of the expected Ly$\alpha$ emission at $z\gtrsim6$.
	\item S/N < 3 in the optical bands blueward of the anticipated Ly$\alpha$ break. Precisely, we use both the HST/ACS F435W and F606W bands for the JADES, UNCOVER, CEERS, and PRIMER datasets, while Subaru/HSC $g$ and $r$ filters are utilized for the COSMOS-Web sources.
	\item The best-fit model for the observed SED is not a star but either a galaxy or a quasar with the inferred $\chi^2_\mathrm{n,\,g}$ and $\chi^2_\mathrm{n,\,q}$ values being < 10. Here, we do not require the candidates to be best fitted by a pure quasar model since their host galaxy emission could dominate the observed SEDs in the lower luminosity regimes, as found in most of JWST-confirmed, $z\gtrsim5$ faint AGNs to date \citep[see, for example,][]{2023arXiv230311946H,2023arXiv230801230M,2023arXiv230905714G}.
	\item The source is located at high-$z$, indicated by the estimated photometric redshift being $z_\mathrm{phot} > 6$, both for the galaxy and quasar models.
	\item The integrated redshift probability at $z > 5.5$ should be more than 90\%, that is, $P(z>5.5) > 0.9$.
\end{enumerate}

Of all sources identified in the combined catalogs, 1370 targets pass our initial criteria and are then visually inspected.
More than half of these candidates are cosmic rays, hot pixels, stray light residues, contaminated by nearby bright sources, moving objects, or other detector artifacts.
During the visual inspection stage, we also discard candidates with extended morphologies, as they could be low-$z$ dusty sources not visible in the ground-based and HST imaging.
Sources with compact shapes with circularized diameter less than 0\farcs5 are preferred because their light is likely dominated by a centralized emission component around the galactic nucleus.

To identify sources that have already been spectroscopically confirmed and published in the literature, we cross-match our candidates to the DJA's JWST sources repository\footnote{\url{https://dawn-cph.github.io/dja/general/jwst-sources}} and the SIMBAD Astronomical Database\footnote{\url{http://simbad.cds.unistra.fr/simbad}} \citep{2000A&AS..143....9W}.
Correspondingly, the current datasets that we have contain at least 36 confirmed broad-line AGNs at $4 \lesssim z \lesssim9$ reported in the literature, for which 11 of them are located at $z\gtrsim6$ \citep{2023arXiv230311946H,2023arXiv230801230M,2023ApJ...953L..29L,2023ApJ...954L...4K,2023A&A...677A.145U,2023arXiv230811610K,2023ApJ...953..180S}.
Our $z_\mathrm{phot}$ estimates for these sources agree with the redshifts derived via spectroscopy considering the estimated uncertainties, indicating a good performance of our SED fitting with \texttt{eazy-py}.
As mentioned before, most of these sources ($\approx$80\%) prefer best-fit SEDs based on the galaxy spectral templates, given the substantial brightness of their host galaxy emission, while the remaining objects opt for the pure quasar models.
Also, these confirmed AGNs often show compact shapes consistent with our selection criteria.
After discarding spurious sources and already published high-$z$ galaxies in the literature, our final selection yields 350 remaining quasar/galaxy candidates (see Table~\ref{tab:preselection}).

\begin{figure}[htb!]
	\centering
	\resizebox{\hsize}{!}{\includegraphics{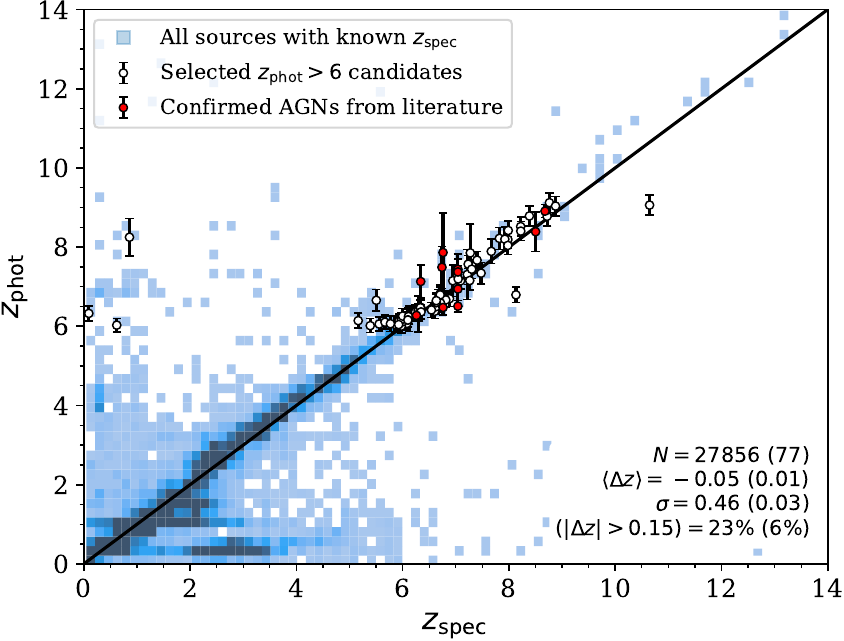}}
	\caption{
		Comparison between $z_\mathrm{phot}$ and $z_\mathrm{spec}$. 
		The number count ($N$), average bias ($|\Delta z|$), scatter ($\sigma$), and outlier fraction ($|\Delta z| > 0.15$) of all sources (blue squares) with available spectroscopic data are reported.
		The region with darker colors corresponds to a higher number of sources within the 2D histogram bins.
		We also show the metrics for a subset that satisfies our high-$z$ selection criteria (white circles with error bars).
		Samples of spectroscopically confirmed AGNs from the literature are depicted with red circles.
	}
	\label{fig:zphot}
\end{figure}

It is noteworthy to mention that the redshift calculated via broadband photometry can exhibit a systematic deviation from the one based on spectroscopy, which we refer to as a systematic offset bias \citep[e.g.,][]{2013MNRAS.432.1483C,2020arXiv200301511N}.
This bias is quantified as $\Delta z = (z_\mathrm{phot} - z_\mathrm{spec})/(1+z_\mathrm{spec})$.
For a subset of 27,856 confirmed AGN/galaxies with available spectroscopic data from the DJA's JWST sources repository, on which we applied our SED modeling, we find that the average bias is $\langle \Delta z \rangle = -0.05$ while its standard deviation is $\sigma = 0.46$. 
The outlier fraction, defined as the fraction of sources with $|\Delta z| > 0.15$, is 23\%.
When we focus on the subset that meets the high-$z$ source criteria outlined in the preceding paragraphs, the corresponding statistics shift to $\langle \Delta z \rangle = 0.01$, $\sigma = 0.03$, and an outlier fraction of 6\%.
While there is a noticeable scatter in the accuracy of $z_\mathrm{phot}$, these results are already sufficient to distinguish between low- and high-$z$ sources, with contamination rates of roughly 5\%--25\% (see Figure~\ref{fig:zphot}).

\clearpage
\begin{figure*}[htb!]
	\centering
	\includegraphics[width=0.46\textwidth]{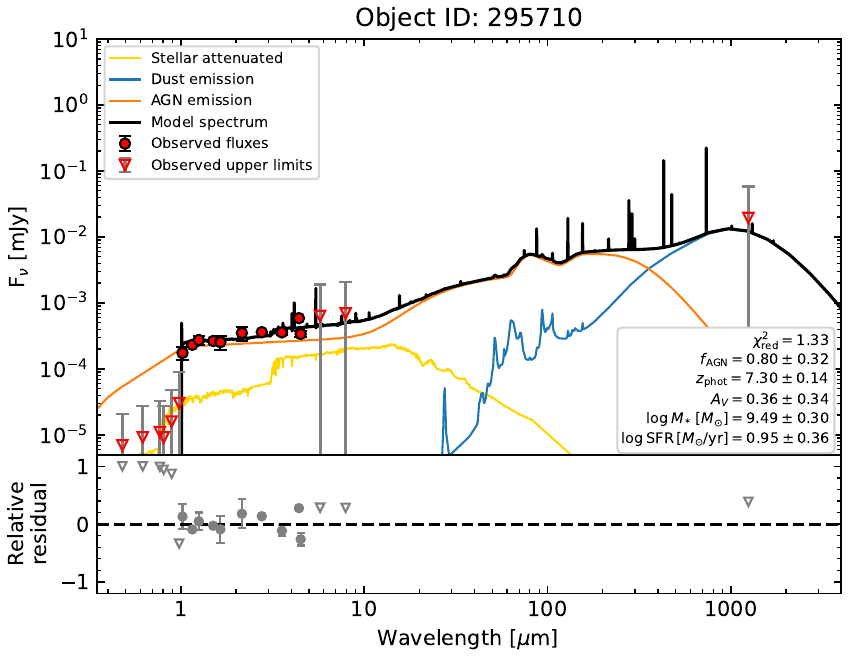}
	\hspace{5pt}
	\includegraphics[width=0.46\textwidth]{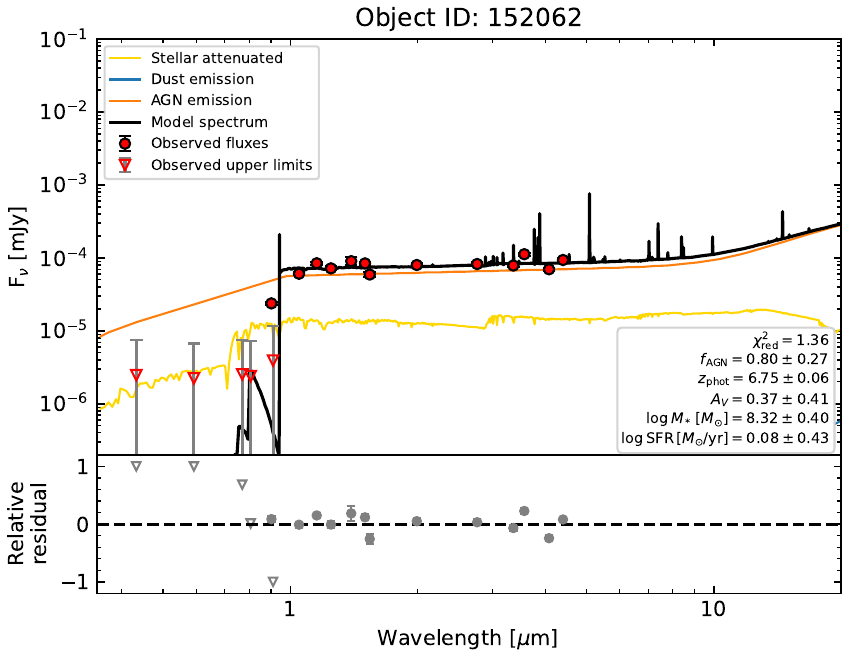}
	\newline\newline
	\includegraphics[width=0.46\textwidth]{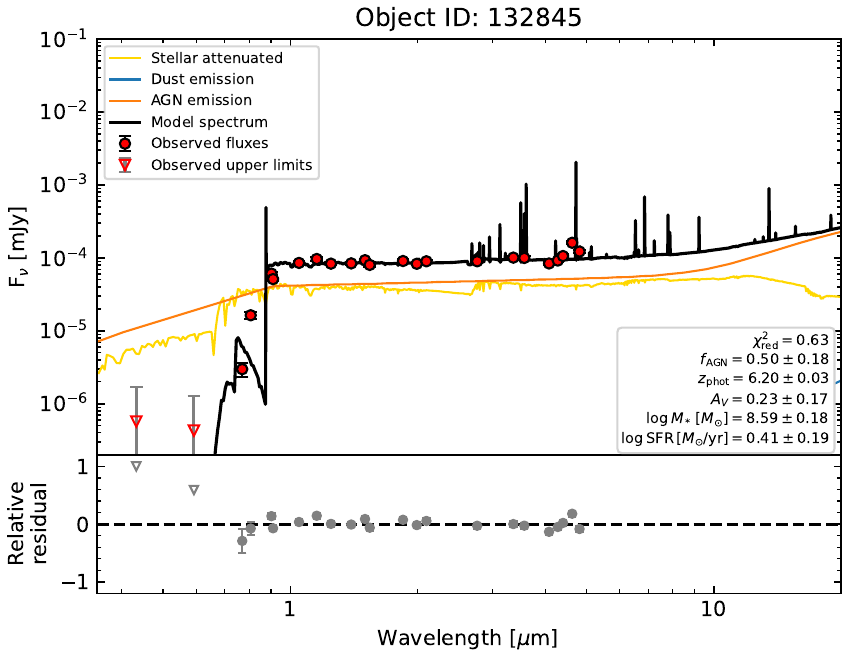}
	\hspace{5pt}
	\includegraphics[width=0.46\textwidth]{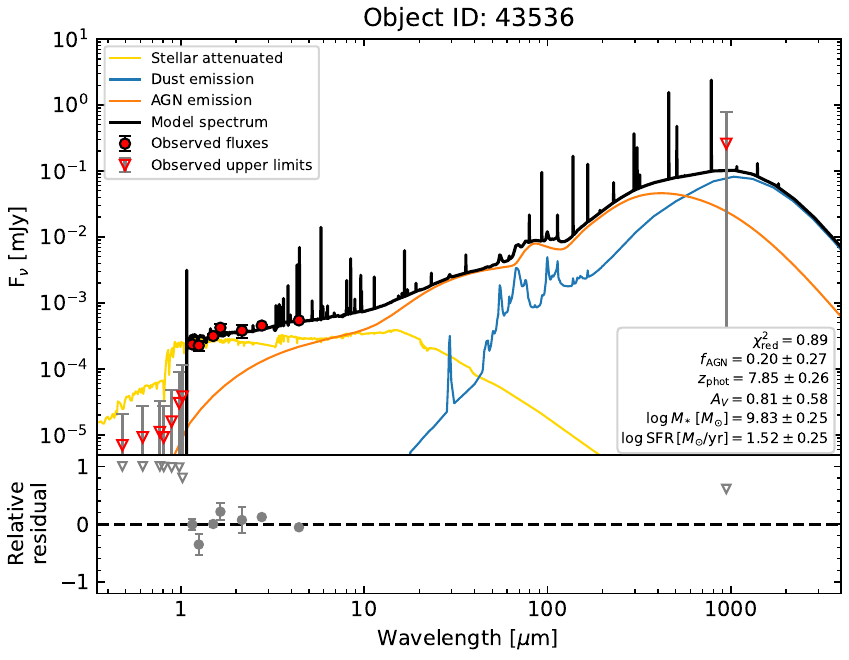}
	\newline\newline
	\includegraphics[width=0.46\textwidth]{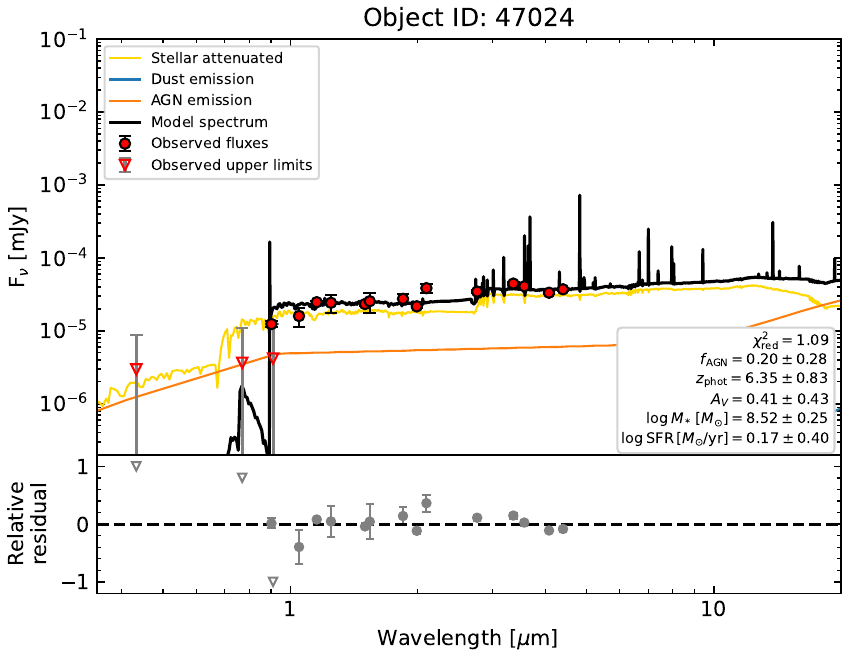}
	\hspace{5pt}
	\includegraphics[width=0.46\textwidth]{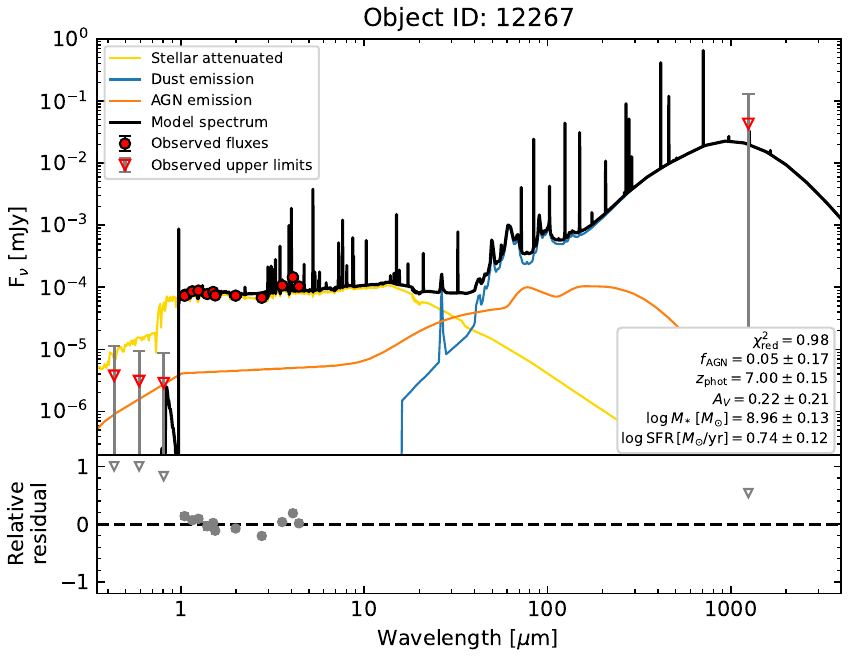}
	\caption{
		Examples of SED fitting with \texttt{CIGALE} with AGN plus galaxy components.
		Observed and upper limit fluxes are shown in the upper part of each panel with red dots and triangles with error bars, respectively.
		The total model spectrum (black) in the observed-frame wavelengths, corrected for the IGM attenuation, is composed of stellar (yellow), dust, and AGN (orange) emissions.
		These decomposed components are shown without adding the IGM absorption model.
		We also report the reduced chi-square value ($\chi^2_\mathrm{red}$), fraction of AGN flux to the total emission ($f_\mathrm{AGN}$) within rest-wavelengths of 0.1--0.7~$\mu$m, photometric redshift ($z_\mathrm{phot}$), dust extinction coefficient ($A_V$), host galaxy stellar mass ($\MStar$), and star formation rate (SFR).
		The lower part of each panel displays the relative residual between the data and the model.
		Sources that are better modeled with no AGN contribution have $f_\mathrm{AGN} \leq 0.05$, while the ones selected as quasar candidates exhibit $f_\mathrm{AGN} \geq 0.2$.
	}
	\label{fig:cigalefit}
\end{figure*}
\clearpage

\subsection{Measurements of the galaxy properties} \label{sec:cigalefit}

After robustly identifying a sample of high-$z$ galaxy and AGN candidates, which should have few interlopers or spurious members, we carry out complementary SED modeling to extract galaxy and AGN parameters from the broad-band SEDs and will also robustly estimate AGN contributions in this sample.
We treat this high-confidence candidate sample as a sample of actual high-$z$ galaxies with variations in AGN contribution between 0\% and 100\%. 
We discuss the validity of this approach in the following sections.

We model the SEDs using the Code Investigating GALaxy Emission \citep[\texttt{CIGALE};][]{2019A&A...622A.103B,2020MNRAS.491..740Y,2022ApJ...927..192Y} package.
Following the default configuration as a reference, we consider a delayed star formation history (SFH) with an e-folding time range of $0.1\leq\tau\leq5$~Gyr and a recent burst, assuming \cite{2003MNRAS.344.1000B} stellar population models along with a \cite{2003PASP..115..763C} initial mass function (IMF) and stellar metallicity of $Z = 0.02$.
Next, nebular emission is approximated using the \cite{2011MNRAS.415.2920I} model, while the dust extinction is added utilizing the combined \cite{2000ApJ...533..682C} and \cite{2002ApJS..140..303L} attenuation laws, dubbed as the modified starburst module in the \texttt{CIGALE} setup.
We set the $E(B-V)$ color excess for both nebular lines and stellar continuum to be between 0.05 and 2.65, which is equivalent to dust attenuation levels of $A_V\approx0.2$--8.2, assuming a ratio of total-to-selective extinction of $R_V = A_V / E(B-V) = 3.1$.
This wide range of attenuation levels is chosen since $z\lesssim5$ dust-enshrouded star-forming galaxies could appear as if they were sources at extremely high redshifts \citep{2023ApJ...943L...9Z,2023arXiv231020675M}.

The ionization parameter, gas metallicity, and electron density of nebular lines are fixed to $\log U = -2$, $Z_\mathrm{gas} = 0.02$, and $n_e=100$~cm$^{-3}$, respectively.
We acknowledge that opting for this choice could introduce an additional uncertainty of up to 5\% on the inferred AGN-to-host galaxy flux ratio, along with $\sim$0.1~dex in the measurements of stellar mass. 
However, its impact on the accuracy of photometric redshift estimations is observed to be minimal.
We also note that $U$, $Z_\mathrm{gas}$, and $n_e$ display higher sensitivity to altering the emission line strengths and lower sensitivity to modifying the continuum shape, indicating broadband photometry data alone, as we used here, would not be enough to constrain them well \citep{2017MNRAS.465.3220K,2019ARA&A..57..511K}.
Given the considerations, the introduced tradeoff is acceptable for achieving a simpler model with significantly faster computation times.

\begin{figure}[htb!]
	\centering
	\resizebox{\hsize}{!}{\includegraphics{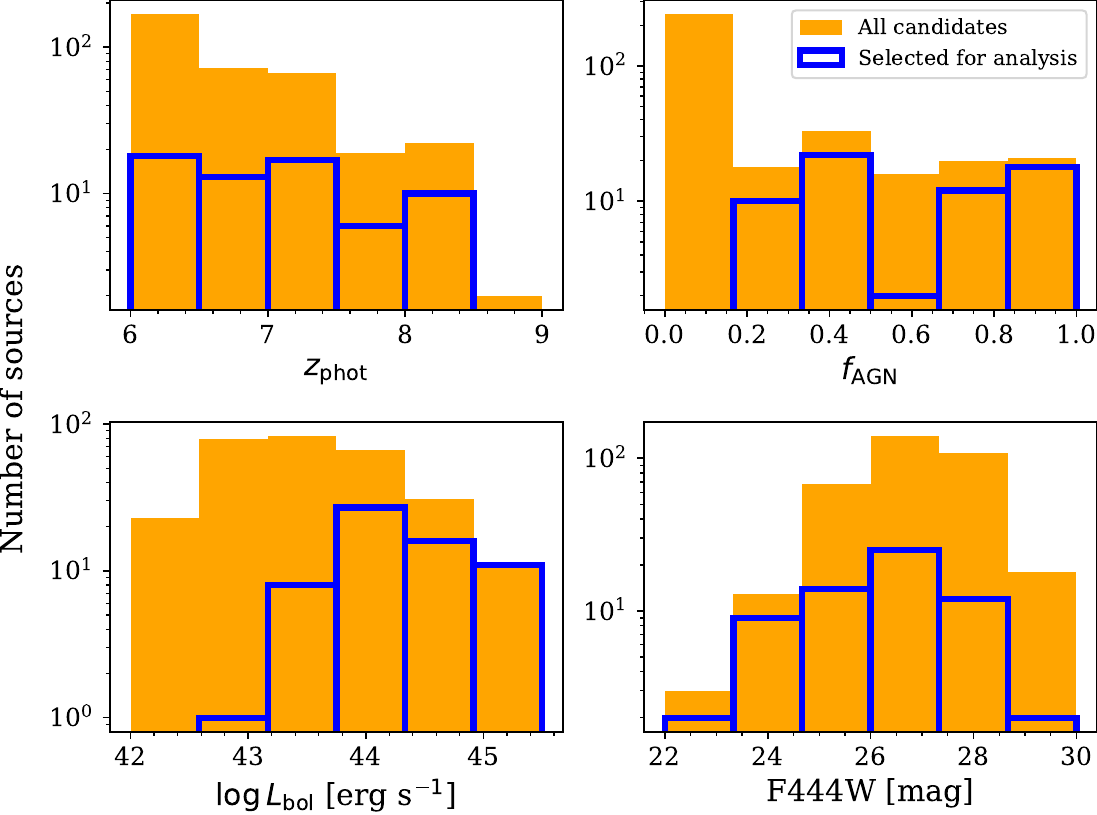}}
	\caption{
		Distributions of the photometric redshift ($z_\mathrm{phot}$), fraction of AGN emission ($f_\mathrm{AGN}$) within the rest-wavelengths of 0.1--0.7~$\mu$m, AGN bolometric luminosity ($L_\mathrm{bol}$), and F444W magnitude of the quasar candidates selected in this work.
		All candidates are shown with the orange histogram, while a subset with $f_\mathrm{AGN}\geq0.2$ chosen for further analysis later in Section~\ref{sec:bh_growth} is colored with blue.
	}
	\label{fig:hist}
\end{figure}

Since we are also interested in assessing how much the AGN emission contributes to the observed total fluxes, we make use of the \texttt{Skirtor2016} model provided by \texttt{CIGALE} on top of the previous SED sets \citep{2012MNRAS.420.2756S,2016MNRAS.458.2288S}. 
This 3D radiative transfer AGN model includes the accretion disk emission on the UV/optical side and the torus plus polar dust emission at infrared (IR) wavelengths.
In addition, the adopted AGN inclination angle could affect the resulting AGN class, namely, obscured or unobscured.
Accordingly, we set this as a free parameter to cover both types.
Following that, the IGM attenuation effect is appended as a function of redshift following the formula from \cite{2006MNRAS.365..807M}.
Finally, the SED models consisting of the galaxy and AGN components are fitted within redshift bins of $0.05 \leq z \leq 16$ using a step size of $\Delta z = 0.05$.
The \texttt{CIGALE} input file will be provided as supplementary data with this paper for reader reference and accessibility.
We refer to the \texttt{CIGALE} documentation\footnote{\url{https://cigale.lam.fr}} for detailed information on all the spectral templates adopted here \citep{2019A&A...622A.103B,2022ApJ...927..192Y}.

Examples of the best-fit SED model made with \texttt{CIGALE} are portrayed in Figure~\ref{fig:cigalefit}.
Correspondingly, the current SED modeling yields posterior distributions of some physical parameters (see Figure~\ref{fig:hist}), such as the AGN fraction of the total emission ($f_\mathrm{AGN}$), host galaxy stellar mass ($M_*$), and SFR averaged over 100 Myr, along with $z_\mathrm{phot}$ and $A_V$.
It should be emphasized that $f_\mathrm{AGN}$ is calculated considering only the rest-frame wavelengths from 0.1 to 0.7~$\mu$m, which is the region constrained by our ground and space-based data while excluding ALMA submillimeter measurements.
Furthermore, this wavelength range covers essential broad emission lines in the quasar SED, such as Ly$\alpha$, H$\beta$, and H$\alpha$.
More details on the generated \texttt{CIGALE} output parameters are discussed in \citep{2019A&A...622A.103B}.
Discussion on the inferred physical characteristics of our quasar candidates will be showcased in the next section.

\section{Results and discussion} \label{sec:result}

\subsection{List of quasar candidates and their number density}

Up to this stage, we have selected 350 candidates of high-$z$ compact sources via our initial photometric cut, visual inspection, and advanced SED modeling with two independent codes.
There will be unresolvable mismatches between observed SEDs and template inputs used for both modeling methods. 
Hence, there is space for nominal AGN components formally compensating for such template mismatch, even for fully nuclear-passive galaxies. 
For the subsequent analysis, we will use a threshold in formal AGN fraction to mitigate this. 

We will only consider candidates with $f_\mathrm{AGN} \geq 0.2$ to ensure the presence of actual AGN contribution to the observed emission.
This threshold level is motivated by a comparative analysis of properties between active and inactive galaxies compiled from the literature, as elaborated in Appendix~\ref{sec:agnvgal}. 
Overall, we anticipate that this cutoff will yield a completeness of approximately 80\% in AGN selection, accompanied by a contamination rate as high as 30\% from normal high-$z$ galaxies.
We further impose a black hole mass ($\MBH$) limit criterion, where $\MBH > 10^5~\MSun$ since confirming the quasar nature below this limit is challenging for numerous reasons.
For instance, the bright host galaxy emission might dilute the quasar light, making the quasar signature hidden from the observers in the optical to NIR regimes \citep[e.g.,][]{2022PASJ...74..689F}.
Furthermore, given the limitation of current observing facilities and the fact that these less massive quasars might only be capable of exhibiting H$\alpha$ with a line width of $\lesssim 100$~km~s$^{-1}$, they will be hard to differentiate from the low-velocity outflows or the narrow-line emissions of their host galaxies \citep{2023arXiv230512492M}.
Details on $\MBH$ estimation will be discussed later in Section~\ref{sec:bh_dist}, but in the end, 64 sources passed these AGN fraction and mass limit criteria of the 350 parent candidates.

As a further note, out of the 11 previously confirmed AGNs at redshifts $z\gtrsim6$ reported by other studies \citep[i.e.,][]{2023arXiv230311946H, 2023arXiv230801230M, 2023ApJ...953L..29L, 2023ApJ...954L...4K, 2023A&A...677A.145U, 2023arXiv230811610K, 2023ApJ...953..180S}, 9 sources met our selection criteria (see Table\ref{tab:qso}). 
These known AGNs have intentionally been excluded from the final sample of the 64 quasar candidates presented here.
These selected sources -- that is, our final quasar candidates -- are then marked as grade~A while the unselected ones are labeled with grade~B.
All of our candidates are listed in Table~\ref{tab:qcand} of Appendix~\ref{sec:qcand}, which contains information on their coordinates, photometry, and derived properties.
Due to the file size constraints, the full table and figures containing the SED fitting results of each source will be exclusively available for online access.

The sky coverages of each survey in the current datasets, for illustration, are approximately 0.28~$\deg^2$, 57~arcmin$^2$, and 49~arcmin$^2$ for COSMOS-Web, JADES, and UNCOVER, respectively (see Table~\ref{tab:preselection}).
Consequently, within the COSMOS-Web field and adopting the luminosity function of \cite{2023arXiv230311946H}, we expect to find around 18 quasars at $z=6$--8 having the UV absolute magnitudes of $M_\mathrm{UV} \lesssim -21$.
With their deeper imaging, JADES and UNCOVER might recover about 12 and 23 sources brighter than $M_\mathrm{UV} \approx -19$, respectively.
Thus, the number of quasar candidates we found seems reasonable since it is within the appropriate range of the empirical predictions.
We note that the luminosity function of \cite{2023arXiv230311946H} is derived based on the recent census of $z\approx4$--7 low-luminosity AGNs ($-18.5 \gtrsim M_\mathrm{UV} \gtrsim -21.5$) detected with the JWST observations.
In contrast, if we take and extrapolate the models from \cite{2018ApJ...869..150M} or \cite{2023ApJ...943...67S} into the fainter regimes, for which they were anchored initially to the bright ($M_\mathrm{UV} \lesssim -22$), unobscured quasar population at $z\sim6$, we anticipate finding only one source in each field.

\begin{figure}[htb!]
	\centering
	\resizebox{\hsize}{!}{\includegraphics{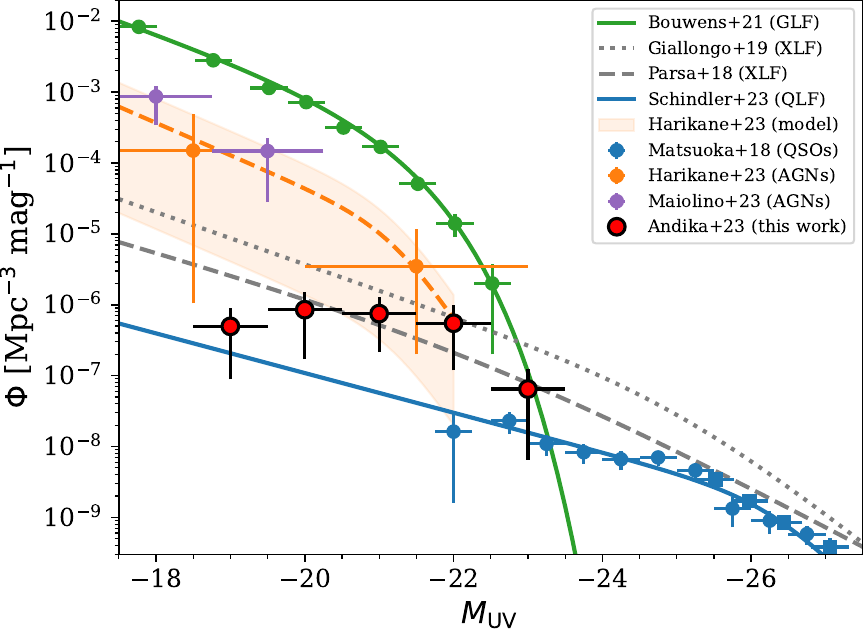}}
	\caption{
		UV luminosity functions of sources at $z\sim6$.
		The number densities of our quasar candidates as a function of UV absolute magnitude are marked with red circles with error bars.
		Blue circles and squares represent the data from \cite{2018ApJ...869..150M} and \cite{2023ApJ...943...67S}, where their associated best-fit quasar luminosity function (QLF) is shown with a blue line.
		The fitted model and observed galaxy luminosity function (GLF) from \cite{2021AJ....162...47B} are displayed with a green line and circles, respectively.
		A sample of JWST-confirmed AGNs from \cite{2023arXiv230311946H} is designated with orange circles, and their double power-law model, along with its uncertainty, is portrayed with an orange dashed line and shaded region.
		We also show the AGN luminosity function at $z=4$--6 reported by \cite{2023arXiv230801230M} with purple circles for comparison.
		The number density of our quasar candidates is higher than the extrapolation of the bright QLF. 
		Nevertheless, it is consistent with the X-ray selected AGN luminosity function (XLF) from \cite{2018MNRAS.474.2904P} and \cite{2019ApJ...884...19G}, which are denoted as dashed and dotted gray lines.
	}
	\label{fig:qlf}
\end{figure}

We present the number density of our quasar candidates in Table~\ref{tab:qlf} and Figure~\ref{fig:qlf}.
Here, we consider a redshift range of $z=6.0$--8.4, and the total solid angle covered by our datasets is around 0.45~$\deg^2$, which corresponds to a survey volume of approximately $8.2\times10^6$~Mpc$^3$.
The $M_\mathrm{UV}$ for our candidates are calculated from the flux observed at the rest-frame wavelength of 1500~\AA, derived based on our best-fit total SED model.
Hence, the reported $M_\mathrm{UV}$ accounts for the total emission from the quasar plus its host galaxy component.
Accordingly, to construct the UV luminosity function, we perform 10$^4$ Monte Carlo draws of our quasar candidates, incorporating their observed $M_\mathrm{UV}$ and $z_\mathrm{phot}$ along with their associated uncertainties. 
These random draws are necessary for instances where sources may fall outside the predefined redshift range or get counted in different magnitude bins across various iterations.
Given that the quasar count depends on the chosen $f_\mathrm{AGN}$ threshold, we also vary this criterion from $f_\mathrm{AGN}\geq0.2$ to 0.9 to take into account additional errors resulting from our selection method.
Our error estimation also accounts for the possible presence of low-$z$ interlopers and inactive galaxies with a contamination rate of up to 30\% (see, for example, Figure~\ref{fig:zphot} and Appendix~\ref{sec:agnvgal}).
We caution that the resulting number density estimation has not been adjusted for survey incompleteness. 

\begin{table}[htb!]
	\caption{Number density of our $6 \lesssim z \lesssim 8$ quasar candidates.}
	\label{tab:qlf}
	\centering
	\small
	\begin{tabular}{ccc}
		\hline\hline
		$M_\mathrm{UV}$ & $\Phi$ & $N$ \\
		$[$mag$]$ & [$10^{-7}$ Mpc$^{-3}$ mag$^{-1}$] & \\
		\hline
		$-23 \pm 0.5$ & $0.65 \pm 0.96$ & $1 \pm 1$  \\
		$-22 \pm 0.5$ & $5.49 \pm 4.31$ & $4 \pm 3$ \\
		$-21 \pm 0.5$ & $7.51 \pm 5.37$ & $6 \pm 4$ \\
		$-20 \pm 0.5$ & $8.52 \pm 6.80$ & $7 \pm 6$ \\
		$-19 \pm 0.5$ & $4.97 \pm 4.07$ & $4 \pm 3$ \\
		\hline
	\end{tabular}
	\tablefoot{
		The columns from left to right are: (1) the UV absolute magnitude bins, (2) the average and standard deviation of the number densities, and (3) the average number of objects obtained from the Monte Carlo draws.
		The reported numbers are not corrected for possible survey incompleteness.
	}
\end{table}

In general, the number density of our quasar candidates exceeds the extrapolated values of the brighter quasar population luminosity function by a factor of $\approx$10 \citep[e.g.,][]{2018ApJ...869..150M,2023ApJ...943...67S}, as shown by the blue line in Figure~\ref{fig:qlf}.
On the other hand, our numbers align with those reported by \cite{2023arXiv230311946H} to some extent; yet, densities at $M_\mathrm{UV} \gtrsim -19$ are largely uncertain given the source faintness and potential incompleteness in our quasar search method.
Interestingly, our samples are consistent with the faint, X-ray-selected AGN luminosity function presented by \cite{2018MNRAS.474.2904P} and \cite{2019ApJ...884...19G}.
The different nature of the bright and faint quasar populations might cause a large discrepancy between the luminosity functions mentioned earlier.
At the same time, many of these faint sources are just being detected with JWST, and it is likely that much remains to be revealed.
Below, we will discuss the constraints on the black hole and host galaxy characteristics of our quasar candidates.

\subsection{Black hole and host galaxy masses} \label{sec:bh_dist}

\begin{figure*}[htb!]
	\centering
	\resizebox{\hsize}{!}{\includegraphics{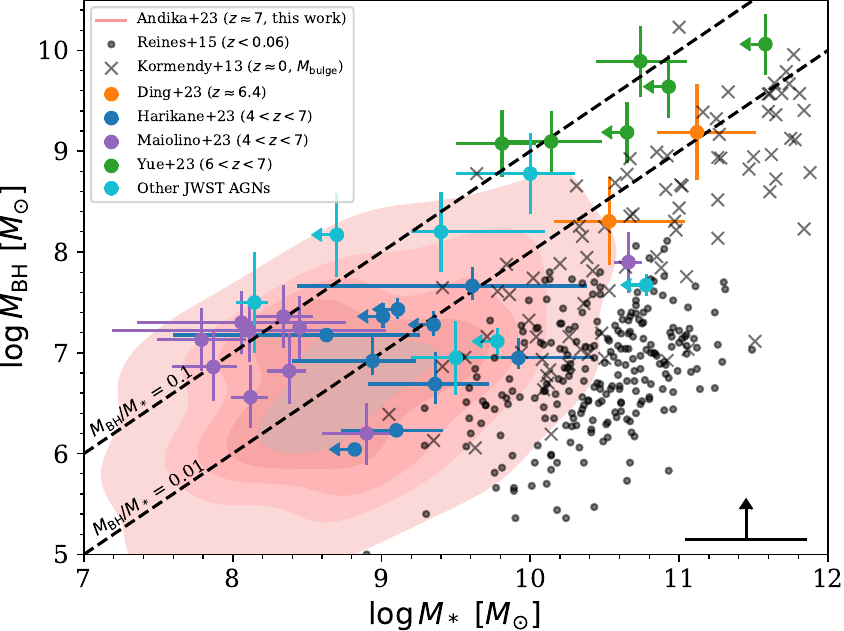}}
	\caption{
		Relation between the black hole mass ($\MBH$) and its host galaxy stellar mass ($\MStar$).
		The red contour represents our quasar candidates at $z\gtrsim6$, where our measurements can only provide lower limits for $M_\mathrm{BH}$, considering Eddington ratio values ranging from 0.1 to 1. 
		The typical statistical errors for $M_*$ are indicated in the lower right corner of the panel. 
		High-$z$ quasar samples with available JWST spectroscopic data from \cite{2023arXiv230311946H}, \cite{2023arXiv230904614Y}, \cite{2023Natur.621...51D}, and \citet[][excluding dual AGNs]{2023arXiv230801230M} are presented with blue, green, orange, and purple circles with error bars.
		Additional AGN samples from \cite{2023ApJ...953L..29L}, \cite{2023A&A...677A.145U}, \cite{2023ApJ...953..180S}, \cite{2023ApJ...954L...4K}, \cite{2023arXiv230811610K}, and \cite{2023ApJ...955L..24G} are indicated with cyan circles.
		The gray dots and crosses are nearby galaxies and AGNs from \cite{2013ARA&A..51..511K} and \cite{2015ApJ...813...82R}.
		The black dashed lines mark the limits where $M_\mathrm{BH}$/$M_*$ equals 0.1 and 0.01.
		Our candidates show a slightly higher $M_\mathrm{BH}$ to $M_*$ ratios than other galaxies at $z\sim0$ with consistent properties compared to high-$z$ low-luminosity quasars.
	}
	\label{fig:mbh_mstar}
\end{figure*}

The distribution of the central black hole mass to the host galaxy's stellar mass ratio -- that is, $\MBH$/$\MStar$ -- is a tracer of the supermassive black hole (SMBH) formation history \citep{2012Sci...337..544V}.
We want to again treat our high-probability quasar candidates as actual quasars and, under that assumption, infer black hole and stellar masses for them. 
To estimate $M_\mathrm{BH}$, we first adopt the canonical normalized accretion rate parameterized by the Eddington ratio, $f_\mathrm{Edd} \equiv \Lbol/\LEdd$, where $\Lbol$ and $\LEdd$ are the bolometric and Eddington luminosities, respectively \citep[e.g.,][]{2022ApJS..263...42W}. 
In this case, $\Lbol$ is calculated by multiplying a bolometric correction factor of 5.15 \citep{2006ApJS..166..470R} with the monochromatic luminosity at the rest-frame wavelength of 3000\,\AA --  that is, $L_{3000}=\lambda L_{\lambda}(3000\,\AA)$ -- derived based on our best-fit AGN SED model obtained in Section~\ref{sec:cigalefit}.
Then, we derive the lower limit $\MBH$ of our quasar candidates, assuming $f_\mathrm{Edd}=1$ and considering that Eddington luminosity can be approximated using:
\begin{equation}
	\label{eq:ledd}
	\LEdd = 1.3 \times 10^{38}\ (\MBH/\MSun)\,\mathrm{erg}\,\mathrm{s}^{-1}.
\end{equation}
As the values derived here represent lower limits, the true $\MBH$ could potentially be significantly higher. 
A comparison between our SED-based $\MBH$ and those determined through broad emission line spectroscopy reveals an actual $\MBH$ that is $\approx$1.6~dex higher, as demonstrated in Appendix~\ref{sec:agnvgal}.
The observed offset is anticipated, given the significant influence of $f_\mathrm{Edd}$ on our $\MBH$ estimates. 
Adjusting the assumed $f_\mathrm{Edd}$ to a much lower value, such as 0.1, results in a 1 dex increase in our data points, bringing them closer to spectroscopic $\MBH$ values.

The inferred $\MBH$/$\MStar$ distribution of our quasar candidates inferred from Equation~\ref{eq:ledd} and Section~\ref{sec:cigalefit} is displayed in Figure~\ref{fig:mbh_mstar}.
This distribution assumes that our quasar candidates may exhibit $f_\mathrm{Edd}$ values ranging from 0.1 to 1 and includes that uncertainty.
While our quasar candidates display a $\MStar$--$\MBH$ distribution slightly higher than that of galaxies at $z\sim0$ \citep[e.g.,][]{2013ARA&A..51..511K,2015ApJ...813...82R}, with properties consistent with observed samples of other high-$z$ low-luminosity AGNs \citep[e.g.,][]{2023arXiv230311946H,2023ApJ...954L...4K}, we emphasize that the derived $\MBH$ values represent lower limits.

Luminous quasars hosting massive black holes tend to reside within galaxies with larger stellar masses, the $\MBH$ to $\MStar$ ratios show a large diversity \citep[see, e.g.,][]{2020ARA&A..58...27I,2023ARA&A..61..373F}.
For instance, $z\gtrsim5$ bright quasars examined by \cite{2023arXiv230904614Y} display $\MBH/\MStar$ reaching as high as 10\%, which is significantly more prominent compared to the sources in the nearby Universe \citep[e.g.,][]{2013ARA&A..51..511K}.
On the other hand, less luminous objects, such as samples of $4 \lesssim z \lesssim 7$ AGNs from \cite{2023arXiv230311946H} are characterized by relatively lower $\MBH/\MStar \sim 1\%$.
To add further support of this diversity, \cite{2023ApJ...953L..29L} reported a presence of a broad-line AGN at $z=8.679$ exhibiting an $\MBH/\MStar \approx 0.3\%$, while, conversely, \cite{2023arXiv230805735F} presented an AGN at $z=7.045$ having  $\MBH/\MStar \gtrsim 3\%$.
Here, we need to note that for all samples, there are strong selection effects at play \citep[e.g.,][]{2022ApJ...931L..11L}, biasing against the ability to see low-luminosity AGN in bright galaxies. 
The exact impact will depend on the selection method but might imply limits by SED preselection, color-color cuts, emission-line strengths, or -- as for our approach -- a minimal required AGN fraction of the total flux. What all methods have in common is that they will preferentially find massive SMBHs.  
With that in mind, the comparison mentioned above implies that the growth of SMBHs at the upper envelope of these actually selected bright quasars may have preceded the star formation in their host galaxies \citep{2023arXiv230811610K,2023arXiv230512492M,2023ApJ...957L...3P}.

Whether the $\MBH$--$\MStar$ relation evolves with redshift is still a subject of debate.
For example, \cite{2018ApJ...867..148C} proposed an increasing SMBH to host mass ratio at higher redshifts, that is, $\MBH/\MStar \propto (1+z)^{1.5}$, which was inferred using an analytical approach to obtain the $\MBH$--$\MStar$ relation that fits the observed quasar luminosity function and SFR density \citep[see also][]{2024arXiv240104159P}.
On the other hand, considering various observable SMBH and host galaxy properties, including mass functions and quasar distributions, \cite{2023MNRAS.518.2123Z} demonstrated that there is no significant evolution of $\MBH$--$\MStar$ up to $z\sim10$ \citep[see also, for example,][]{2020ApJ...889...32S,2020ApJ...888...37D,2021ApJ...922..142L}.
In addition, in flux-limited surveys, quasars harboring overmassive black holes -- e.g., $\MBH/\MStar > 0.01$ -- could dominate the picked-up samples due to selection effects \citep{2007ApJ...670..249L}.
As seen in Figure~\ref{fig:mbh_mstar}, luminous quasars investigated by \cite{2023arXiv230904614Y}, \cite{2023A&A...677A.145U}, and \cite{2023ApJ...953..180S} lie way above the local $\MBH$--$\MStar$ relation, indicating a potential bias mentioned earlier.
This bias might occur because larger SMBH masses could produce higher quasar luminosities, which are more accessible to locate in flux-limited observations.


\subsection{Possible pathways for SMBH growth} \label{sec:bh_growth}

The significant diversity observed in the most distant SMBHs and their host galaxies might suggest a range of distinct growth histories and progenitors, which we will discuss further here.
While the exact seeding mechanisms remain elusive, it is generally accepted that early SMBHs might originate from at least two types of progenitors: (i) light seeds arising from the remnants of Population III stars having masses of $\approx$10--100~$\MSun$ and (ii) heavy seeds with a mass range of $10^4$--$10^6~\MSun$ produced by the collapse of primordial gas or dense star clusters \citep{2020ARA&A..58...27I}.

Here, we want to trace back the growth of our quasar candidates following the method presented by \cite{2022MNRAS.509.1885P}.
As the first step, we describe the connection between the initial seed mass $\MSeed$ and the accumulated black hole mass $\MBH$ at a specific cosmic time $t$ using the relation:
\begin{equation}
	\MBH(t) = \MSeed\ \exp\left(f_\mathrm{Edd}\ D\ \frac{1 - \epsilon}{\epsilon}\ \frac{\Delta t}{t_\mathrm{Edd}}\right).
	\label{eq:seed_growth}
\end{equation}
In this case, $t_\mathrm{Edd}$ is a mixture of constants where its typical value is $\approx$450~Myr \citep{2022MNRAS.509.1885P}, $f_\mathrm{Edd}$ is the average Eddington ratio across the accretion time interval of $\Delta t$, and $\epsilon$ is the mean radiative efficiency over the $\Delta t$.
The time interval is expressed as $\Delta t = t - t_\mathrm{seed}$, where $t_\mathrm{seed}$ corresponds to the black hole seeding epoch.
Assuming that a black hole seed is assembled at $z=25$, this would equal a cosmic time of $t_\mathrm{seed}\approx130$~Myr.
For an accretion mode following the thin disk model, $\epsilon$ ranges from 0.34 down to 0.057, depending on whether the central black hole is maximally rotating or nonrotating
\citep{2019arXiv191104305F,2020ApJ...895...95P,2020ApJ...903...85A}.
Then, the fraction of the quasar lifetime for which the accretion occurs is parametrized with the duty cycle $D$.
Unfortunately, $f_\mathrm{Edd}$ and $D$ are degenerate, meaning that one can obtain an identical value of $\MBH$ by combining different values of both growth parameters.
Due to this reason, we assume that the sources are actively accreting throughout their entire lifetime so that $D$ can be set to unity for simplicity.

We subsequently simulate the SMBH growth using a simple Monte Carlo strategy with 2000 realizations, exploiting Equation~\ref{eq:seed_growth} as the target function.
Starting from smaller seeds, we aim to match the masses of our quasar candidates that are, on average, within $\MBH = 10^5$--$10^8$~$\MSun$, at a median redshift of $z=6.7$.
Three essential parameters control our growth model, that is, $\MSeed$, $f_\mathrm{Edd}$, and $\epsilon$.
Correspondingly, we adopt a flat prior of $\log \MSeed \in [1, 6]~\MSun$, covering both the light and heavy seed mass regimes, and adjust the radiative efficiency in a physical range of $\epsilon \in [0.057, 0.34]$.
Considering that (i) many high-$z$ quasars detected so far are showing instantaneous accretion rates below or around the Eddington limit \citep[e.g.,][]{2023ARA&A..61..373F} and (ii) super- or hyper-Eddington accretion periods are typically short-lived ($\Delta t \sim 0.1 $~Myr), we adopt the Eddington ratio to be uniformly distributed within $f_\mathrm{Edd} \in [0, 1]$ \citep[e.g.,][]{2023arXiv230814986F}.
While $\MSeed$ is constant since the black hole is only seeded one time, $\epsilon$ and $f_\mathrm{Edd}$, on the other hand, may change over the period between the seed formation until it is detected at a later time.
Thus, these two parameters should be viewed as averages over the quasar lifetime.  

\begin{figure}[htb!]
	\centering
	\resizebox{\hsize}{!}{\includegraphics{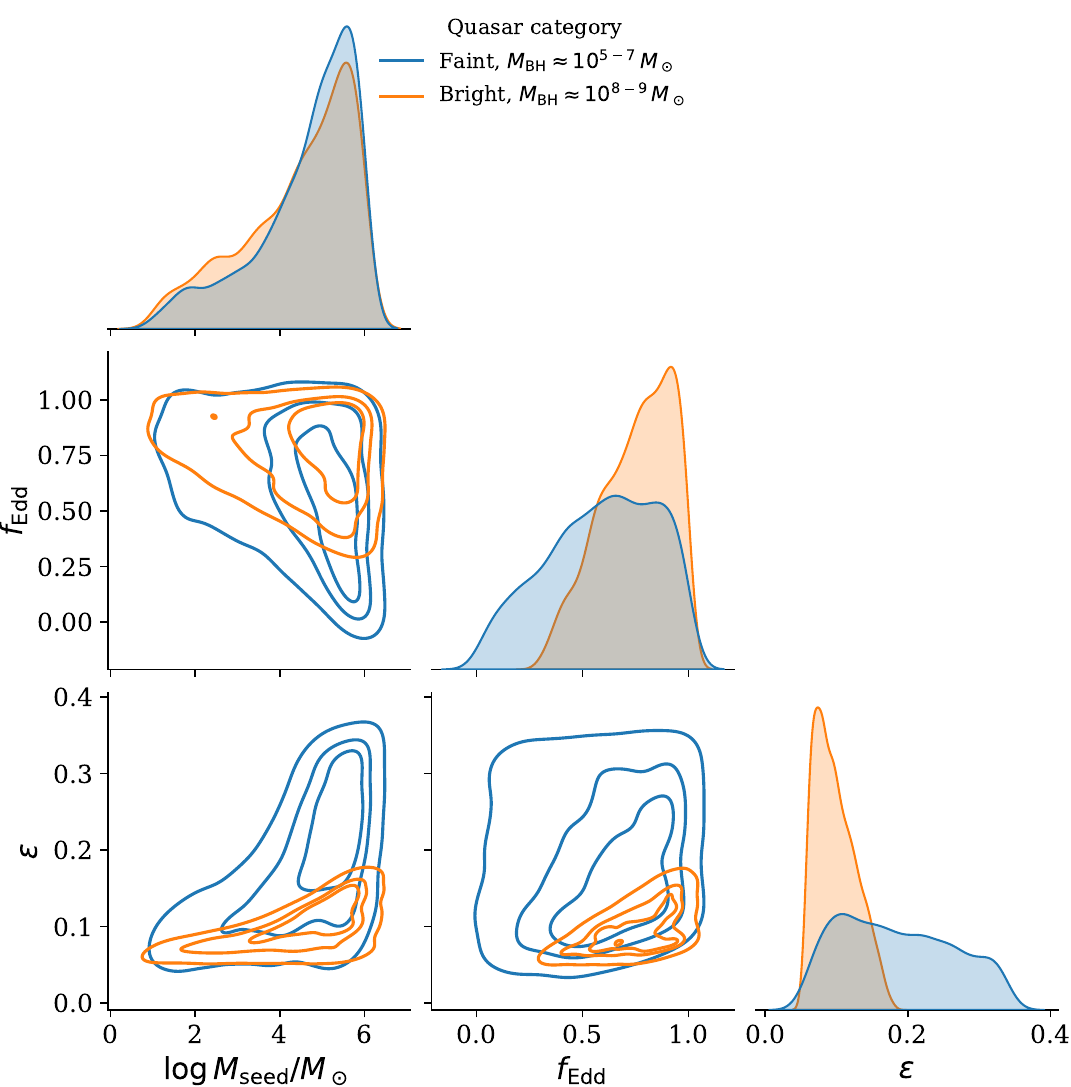}}
	\caption{
		Combination of $\MSeed$, $f_\mathrm{Edd}$, $\epsilon$ values that can produce the observed $\MBH$ of high-$z$ quasars.
		The parameter distributions of our quasar candidates and sources from \cite{2023ARA&A..61..373F} are depicted in blue and orange colors, respectively.
		Assuming a thin disk model and Eddington-limited accretion, larger seed masses with $\MSeed > 10^4~\MSun$ are the preferred channel for growing the SMBHs.
	}
	\label{fig:seed_mc}
\end{figure}

\begin{figure}[htb!]
	\centering
	\resizebox{\hsize}{!}{\includegraphics{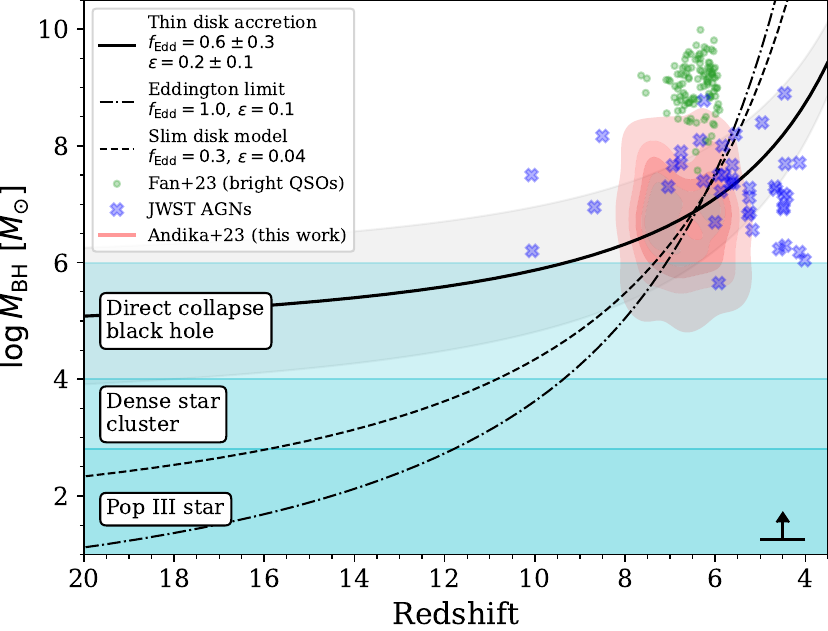}}
	\caption{
		Black hole mass growth as a function of redshift.
		The red contour represents the expected lower limit masses of the quasar candidates identified in this study.
		Additionally, the typical photometric redshift uncertainty for these candidates is illustrated in the lower right corner of the panel.
		Green circles depict the bright quasar samples from \cite{2023ARA&A..61..373F} while blue crosses display broad-line AGNs from the literature that have been observed with JWST spectroscopy (see main text).
		The cyan-shaded region shows the mass range of different progenitors.
		The solid black line and region show an evolutionary track along with its posterior distribution, assuming a thin disk accretion and heavy seed progenitors.
		On the other hand, cases where we use a thin disk accretion at the Eddington limit and a slim disk model to grow light seeds into SMBHs are shown with dash-dotted and dashed black lines.
	}
	\label{fig:mbh_growth}
\end{figure}

The combination of parameters that permits the formation of the central SMBHs we assume are residing in our quasar candidates is presented in Figure~\ref{fig:seed_mc}. 
At the same time, the associated growth track is provided in Figure~\ref{fig:mbh_growth}.
The majority of bright quasars from \cite{2023ARA&A..61..373F} occupy the region where $\MBH\gtrsim10^8~\MSun$ while our candidates, as well as some JWST-confirmed AGNs, reside in the lower mass side. \citep[e.g.,][]{2023ApJ...955L..24G,2023A&A...677A.145U,2023ApJ...953..180S,2023ApJ...953L..29L,2023ApJ...954L...4K,2023arXiv230311946H,2023arXiv230801230M,2023arXiv230905714G}.
Larger seeds with $\MSeed > 10^4~\MSun$ seem to be the preferred progenitor to develop these SMBHs by $z\approx7$.
In particular, most of our quasar candidates might have arisen from the black hole seeds as big as $\MSeed \sim 10^5~\MSun$, assuming the values of $f_\mathrm{Edd}=0.6\pm0.3$ and $\epsilon = 0.2\pm0.1$.
If short super-Eddington episodes occur during their evolution, the required progenitor mass could be lower, indicating that dense star cluster seeds could also be the ancestors of our quasar candidates.
Distinguishing between the formation through direct collapse black hole or dense star cluster channels is complicated, given the necessity of more precise measurements of the SMBH and host galaxy masses along with the gas metallicity, denoting that extra spectroscopic data are needed \citep{2023MNRAS.521..241V}.

After that, we run similar modeling as a comparison, but now targeting the bright quasars with $\MBH \approx 10^8$--$10^9~\MSun$ compiled by \cite{2023ARA&A..61..373F}.
As a result, this population also gives preference for heavy seeds with the Eddington ratio pushed higher to $f_\mathrm{Edd} = 0.78\pm0.17$ and radiative efficiency going down to $\epsilon = 0.09\pm0.03$.
We note that the parameter space occupied by this population is tighter than our less luminous quasar candidates, showing that detecting larger SMBHs at the farthest accessible distances could shrink the viable growth parameters and modes significantly.
Furthermore, our simple calculation confirms that as long as the radiative efficiency is at the lower end of the range accommodated by the thin disk model and the accretion is not dominated by super-Eddington episodes, it is less likely to yield SMBHs from the light seeds.

Maturing the light seeds in a short amount of time is complicated as this process would require the growth dominated with the Eddington-limited ($f_\mathrm{Edd}=1$) or even super-Eddington ($f_\mathrm{Edd}>1$) accretion to match the $z\approx6$--7 quasar mass distribution.
However, assembling such enormous masses and sustaining high accretion rates will be challenging, given the intricacy created by the enhanced stellar feedback from the host galaxies.
For example, a vast number of supernova explosions will happen during the rapid mass build-up, resulting in intense heating and mixing of the gas, making the accretion inefficient and more likely to be sub-Eddington \citep{2023ApJ...953L..29L}.
The only way to develop the light seeds into SMBHs is probably to adopt a hypothetical slim disk scenario, which lowers the radiative efficiency to $\epsilon = 0.04$ \citep{1988ApJ...332..646A,2000PASJ...52..499M,2015MNRAS.452.1922P,2015ApJ...804..148V}.
With just a mild accretion of $f_\mathrm{Edd} = 0.3$, for example, this channel could already produce $\gtrsim 10^6~\MSun$ black holes by $z\approx7$ as shown in Figure~\ref{fig:mbh_growth}.
Whether this mass accumulation channel dominates the high-$z$ quasar population is still debatable.
Therefore, further study to understand the typical accretion mode and the interplay between the growth parameters of early black holes is vital to constrain the evolution of these intriguing sources.

\section{Summary and conclusion} \label{sec:conclusion}

We have presented 350 candidates of compact galaxies, of which 64 show a high probability of being quasars at $z\gtrsim6$, selected by exploiting the rich multiband dataset of COSMOS-Web, as well as the JADES, UNCOVER, CEERS, and PRIMER projects.
These surveys consist primarily of JWST/NIRCam observations. 
The subsequent photometric catalog creation incorporated ancillary data from HST and other ground-based surveys.
Accordingly, our search strategy consists of two primary steps: photometric cut on catalog-level information and SED fitting to separate the quasars from other contaminating sources.
While the initial goals of the SED fitting are to classify and estimate the photometric redshift of each candidate, we also assess their associated physical properties under the assumption that they indeed are quasars, including the SMBH and host galaxy's stellar masses, as well as the fraction of AGN emission.

Our quasar candidates exhibit features consistent with the low-luminosity AGN population, where they potentially host less massive SMBHs with $\MBH \approx 10^5$--$10^8~\MSun$ residing in galaxies having $M_* \approx 10^8$--$10^{10}~\MSun$.
Furthermore, these sources display $\MBH$--$\MStar$ distribution that is slightly higher than those of galaxies at $z\sim0$ \citep[e.g.,][]{2013ARA&A..51..511K,2015ApJ...813...82R}, or in other words, their SMBHs tend to be overmassive compared to their hosts.
However, we stress that all quasars identified in these surveys are naturally biased to high $\MBH$/$\MStar$-ratios. 
This means they are not representative of the underlying population but preferentially form the upper envelope of the distribution.

With this in mind, we then run a simple Monte Carlo simulation to explain how these SMBHs accumulate their mass by the time they are detected.
Larger seeds from the direct collapse scenario, with $\MSeed > 10^4~\MSun$, appear to be the favored origins to develop these SMBHs by $z\sim7$.
Notably, most of our quasar candidates might have emerged from the black hole seeds as large as $\MSeed \sim 10^5~\MSun$, considering the values of $f_\mathrm{Edd}=0.6\pm0.3$ and $\epsilon = 0.2\pm0.1$ -- that is, the Eddington limited accretion in thin disk model.
If brief super-Eddington events arise during their growth, the required progenitor mass could be smaller, implying that dense star cluster seeds could also be the ancestors of our quasar candidates.

As we have offered the most promising and robust high-$z$ quasar candidates in this paper, further confirmation is vital to uncover their true nature.
For example, spectroscopy with JWST would be the best opportunity to acquire the rest-frame UV/optical spectra of these quasars, allowing the detection of broad emission lines to get more precise SMBH mass measurements and gas-phase metallicity.
In addition to that, we can probe the cold molecular gas, tracing the galaxy dynamics and star formation activity, with ALMA.
With all of that being said, the samples presented in this work are ideal laboratories for dissecting the nature of the first galaxies and SMBHs formed during the reionization era.

\begin{acknowledgements}
	
We express our gratitude to the referee, Chiara Feruglio, for the constructive and insightful comments.
We thank Sherry Suyu and Dian Triani for the helpful feedback and fruitful discussions, which have significantly enhanced the quality of this paper.
This research is supported in part by the Excellence Cluster ORIGINS, which is funded by the Deutsche Forschungsgemeinschaft (DFG, German Research Foundation) under Germany's Excellence Strategy -- EXC-2094 -- 390783311.
JS is supported by JSPS KAKENHI (JP22H01262), the World Premier International Research Center Initiative (WPI), MEXT, Japan and the JSPS Core-to-Core Program (grant number: JPJSCCA20210003).
MH acknowledges funding from the Swiss National Science Foundation (SNF) via a PRIMA Grant PR00P2 193577 ``From cosmic dawn to high noon: the role of black holes for young galaxies.''
SG acknowledges financial support from the Villum Young Investigator grants 37440 and 13160 and the Cosmic Dawn Center (DAWN), funded by the Danish National Research Foundation (DNRF) under grant No. 140.
Some of the data products presented herein were retrieved from the Dawn JWST Archive (DJA). DJA is an initiative of the Cosmic Dawn Center.
BT acknowledges support from the European Research Council (ERC) under the European Union's Horizon 2020 research and innovation program (grant agreement 950533) and from the Israel Science Foundation (grant 1849/19).
The Flatiron Institute is supported by the Simons Foundation.
This research has made use of the SIMBAD database, operated at CDS, Strasbourg, France.
This research is based on observations made with the NASA/ESA Hubble Space Telescope obtained from the Space Telescope Science Institute, which is operated by the Association of Universities for Research in Astronomy, Inc., under NASA contract NAS 5-26555.
This work is based in part on observations made with the NASA/ESA/CSA James Webb Space Telescope. The data were obtained from the Mikulski Archive for Space Telescopes at the Space Telescope Science Institute, which is operated by the Association of Universities for Research in Astronomy, Inc., under NASA contract NAS 5-03127 for JWST.

\end{acknowledgements}

\tiny{
	\noindent
	\textit{Facilities.} ALMA, ESO:VISTA (VIRCAM), HST (ACS, WFC3), JWST (NIRCam), Spitzer (IRAC), Subaru (HSC).
}

\vspace{1mm}

\tiny{
	\noindent
	\textit{Software.}
	Astropy \citep{2013A&A...558A..33A,2018AJ....156..123A},
	CIGALE \citep{2019A&A...622A.103B,2020MNRAS.491..740Y,2022ApJ...927..192Y},
	Dask \citep{matthew_rocklin-proc-scipy-2015},
	eazy-py \citep{2008ApJ...686.1503B},
	grizli \citep{2022zndo...6672538B},
	Matplotlib \citep{2021zndo....592536C},
	msaexp \citep{2022zndo...7299500B},
	NumPy \citep{2020Natur.585..357H},
	Pandas \citep{2022zndo...3509134R},
	Seaborn \citep{2021JOSS....6.3021W},
	TOPCAT \citep{2005ASPC..347...29T}.
}

%
   \bibliographystyle{aa} 
   \bibliography{biblio} 

\begin{thebibliography}{171}
\expandafter\ifx\csname natexlab\endcsname\relax\def\natexlab#1{#1}\fi

\bibitem[{{Abramowicz} {et~al.}(1988){Abramowicz}, {Czerny}, {Lasota}, \&
  {Szuszkiewicz}}]{1988ApJ...332..646A}
{Abramowicz}, M.~A., {Czerny}, B., {Lasota}, J.~P., \& {Szuszkiewicz}, E. 1988,
  \apj, 332, 646

\bibitem[{{Aihara} {et~al.}(2022){Aihara}, {AlSayyad}, {Ando}, {Armstrong},
  {Bosch}, {Egami}, {Furusawa}, {Furusawa}, {Harasawa}, {Harikane}, {Hsieh},
  {Ikeda}, {Ito}, {Iwata}, {Kodama}, {Koike}, {Kokubo}, {Komiyama}, {Li},
  {Liang}, {Lin}, {Lupton}, {Lust}, {MacArthur}, {Mawatari}, {Mineo},
  {Miyatake}, {Miyazaki}, {More}, {Morishima}, {Murayama}, {Nakajima},
  {Nakata}, {Nishizawa}, {Oguri}, {Okabe}, {Okura}, {Ono}, {Osato}, {Ouchi},
  {Pan}, {Plazas Malag{\'o}n}, {Price}, {Reed}, {Rykoff}, {Shibuya},
  {Simunovic}, {Strauss}, {Sugimori}, {Suto}, {Suzuki}, {Takada}, {Takagi},
  {Takata}, {Takita}, {Tanaka}, {Tang}, {Taranu}, {Terai}, {Toba}, {Turner},
  {Uchiyama}, {Vijarnwannaluk}, {Waters}, {Yamada}, {Yamamoto}, \&
  {Yamashita}}]{2022PASJ...74..247A}
{Aihara}, H., {AlSayyad}, Y., {Ando}, M., {et~al.} 2022, \pasj, 74, 247

\bibitem[{{Alexander} \& {Natarajan}(2014)}]{2014Sci...345.1330A}
{Alexander}, T. \& {Natarajan}, P. 2014, Science, 345, 1330

\bibitem[{{Ananna} {et~al.}(2017){Ananna}, {Salvato}, {LaMassa}, {Urry},
  {Cappelluti}, {Cardamone}, {Civano}, {Farrah}, {Gilfanov}, {Glikman},
  {Hamilton}, {Kirkpatrick}, {Lanzuisi}, {Marchesi}, {Merloni}, {Nandra},
  {Natarajan}, {Richards}, \& {Timlin}}]{2017ApJ...850...66A}
{Ananna}, T.~T., {Salvato}, M., {LaMassa}, S., {et~al.} 2017, \apj, 850, 66

\bibitem[{{Ananna} {et~al.}(2020){Ananna}, {Urry}, {Treister}, {Hickox},
  {Shankar}, {Ricci}, {Cappelluti}, {Marchesi}, \&
  {Turner}}]{2020ApJ...903...85A}
{Ananna}, T.~T., {Urry}, C.~M., {Treister}, E., {et~al.} 2020, \apj, 903, 85

\bibitem[{{Andika} {et~al.}(2022){Andika}, {Jahnke}, {Ba{\~n}ados}, {Bosman},
  {Davies}, {Eilers}, {Farina}, {Onoue}, \& {van der
  Wel}}]{2022AJ....163..251A}
{Andika}, I.~T., {Jahnke}, K., {Ba{\~n}ados}, E., {et~al.} 2022, \aj, 163, 251

\bibitem[{{Andika} {et~al.}(2020){Andika}, {Jahnke}, {Onoue}, {Ba{\~n}ados},
  {Mazzucchelli}, {Novak}, {Eilers}, {Venemans}, {Schindler}, {Walter},
  {Neeleman}, {Simcoe}, {Decarli}, {Farina}, {Marian}, {Pensabene}, {Cooper},
  \& {Rojas}}]{2020ApJ...903...34A}
{Andika}, I.~T., {Jahnke}, K., {Onoue}, M., {et~al.} 2020, \apj, 903, 34

\bibitem[{{Andika} {et~al.}(2023{\natexlab{a}}){Andika}, {Jahnke}, {van der
  Wel}, {Ba{\~n}ados}, {Bosman}, {Davies}, {Eilers}, {Jaelani}, {Mazzucchelli},
  {Onoue}, \& {Schindler}}]{2023ApJ...943..150A}
{Andika}, I.~T., {Jahnke}, K., {van der Wel}, A., {et~al.} 2023{\natexlab{a}},
  \apj, 943, 150

\bibitem[{{Andika} {et~al.}(2023{\natexlab{b}}){Andika}, {Suyu},
  {Ca{\~n}ameras}, {Melo}, {Schuldt}, {Shu}, {Eilers}, {Jaelani}, \&
  {Yue}}]{2023A&A...678A.103A}
{Andika}, I.~T., {Suyu}, S.~H., {Ca{\~n}ameras}, R., {et~al.}
  2023{\natexlab{b}}, \aap, 678, A103

\bibitem[{{Arrabal Haro} {et~al.}(2023){Arrabal Haro}, {Dickinson},
  {Finkelstein}, {Fujimoto}, {Fern{\'a}ndez}, {Kartaltepe}, {Jung}, {Cole},
  {Burgarella}, {Chworowsky}, {Hutchison}, {Morales}, {Papovich}, {Simons},
  {Amor{\'\i}n}, {Backhaus}, {Bagley}, {Bisigello}, {Calabr{\`o}},
  {Castellano}, {Cleri}, {Dav{\'e}}, {Dekel}, {Ferguson}, {Fontana}, {Gawiser},
  {Giavalisco}, {Harish}, {Hathi}, {Hirschmann}, {Holwerda}, {Huertas-Company},
  {Koekemoer}, {Larson}, {Lucas}, {Mobasher}, {P{\'e}rez-Gonz{\'a}lez},
  {Pirzkal}, {Rose}, {Santini}, {Trump}, {de la Vega}, {Wang}, {Weiner},
  {Wilkins}, {Yang}, {Yung}, \& {Zavala}}]{2023ApJ...951L..22A}
{Arrabal Haro}, P., {Dickinson}, M., {Finkelstein}, S.~L., {et~al.} 2023,
  \apjl, 951, L22

\bibitem[{{Astropy Collaboration} {et~al.}(2018){Astropy Collaboration},
  {Price-Whelan}, {Sip{\H{o}}cz}, {G{\"u}nther}, {Lim}, {Crawford}, {Conseil},
  {Shupe}, {Craig}, {Dencheva}, {Ginsburg}, {VanderPlas}, {Bradley},
  {P{\'e}rez-Su{\'a}rez}, {de Val-Borro}, {Aldcroft}, {Cruz}, {Robitaille},
  {Tollerud}, {Ardelean}, {Babej}, {Bach}, {Bachetti}, {Bakanov}, {Bamford},
  {Barentsen}, {Barmby}, {Baumbach}, {Berry}, {Biscani}, {Boquien}, {Bostroem},
  {Bouma}, {Brammer}, {Bray}, {Breytenbach}, {Buddelmeijer}, {Burke},
  {Calderone}, {Cano Rodr{\'\i}guez}, {Cara}, {Cardoso}, {Cheedella}, {Copin},
  {Corrales}, {Crichton}, {D'Avella}, {Deil}, {Depagne}, {Dietrich}, {Donath},
  {Droettboom}, {Earl}, {Erben}, {Fabbro}, {Ferreira}, {Finethy}, {Fox},
  {Garrison}, {Gibbons}, {Goldstein}, {Gommers}, {Greco}, {Greenfield},
  {Groener}, {Grollier}, {Hagen}, {Hirst}, {Homeier}, {Horton}, {Hosseinzadeh},
  {Hu}, {Hunkeler}, {Ivezi{\'c}}, {Jain}, {Jenness}, {Kanarek}, {Kendrew},
  {Kern}, {Kerzendorf}, {Khvalko}, {King}, {Kirkby}, {Kulkarni}, {Kumar},
  {Lee}, {Lenz}, {Littlefair}, {Ma}, {Macleod}, {Mastropietro}, {McCully},
  {Montagnac}, {Morris}, {Mueller}, {Mumford}, {Muna}, {Murphy}, {Nelson},
  {Nguyen}, {Ninan}, {N{\"o}the}, {Ogaz}, {Oh}, {Parejko}, {Parley}, {Pascual},
  {Patil}, {Patil}, {Plunkett}, {Prochaska}, {Rastogi}, {Reddy Janga},
  {Sabater}, {Sakurikar}, {Seifert}, {Sherbert}, {Sherwood-Taylor}, {Shih},
  {Sick}, {Silbiger}, {Singanamalla}, {Singer}, {Sladen}, {Sooley},
  {Sornarajah}, {Streicher}, {Teuben}, {Thomas}, {Tremblay}, {Turner},
  {Terr{\'o}n}, {van Kerkwijk}, {de la Vega}, {Watkins}, {Weaver}, {Whitmore},
  {Woillez}, {Zabalza}, \& {Astropy Contributors}}]{2018AJ....156..123A}
{Astropy Collaboration}, {Price-Whelan}, A.~M., {Sip{\H{o}}cz}, B.~M., {et~al.}
  2018, \aj, 156, 123

\bibitem[{{Astropy Collaboration} {et~al.}(2013){Astropy Collaboration},
  {Robitaille}, {Tollerud}, {Greenfield}, {Droettboom}, {Bray}, {Aldcroft},
  {Davis}, {Ginsburg}, {Price-Whelan}, {Kerzendorf}, {Conley}, {Crighton},
  {Barbary}, {Muna}, {Ferguson}, {Grollier}, {Parikh}, {Nair}, {Unther},
  {Deil}, {Woillez}, {Conseil}, {Kramer}, {Turner}, {Singer}, {Fox}, {Weaver},
  {Zabalza}, {Edwards}, {Azalee Bostroem}, {Burke}, {Casey}, {Crawford},
  {Dencheva}, {Ely}, {Jenness}, {Labrie}, {Lim}, {Pierfederici}, {Pontzen},
  {Ptak}, {Refsdal}, {Servillat}, \& {Streicher}}]{2013A&A...558A..33A}
{Astropy Collaboration}, {Robitaille}, T.~P., {Tollerud}, E.~J., {et~al.} 2013,
  \aap, 558, A33

\bibitem[{{Ba{\~n}ados} {et~al.}(2019){Ba{\~n}ados}, {Novak}, {Neeleman},
  {Walter}, {Decarli}, {Venemans}, {Mazzucchelli}, {Carilli}, {Wang}, {Fan},
  {Farina}, \& {Rix}}]{2019ApJ...881L..23B}
{Ba{\~n}ados}, E., {Novak}, M., {Neeleman}, M., {et~al.} 2019, \apjl, 881, L23

\bibitem[{{Bagley} {et~al.}(2023){Bagley}, {Finkelstein}, {Koekemoer},
  {Ferguson}, {Arrabal Haro}, {Dickinson}, {Kartaltepe}, {Papovich},
  {P{\'e}rez-Gonz{\'a}lez}, {Pirzkal}, {Somerville}, {Willmer}, {Yang}, {Yung},
  {Fontana}, {Grazian}, {Grogin}, {Hirschmann}, {Kewley}, {Kirkpatrick},
  {Kocevski}, {Lotz}, {Medrano}, {Morales}, {Pentericci}, {Ravindranath},
  {Trump}, {Wilkins}, {Calabr{\`o}}, {Cooper}, {Costantin}, {de la Vega},
  {Hilbert}, {Hutchison}, {Larson}, {Lucas}, {McGrath}, {Ryan}, {Wang}, \&
  {Wuyts}}]{2023ApJ...946L..12B}
{Bagley}, M.~B., {Finkelstein}, S.~L., {Koekemoer}, A.~M., {et~al.} 2023,
  \apjl, 946, L12

\bibitem[{{Bagley} {et~al.}(2022){Bagley}, {Finkelstein}, {Rojas-Ruiz},
  {Diekmann}, {Finkelstein}, {Song}, {Papovich}, {Somerville}, {Baronchelli},
  \& {Dai}}]{2022arXiv220512980B}
{Bagley}, M.~B., {Finkelstein}, S.~L., {Rojas-Ruiz}, S., {et~al.} 2022, arXiv
  e-prints, arXiv:2205.12980

\bibitem[{{Begelman} \& {Volonteri}(2017)}]{2017MNRAS.464.1102B}
{Begelman}, M.~C. \& {Volonteri}, M. 2017, \mnras, 464, 1102

\bibitem[{{Bertin} {et~al.}(2020){Bertin}, {Schefer}, {Apostolakos},
  {{\'A}lvarez-Ayll{\'o}n}, {Dubath}, \& {K{\"u}mmel}}]{2020ASPC..527..461B}
{Bertin}, E., {Schefer}, M., {Apostolakos}, N., {et~al.} 2020, in Astronomical
  Society of the Pacific Conference Series, Vol. 527, Astronomical Data
  Analysis Software and Systems XXIX, ed. R.~{Pizzo}, E.~R. {Deul}, J.~D.
  {Mol}, J.~{de Plaa}, \& H.~{Verkouter}, 461

\bibitem[{{Bertin} {et~al.}(2022){Bertin}, {Schefer}, {Apostolakos},
  {{\'A}lvarez-Ayll{\'o}n}, {Dubath}, \& {K{\"u}mmel}}]{2022ascl.soft12018B}
{Bertin}, E., {Schefer}, M., {Apostolakos}, N., {et~al.} 2022,
  {SourceXtractor++: Extracts sources from astronomical images}, Astrophysics
  Source Code Library, record ascl:2212.018

\bibitem[{{Bezanson} {et~al.}(2022){Bezanson}, {Labbe}, {Whitaker}, {Leja},
  {Price}, {Franx}, {Brammer}, {Marchesini}, {Zitrin}, {Wang}, {Weaver},
  {Furtak}, {Atek}, {Coe}, {Cutler}, {Dayal}, {van Dokkum}, {Feldmann},
  {Forster Schreiber}, {Fujimoto}, {Geha}, {Glazebrook}, {de Graaff}, {Greene},
  {Juneau}, {Kassin}, {Kriek}, {Khullar}, {Maseda}, {Mowla}, {Muzzin},
  {Nanayakkara}, {Nelson}, {Oesch}, {Pacifici}, {Pan}, {Papovich}, {Setton},
  {Shapley}, {Smit}, {Stefanon}, {Taylor}, \& {Williams}}]{2022arXiv221204026B}
{Bezanson}, R., {Labbe}, I., {Whitaker}, K.~E., {et~al.} 2022, arXiv e-prints,
  arXiv:2212.04026

\bibitem[{{Boekholt} {et~al.}(2018){Boekholt}, {Schleicher}, {Fellhauer},
  {Klessen}, {Reinoso}, {Stutz}, \& {Haemmerl{\'e}}}]{2018MNRAS.476..366B}
{Boekholt}, T.~C.~N., {Schleicher}, D.~R.~G., {Fellhauer}, M., {et~al.} 2018,
  \mnras, 476, 366

\bibitem[{{Boquien} {et~al.}(2019){Boquien}, {Burgarella}, {Roehlly}, {Buat},
  {Ciesla}, {Corre}, {Inoue}, \& {Salas}}]{2019A&A...622A.103B}
{Boquien}, M., {Burgarella}, D., {Roehlly}, Y., {et~al.} 2019, \aap, 622, A103

\bibitem[{{Bouwens} {et~al.}(2021){Bouwens}, {Oesch}, {Stefanon},
  {Illingworth}, {Labb{\'e}}, {Reddy}, {Atek}, {Montes}, {Naidu},
  {Nanayakkara}, {Nelson}, \& {Wilkins}}]{2021AJ....162...47B}
{Bouwens}, R.~J., {Oesch}, P.~A., {Stefanon}, M., {et~al.} 2021, \aj, 162, 47

\bibitem[{{Brammer}(2023)}]{2022zndo...7299500B}
{Brammer}, G. 2023, {msaexp: NIRSpec analyis tools}, Zenodo

\bibitem[{{Brammer} {et~al.}(2022){Brammer}, {Strait}, {Matharu}, \&
  {Momcheva}}]{2022zndo...6672538B}
{Brammer}, G., {Strait}, V., {Matharu}, J., \& {Momcheva}, I. 2022, {grizli},
  Zenodo

\bibitem[{{Brammer} {et~al.}(2008){Brammer}, {van Dokkum}, \&
  {Coppi}}]{2008ApJ...686.1503B}
{Brammer}, G.~B., {van Dokkum}, P.~G., \& {Coppi}, P. 2008, \apj, 686, 1503

\bibitem[{{Bromm} \& {Loeb}(2003)}]{2003ApJ...596...34B}
{Bromm}, V. \& {Loeb}, A. 2003, \apj, 596, 34

\bibitem[{{Bruzual} \& {Charlot}(2003)}]{2003MNRAS.344.1000B}
{Bruzual}, G. \& {Charlot}, S. 2003, \mnras, 344, 1000

\bibitem[{{Bunker} {et~al.}(2023){Bunker}, {Cameron}, {Curtis-Lake},
  {Jakobsen}, {Carniani}, {Curti}, {Witstok}, {Maiolino}, {D'Eugenio},
  {Looser}, {Willott}, {Bonaventura}, {Hainline}, {Uebler}, {Willmer},
  {Saxena}, {Smit}, {Alberts}, {Arribas}, {Baker}, {Baum}, {Bhatawdekar},
  {Bowler}, {Boyett}, {Charlot}, {Chen}, {Chevallard}, {Circosta}, {DeCoursey},
  {de Graaff}, {Egami}, {Eisenstein}, {Endsley}, {Ferruit}, {Giardino},
  {Hausen}, {Helton}, {Hviding}, {Ji}, {Johnson}, {Jones}, {Kumari}, {Laseter},
  {Luetzgendorf}, {Maseda}, {Nelson}, {Parlanti}, {Perna}, {Rawle}, {Rix},
  {Rieke}, {Robertson}, {Rodriguez Del Pino}, {Sandles}, {Scholtz}, {Sharpe},
  {Skarbinski}, {Stark}, {Sun}, {Tacchella}, {Topping}, {Villanueva},
  {Wallace}, {Williams}, \& {Woodrum}}]{2023arXiv230602467B}
{Bunker}, A.~J., {Cameron}, A.~J., {Curtis-Lake}, E., {et~al.} 2023, arXiv
  e-prints, arXiv:2306.02467

\bibitem[{{Bushouse} {et~al.}(2022){Bushouse}, {Eisenhamer}, {Dencheva},
  {Davies}, {Greenfield}, {Morrison}, {Hodge}, {Simon}, {Grumm}, {Droettboom},
  {Slavich}, {Sosey}, {Pauly}, {Miller}, {Jedrzejewski}, {Hack}, {Davis},
  {Crawford}, {Law}, {Gordon}, {Regan}, {Cara}, {MacDonald}, {Bradley},
  {Shanahan}, {Jamieson}, {Teodoro}, \& {Williams}}]{2022zndo...7325378B}
{Bushouse}, H., {Eisenhamer}, J., {Dencheva}, N., {et~al.} 2022, {JWST
  Calibration Pipeline}, Zenodo

\bibitem[{{Calzetti} {et~al.}(2000){Calzetti}, {Armus}, {Bohlin}, {Kinney},
  {Koornneef}, \& {Storchi-Bergmann}}]{2000ApJ...533..682C}
{Calzetti}, D., {Armus}, L., {Bohlin}, R.~C., {et~al.} 2000, \apj, 533, 682

\bibitem[{{Caplar} {et~al.}(2018){Caplar}, {Lilly}, \&
  {Trakhtenbrot}}]{2018ApJ...867..148C}
{Caplar}, N., {Lilly}, S.~J., \& {Trakhtenbrot}, B. 2018, \apj, 867, 148

\bibitem[{{Carrasco Kind} \& {Brunner}(2013)}]{2013MNRAS.432.1483C}
{Carrasco Kind}, M. \& {Brunner}, R.~J. 2013, \mnras, 432, 1483

\bibitem[{{Casey} {et~al.}(2023){Casey}, {Kartaltepe}, {Drakos}, {Franco},
  {Harish}, {Paquereau}, {Ilbert}, {Rose}, {Cox}, {Nightingale}, {Robertson},
  {Silverman}, {Koekemoer}, {Massey}, {McCracken}, {Rhodes}, {Akins}, {Allen},
  {Amvrosiadis}, {Arango-Toro}, {Bagley}, {Bongiorno}, {Capak}, {Champagne},
  {Chartab}, {Ch{\'a}vez Ortiz}, {Chworowsky}, {Cooke}, {Cooper}, {Darvish},
  {Ding}, {Faisst}, {Finkelstein}, {Fujimoto}, {Gentile}, {Gillman}, {Gould},
  {Gozaliasl}, {Hayward}, {He}, {Hemmati}, {Hirschmann}, {Jahnke}, {Jin},
  {Khostovan}, {Kokorev}, {Lambrides}, {Laigle}, {Larson}, {Leung}, {Liu},
  {Liaudat}, {Long}, {Magdis}, {Mahler}, {Mainieri}, {Manning}, {Maraston},
  {Martin}, {McCleary}, {McKinney}, {McPartland}, {Mobasher}, {Pattnaik},
  {Renzini}, {Rich}, {Sanders}, {Sattari}, {Scognamiglio}, {Scoville}, {Sheth},
  {Shuntov}, {Sparre}, {Suzuki}, {Talia}, {Toft}, {Trakhtenbrot}, {Urry},
  {Valentino}, {Vanderhoof}, {Vardoulaki}, {Weaver}, {Whitaker}, {Wilkins},
  {Yang}, \& {Zavala}}]{2023ApJ...954...31C}
{Casey}, C.~M., {Kartaltepe}, J.~S., {Drakos}, N.~E., {et~al.} 2023, \apj, 954,
  31

\bibitem[{{Caswell} {et~al.}(2021){Caswell}, {Droettboom}, {Lee}, {Sales De
  Andrade}, {Hoffmann}, {Hunter}, {Klymak}, {Firing}, {Stansby}, {Varoquaux},
  {Hedegaard Nielsen}, {Root}, {May}, {Elson}, {Sepp{\"a}nen}, {Dale}, {Lee},
  {McDougall}, {Straw}, {Hobson}, {Hannah}, {Gohlke}, {Vincent}, {Yu}, {Ma},
  {Silvester}, {Moad}, {Kniazev}, {Ernest}, \& {Ivanov}}]{2021zndo....592536C}
{Caswell}, T.~A., {Droettboom}, M., {Lee}, A., {et~al.} 2021,
  {matplotlib/matplotlib: REL: v3.5.1}, Zenodo

\bibitem[{{Chabrier}(2003)}]{2003PASP..115..763C}
{Chabrier}, G. 2003, \pasp, 115, 763

\bibitem[{{Conroy} \& {Gunn}(2010)}]{2010ApJ...712..833C}
{Conroy}, C. \& {Gunn}, J.~E. 2010, \apj, 712, 833

\bibitem[{{Conroy} {et~al.}(2009){Conroy}, {Gunn}, \&
  {White}}]{2009ApJ...699..486C}
{Conroy}, C., {Gunn}, J.~E., \& {White}, M. 2009, \apj, 699, 486

\bibitem[{{Conroy} {et~al.}(2010){Conroy}, {White}, \&
  {Gunn}}]{2010ApJ...708...58C}
{Conroy}, C., {White}, M., \& {Gunn}, J.~E. 2010, \apj, 708, 58

\bibitem[{{Ding} {et~al.}(2023){Ding}, {Onoue}, {Silverman}, {Matsuoka},
  {Izumi}, {Strauss}, {Jahnke}, {Phillips}, {Li}, {Volonteri}, {Haiman},
  {Andika}, {Aoki}, {Baba}, {Bieri}, {Bosman}, {Bottrell}, {Eilers},
  {Fujimoto}, {Habouzit}, {Imanishi}, {Inayoshi}, {Iwasawa}, {Kashikawa},
  {Kawaguchi}, {Kohno}, {Lee}, {Lupi}, {Lyu}, {Nagao}, {Overzier}, {Schindler},
  {Schramm}, {Shimasaku}, {Toba}, {Trakhtenbrot}, {Trebitsch}, {Treu},
  {Umehata}, {Venemans}, {Vestergaard}, {Walter}, {Wang}, \&
  {Yang}}]{2023Natur.621...51D}
{Ding}, X., {Onoue}, M., {Silverman}, J.~D., {et~al.} 2023, \nat, 621, 51

\bibitem[{{Ding} {et~al.}(2020){Ding}, {Silverman}, {Treu}, {Schulze},
  {Schramm}, {Birrer}, {Park}, {Jahnke}, {Bennert}, {Kartaltepe}, {Koekemoer},
  {Malkan}, \& {Sanders}}]{2020ApJ...888...37D}
{Ding}, X., {Silverman}, J., {Treu}, T., {et~al.} 2020, \apj, 888, 37

\bibitem[{{Dubois} {et~al.}(2014){Dubois}, {Volonteri}, \&
  {Silk}}]{2014MNRAS.440.1590D}
{Dubois}, Y., {Volonteri}, M., \& {Silk}, J. 2014, \mnras, 440, 1590

\bibitem[{{Duncan} {et~al.}(2021){Duncan}, {Kondapally}, {Brown}, {Bonato},
  {Best}, {R{\"o}ttgering}, {Bondi}, {Bowler}, {Cochrane}, {G{\"u}rkan},
  {Hardcastle}, {Jarvis}, {Kunert-Bajraszewska}, {Leslie}, {Ma{\l}ek},
  {Morabito}, {O'Sullivan}, {Prandoni}, {Sabater}, {Shimwell}, {Smith}, {Wang},
  {Wo{\l}owska}, \& {Tasse}}]{2021A&A...648A...4D}
{Duncan}, K.~J., {Kondapally}, R., {Brown}, M.~J.~I., {et~al.} 2021, \aap, 648,
  A4

\bibitem[{{Dunlop} {et~al.}(2021){Dunlop}, {Abraham}, {Ashby}, {Bagley},
  {Best}, {Bongiorno}, {Bouwens}, {Bowler}, {Brammer}, {Bremer}, {Calabro'},
  {Carnall}, {Castellano}, {Cirasuolo}, {Conselice}, {Cullen}, {Dave}, {Dayal},
  {Dekel}, {Dickinson}, {Duncan}, {Elbaz}, {Ellis}, {Ferguson}, {Ferrara},
  {Finkelstein}, {Fontana}, {Furlanetto}, {Fynbo}, {Gallerani}, {Gardner},
  {Giavalisco}, {Grazian}, {Grogin}, {Harikane}, {Hopkins}, {Ilbert},
  {Illingworth}, {Juneau}, {Jung}, {Kartaltepe}, {Kassin}, {Kauffmann},
  {Khochfar}, {Kirkpatrick}, {Kocevski}, {Koekemoer}, {Labbe}, {Laporte},
  {Larson}, {Lucas}, {Magee}, {Mason}, {McCracken}, {McLeod}, {McLure},
  {Merlin}, {Mesinger}, {Milvang-Jensen}, {Newman}, {Oesch}, {Ouchi},
  {Pacifici}, {Papovich}, {Peacock}, {Peeples}, {Pentericci}, {Perez-Gonzalez},
  {Pirzkal}, {Pope}, {Pye}, {Reddy}, {Robertson}, {Salvato}, {Santini},
  {Schaerer}, {Shapley}, {Simons}, {Smit}, {Smith}, {Snyder}, {Somerville},
  {Stanway}, {Stefanon}, {Tasca}, {Tikkanen}, {Tresse}, {Trump}, {Whitaker},
  {Wilkins}, {Wright}, {Wyithe}, {van Dokkum}, \& {van der
  Werf}}]{2021jwst.prop.1837D}
{Dunlop}, J.~S., {Abraham}, R.~G., {Ashby}, M. L.~N., {et~al.} 2021, {PRIMER:
  Public Release IMaging for Extragalactic Research}, JWST Proposal. Cycle 1,
  ID. \#1837

\bibitem[{{Eisenstein} {et~al.}(2023){Eisenstein}, {Willott}, {Alberts},
  {Arribas}, {Bonaventura}, {Bunker}, {Cameron}, {Carniani}, {Charlot},
  {Curtis-Lake}, {D'Eugenio}, {Endsley}, {Ferruit}, {Giardino}, {Hainline},
  {Hausen}, {Jakobsen}, {Johnson}, {Maiolino}, {Rieke}, {Rieke}, {Rix},
  {Robertson}, {Stark}, {Tacchella}, {Williams}, {Willmer}, {Baker}, {Baum},
  {Bhatawdekar}, {Boyett}, {Chen}, {Chevallard}, {Circosta}, {Curti},
  {Danhaive}, {DeCoursey}, {de Graaff}, {Dressler}, {Egami}, {Helton},
  {Hviding}, {Ji}, {Jones}, {Kumari}, {L{\"u}tzgendorf}, {Laseter}, {Looser},
  {Lyu}, {Maseda}, {Nelson}, {Parlanti}, {Perna}, {Pusk{\'a}s}, {Rawle},
  {Rodr{\'\i}guez Del Pino}, {Sandles}, {Saxena}, {Scholtz}, {Sharpe},
  {Shivaei}, {Silcock}, {Simmonds}, {Skarbinski}, {Smit}, {Stone}, {Suess},
  {Sun}, {Tang}, {Topping}, {{\"U}bler}, {Villanueva}, {Wallace}, {Whitler},
  {Witstok}, \& {Woodrum}}]{2023arXiv230602465E}
{Eisenstein}, D.~J., {Willott}, C., {Alberts}, S., {et~al.} 2023, arXiv
  e-prints, arXiv:2306.02465

\bibitem[{{Erb} {et~al.}(2010){Erb}, {Pettini}, {Shapley}, {Steidel}, {Law}, \&
  {Reddy}}]{2010ApJ...719.1168E}
{Erb}, D.~K., {Pettini}, M., {Shapley}, A.~E., {et~al.} 2010, \apj, 719, 1168

\bibitem[{{Euclid Collaboration} {et~al.}(2022){Euclid Collaboration},
  {Moneti}, {McCracken}, {Shuntov}, {Kauffmann}, {Capak}, {Davidzon}, {Ilbert},
  {Scarlata}, {Toft}, {Weaver}, {Chary}, {Cuby}, {Faisst}, {Masters},
  {McPartland}, {Mobasher}, {Sanders}, {Scaramella}, {Stern}, {Szapudi},
  {Teplitz}, {Zalesky}, {Amara}, {Auricchio}, {Bodendorf}, {Bonino},
  {Branchini}, {Brau-Nogue}, {Brescia}, {Brinchmann}, {Capobianco}, {Carbone},
  {Carretero}, {Castander}, {Castellano}, {Cavuoti}, {Cimatti}, {Cledassou},
  {Congedo}, {Conselice}, {Conversi}, {Copin}, {Corcione}, {Costille},
  {Cropper}, {Da Silva}, {Degaudenzi}, {Douspis}, {Dubath}, {Duncan}, {Dupac},
  {Dusini}, {Farrens}, {Ferriol}, {Fosalba}, {Frailis}, {Franceschi}, {Fumana},
  {Garilli}, {Gillis}, {Giocoli}, {Granett}, {Grazian}, {Grupp}, {Haugan},
  {Hoekstra}, {Holmes}, {Hormuth}, {Hudelot}, {Jahnke}, {Kermiche},
  {Kiessling}, {Kilbinger}, {Kitching}, {Kohley}, {K{\"u}mmel}, {Kunz},
  {Kurki-Suonio}, {Ligori}, {Lilje}, {Lloro}, {Maiorano}, {Mansutti},
  {Marggraf}, {Markovic}, {Marulli}, {Massey}, {Maurogordato}, {Meneghetti},
  {Merlin}, {Meylan}, {Moresco}, {Moscardini}, {Munari}, {Niemi}, {Padilla},
  {Paltani}, {Pasian}, {Pedersen}, {Pires}, {Poncet}, {Popa}, {Pozzetti},
  {Raison}, {Rebolo}, {Rhodes}, {Rix}, {Roncarelli}, {Rossetti}, {Saglia},
  {Schneider}, {Secroun}, {Seidel}, {Serrano}, {Sirignano}, {Sirri}, {Stanco},
  {Tallada-Cresp{\'\i}}, {Taylor}, {Tereno}, {Toledo-Moreo}, {Torradeflot},
  {Wang}, {Welikala}, {Weller}, {Zamorani}, {Zoubian}, {Andreon}, {Bardelli},
  {Camera}, {Graci{\'a}-Carpio}, {Medinaceli}, {Mei}, {Polenta}, {Romelli},
  {Sureau}, {Tenti}, {Vassallo}, {Zacchei}, {Zucca}, {Baccigalupi},
  {Balaguera-Antol{\'\i}nez}, {Bernardeau}, {Biviano}, {Bolzonella}, {Bozzo},
  {Burigana}, {Cabanac}, {Cappi}, {Carvalho}, {Casas}, {Castignani},
  {Colodro-Conde}, {Coupon}, {Courtois}, {Di Ferdinando}, {Farina}, {Finelli},
  {Flose-Reimberg}, {Fotopoulou}, {Galeotta}, {Ganga}, {Garcia-Bellido},
  {Gaztanaga}, {Gozaliasl}, {Hook}, {Joachimi}, {Kansal}, {Keihanen},
  {Kirkpatrick}, {Lindholm}, {Mainetti}, {Maino}, {Maoli}, {Martinelli},
  {Martinet}, {Maturi}, {Metcalf}, {Morgante}, {Morisset}, {Nucita},
  {Patrizii}, {Potter}, {Renzi}, {Riccio}, {S{\'a}nchez}, {Sapone}, {Schirmer},
  {Schultheis}, {Scottez}, {Sefusatti}, {Teyssier}, {Tubio}, {Tutusaus},
  {Valiviita}, {Viel}, \& {Hildebrandt}}]{2022A&A...658A.126E}
{Euclid Collaboration}, {Moneti}, A., {McCracken}, H.~J., {et~al.} 2022, \aap,
  658, A126

\bibitem[{{Fabian} \& {Lasenby}(2019)}]{2019arXiv191104305F}
{Fabian}, A.~C. \& {Lasenby}, A.~N. 2019, arXiv e-prints, arXiv:1911.04305

\bibitem[{{Fan} {et~al.}(2023){Fan}, {Ba{\~n}ados}, \&
  {Simcoe}}]{2023ARA&A..61..373F}
{Fan}, X., {Ba{\~n}ados}, E., \& {Simcoe}, R.~A. 2023, \araa, 61, 373

\bibitem[{{Finkelstein} {et~al.}(2023){Finkelstein}, {Bagley}, {Ferguson},
  {Wilkins}, {Kartaltepe}, {Papovich}, {Yung}, {Arrabal Haro}, {Behroozi},
  {Dickinson}, {Kocevski}, {Koekemoer}, {Larson}, {Le Bail}, {Morales},
  {P{\'e}rez-Gonz{\'a}lez}, {Burgarella}, {Dav{\'e}}, {Hirschmann},
  {Somerville}, {Wuyts}, {Bromm}, {Casey}, {Fontana}, {Fujimoto}, {Gardner},
  {Giavalisco}, {Grazian}, {Grogin}, {Hathi}, {Hutchison}, {Jha}, {Jogee},
  {Kewley}, {Kirkpatrick}, {Long}, {Lotz}, {Pentericci}, {Pierel}, {Pirzkal},
  {Ravindranath}, {Ryan}, {Trump}, {Yang}, {Bhatawdekar}, {Bisigello}, {Buat},
  {Calabr{\`o}}, {Castellano}, {Cleri}, {Cooper}, {Croton}, {Daddi}, {Dekel},
  {Elbaz}, {Franco}, {Gawiser}, {Holwerda}, {Huertas-Company}, {Jaskot},
  {Leung}, {Lucas}, {Mobasher}, {Pandya}, {Tacchella}, {Weiner}, \&
  {Zavala}}]{2023ApJ...946L..13F}
{Finkelstein}, S.~L., {Bagley}, M.~B., {Ferguson}, H.~C., {et~al.} 2023, \apjl,
  946, L13

\bibitem[{{Fitriana} \& {Murayama}(2022)}]{2022PASJ...74..689F}
{Fitriana}, I.~K. \& {Murayama}, T. 2022, \pasj, 74, 689

\bibitem[{{Fitzpatrick}(1999)}]{1999PASP..111...63F}
{Fitzpatrick}, E.~L. 1999, \pasp, 111, 63

\bibitem[{{Fragione} \& {Pacucci}(2023)}]{2023arXiv230814986F}
{Fragione}, G. \& {Pacucci}, F. 2023, arXiv e-prints, arXiv:2308.14986

\bibitem[{{Fujimoto} {et~al.}(2023{\natexlab{a}}){Fujimoto}, {Arrabal Haro},
  {Dickinson}, {Finkelstein}, {Kartaltepe}, {Larson}, {Burgarella}, {Bagley},
  {Behroozi}, {Chworowsky}, {Hirschmann}, {Trump}, {Wilkins}, {Yung},
  {Koekemoer}, {Papovich}, {Pirzkal}, {Ferguson}, {Fontana}, {Grogin},
  {Grazian}, {Kewley}, {Kocevski}, {Lotz}, {Pentericci}, {Ravindranath},
  {Somerville}, {Wilkins}, {Amor{\'\i}n}, {Backhaus}, {Calabr{\`o}}, {Casey},
  {Cooper}, {Fern{\'a}ndez}, {Franco}, {Giavalisco}, {Hathi}, {Harish},
  {Hutchison}, {Iyer}, {Jung}, {Lucas}, \& {Zavala}}]{2023ApJ...949L..25F}
{Fujimoto}, S., {Arrabal Haro}, P., {Dickinson}, M., {et~al.}
  2023{\natexlab{a}}, \apjl, 949, L25

\bibitem[{{Fujimoto} {et~al.}(2023{\natexlab{b}}){Fujimoto}, {Bezanson},
  {Labbe}, {Brammer}, {Price}, {Wang}, {Weaver}, {Fudamoto}, {Oesch},
  {Williams}, {Dayal}, {Feldmann}, {Greene}, {Leja}, {Whitaker}, {Zitrin},
  {Cutler}, {Furtak}, {Pan}, {Chemerynska}, {Kokorev}, {Miller}, {Atek}, {van
  Dokkum}, {Juneau}, {Kassin}, {Khullar}, {Marchesini}, {Maseda}, {Nelson},
  {Setton}, \& {Smit}}]{2023arXiv230907834F}
{Fujimoto}, S., {Bezanson}, R., {Labbe}, I., {et~al.} 2023{\natexlab{b}}, arXiv
  e-prints, arXiv:2309.07834

\bibitem[{{Fujimoto} {et~al.}(2023{\natexlab{c}}){Fujimoto}, {Wang}, {Weaver},
  {Kokorev}, {Atek}, {Bezanson}, {Labbe}, {Brammer}, {Greene}, {Chemerynska},
  {Dayal}, {de Graaff}, {Furtak}, {Oesch}, {Setton}, {Price}, {Miller},
  {Williams}, {Whitaker}, {Zitrin}, {Cutler}, {Leja}, {Pan}, {Coe}, {van
  Dokkum}, {Feldmann}, {Fudamoto}, {Goulding}, {Khullar}, {Marchesini},
  {Maseda}, {Nanayakkara}, {Nelson}, {Smit}, {Stefanon}, \&
  {Weibel}}]{2023arXiv230811609F}
{Fujimoto}, S., {Wang}, B., {Weaver}, J., {et~al.} 2023{\natexlab{c}}, arXiv
  e-prints, arXiv:2308.11609

\bibitem[{{Furtak} {et~al.}(2023){Furtak}, {Labb{\'e}}, {Zitrin}, {Greene},
  {Dayal}, {Chemerynska}, {Kokorev}, {Miller}, {Goulding}, {Bezanson},
  {Brammer}, {Cutler}, {Leja}, {Pan}, {Price}, {Wang}, {Weaver}, {Whitaker},
  {Atek}, {Bogd{\'a}n}, {Charlot}, {Curtis-Lake}, {van Dokkum}, {Endsley},
  {Fudamoto}, {Fujimoto}, {de Graaff}, {Glazebrook}, {Juneau}, {Marchesini},
  {Maseda}, {Nelson}, {Oesch}, {Plat}, {Setton}, {Stark}, \&
  {Williams}}]{2023arXiv230805735F}
{Furtak}, L.~J., {Labb{\'e}}, I., {Zitrin}, A., {et~al.} 2023, arXiv e-prints,
  arXiv:2308.05735

\bibitem[{{Gaia Collaboration} {et~al.}(2023){Gaia Collaboration}, {Vallenari},
  {Brown}, {Prusti}, {de Bruijne}, {Arenou}, {Babusiaux}, {Biermann},
  {Creevey}, {Ducourant}, {Evans}, {Eyer}, {Guerra}, {Hutton}, {Jordi},
  {Klioner}, {Lammers}, {Lindegren}, {Luri}, {Mignard}, {Panem}, {Pourbaix},
  {Randich}, {Sartoretti}, {Soubiran}, {Tanga}, {Walton}, {Bailer-Jones},
  {Bastian}, {Drimmel}, {Jansen}, {Katz}, {Lattanzi}, {van Leeuwen}, {Bakker},
  {Cacciari}, {Casta{\~n}eda}, {De Angeli}, {Fabricius}, {Fouesneau},
  {Fr{\'e}mat}, {Galluccio}, {Guerrier}, {Heiter}, {Masana}, {Messineo},
  {Mowlavi}, {Nicolas}, {Nienartowicz}, {Pailler}, {Panuzzo}, {Riclet}, {Roux},
  {Seabroke}, {Sordo}, {Th{\'e}venin}, {Gracia-Abril}, {Portell}, {Teyssier},
  {Altmann}, {Andrae}, {Audard}, {Bellas-Velidis}, {Benson}, {Berthier},
  {Blomme}, {Burgess}, {Busonero}, {Busso}, {C{\'a}novas}, {Carry}, {Cellino},
  {Cheek}, {Clementini}, {Damerdji}, {Davidson}, {de Teodoro}, {Nu{\~n}ez
  Campos}, {Delchambre}, {Dell'Oro}, {Esquej}, {Fern{\'a}ndez-Hern{\'a}ndez},
  {Fraile}, {Garabato}, {Garc{\'\i}a-Lario}, {Gosset}, {Haigron}, {Halbwachs},
  {Hambly}, {Harrison}, {Hern{\'a}ndez}, {Hestroffer}, {Hodgkin}, {Holl},
  {Jan{\ss}en}, {Jevardat de Fombelle}, {Jordan}, {Krone-Martins}, {Lanzafame},
  {L{\"o}ffler}, {Marchal}, {Marrese}, {Moitinho}, {Muinonen}, {Osborne},
  {Pancino}, {Pauwels}, {Recio-Blanco}, {Reyl{\'e}}, {Riello}, {Rimoldini},
  {Roegiers}, {Rybizki}, {Sarro}, {Siopis}, {Smith}, {Sozzetti}, {Utrilla},
  {van Leeuwen}, {Abbas}, {{\'A}brah{\'a}m}, {Abreu Aramburu}, {Aerts},
  {Aguado}, {Ajaj}, {Aldea-Montero}, {Altavilla}, {{\'A}lvarez}, {Alves},
  {Anders}, {Anderson}, {Anglada Varela}, {Antoja}, {Baines}, {Baker},
  {Balaguer-N{\'u}{\~n}ez}, {Balbinot}, {Balog}, {Barache}, {Barbato},
  {Barros}, {Barstow}, {Bartolom{\'e}}, {Bassilana}, {Bauchet}, {Becciani},
  {Bellazzini}, {Berihuete}, {Bernet}, {Bertone}, {Bianchi}, {Binnenfeld},
  {Blanco-Cuaresma}, {Blazere}, {Boch}, {Bombrun}, {Bossini}, {Bouquillon},
  {Bragaglia}, {Bramante}, {Breedt}, {Bressan}, {Brouillet}, {Brugaletta},
  {Bucciarelli}, {Burlacu}, {Butkevich}, {Buzzi}, {Caffau}, {Cancelliere},
  {Cantat-Gaudin}, {Carballo}, {Carlucci}, {Carnerero}, {Carrasco},
  {Casamiquela}, {Castellani}, {Castro-Ginard}, {Chaoul}, {Charlot}, {Chemin},
  {Chiaramida}, {Chiavassa}, {Chornay}, {Comoretto}, {Contursi}, {Cooper},
  {Cornez}, {Cowell}, {Crifo}, {Cropper}, {Crosta}, {Crowley}, {Dafonte},
  {Dapergolas}, {David}, {David}, {de Laverny}, {De Luise}, {De March}, {De
  Ridder}, {de Souza}, {de Torres}, {del Peloso}, {del Pozo}, {Delbo},
  {Delgado}, {Delisle}, {Demouchy}, {Dharmawardena}, {Di Matteo}, {Diakite},
  {Diener}, {Distefano}, {Dolding}, {Edvardsson}, {Enke}, {Fabre}, {Fabrizio},
  {Faigler}, {Fedorets}, {Fernique}, {Fienga}, {Figueras}, {Fournier},
  {Fouron}, {Fragkoudi}, {Gai}, {Garcia-Gutierrez}, {Garcia-Reinaldos},
  {Garc{\'\i}a-Torres}, {Garofalo}, {Gavel}, {Gavras}, {Gerlach}, {Geyer},
  {Giacobbe}, {Gilmore}, {Girona}, {Giuffrida}, {Gomel}, {Gomez},
  {Gonz{\'a}lez-N{\'u}{\~n}ez}, {Gonz{\'a}lez-Santamar{\'\i}a},
  {Gonz{\'a}lez-Vidal}, {Granvik}, {Guillout}, {Guiraud},
  {Guti{\'e}rrez-S{\'a}nchez}, {Guy}, {Hatzidimitriou}, {Hauser}, {Haywood},
  {Helmer}, {Helmi}, {Sarmiento}, {Hidalgo}, {Hilger}, {H{\l}adczuk}, {Hobbs},
  {Holland}, {Huckle}, {Jardine}, {Jasniewicz}, {Jean-Antoine Piccolo},
  {Jim{\'e}nez-Arranz}, {Jorissen}, {Juaristi Campillo}, {Julbe}, {Karbevska},
  {Kervella}, {Khanna}, {Kontizas}, {Kordopatis}, {Korn}, {K{\'o}sp{\'a}l},
  {Kostrzewa-Rutkowska}, {Kruszy{\'n}ska}, {Kun}, {Laizeau}, {Lambert},
  {Lanza}, {Lasne}, {Le Campion}, {Lebreton}, {Lebzelter}, {Leccia}, {Leclerc},
  {Lecoeur-Taibi}, {Liao}, {Licata}, {Lindstr{\o}m}, {Lister}, {Livanou},
  {Lobel}, {Lorca}, {Loup}, {Madrero Pardo}, {Magdaleno Romeo}, {Managau},
  {Mann}, {Manteiga}, {Marchant}, {Marconi}, {Marcos}, {Marcos Santos},
  {Mar{\'\i}n Pina}, {Marinoni}, {Marocco}, {Marshall}, {Martin Polo},
  {Mart{\'\i}n-Fleitas}, {Marton}, {Mary}, {Masip}, {Massari},
  {Mastrobuono-Battisti}, {Mazeh}, {McMillan}, {Messina}, {Michalik}, {Millar},
  {Mints}, {Molina}, {Molinaro}, {Moln{\'a}r}, {Monari}, {Mongui{\'o}},
  {Montegriffo}, {Montero}, {Mor}, {Mora}, {Morbidelli}, {Morel}, {Morris},
  {Muraveva}, {Murphy}, {Musella}, {Nagy}, {Noval}, {Oca{\~n}a}, {Ogden},
  {Ordenovic}, {Osinde}, {Pagani}, {Pagano}, {Palaversa}, {Palicio},
  {Pallas-Quintela}, {Panahi}, {Payne-Wardenaar}, {Pe{\~n}alosa Esteller},
  {Penttil{\"a}}, {Pichon}, {Piersimoni}, {Pineau}, {Plachy}, {Plum}, {Poggio},
  {Pr{\v{s}}a}, {Pulone}, {Racero}, {Ragaini}, {Rainer}, {Raiteri}, {Rambaux},
  {Ramos}, {Ramos-Lerate}, {Re Fiorentin}, {Regibo}, {Richards}, {Rios Diaz},
  {Ripepi}, {Riva}, {Rix}, {Rixon}, {Robichon}, {Robin}, {Robin}, {Roelens},
  {Rogues}, {Rohrbasser}, {Romero-G{\'o}mez}, {Rowell}, {Royer}, {Ruz Mieres},
  {Rybicki}, {Sadowski}, {S{\'a}ez N{\'u}{\~n}ez}, {Sagrist{\`a} Sell{\'e}s},
  {Sahlmann}, {Salguero}, {Samaras}, {Sanchez Gimenez}, {Sanna},
  {Santove{\~n}a}, {Sarasso}, {Schultheis}, {Sciacca}, {Segol}, {Segovia},
  {S{\'e}gransan}, {Semeux}, {Shahaf}, {Siddiqui}, {Siebert}, {Siltala},
  {Silvelo}, {Slezak}, {Slezak}, {Smart}, {Snaith}, {Solano}, {Solitro},
  {Souami}, {Souchay}, {Spagna}, {Spina}, {Spoto}, {Steele},
  {Steidelm{\"u}ller}, {Stephenson}, {S{\"u}veges}, {Surdej}, {Szabados},
  {Szegedi-Elek}, {Taris}, {Taylor}, {Teixeira}, {Tolomei}, {Tonello}, {Torra},
  {Torra}, {Torralba Elipe}, {Trabucchi}, {Tsounis}, {Turon}, {Ulla}, {Unger},
  {Vaillant}, {van Dillen}, {van Reeven}, {Vanel}, {Vecchiato}, {Viala},
  {Vicente}, {Voutsinas}, {Weiler}, {Wevers}, {Wyrzykowski}, {Yoldas}, {Yvard},
  {Zhao}, {Zorec}, {Zucker}, \& {Zwitter}}]{2023A&A...674A...1G}
{Gaia Collaboration}, {Vallenari}, A., {Brown}, A.~G.~A., {et~al.} 2023, \aap,
  674, A1

\bibitem[{{Giallongo} {et~al.}(2019){Giallongo}, {Grazian}, {Fiore}, {Kodra},
  {Urrutia}, {Castellano}, {Cristiani}, {Dickinson}, {Fontana}, {Menci},
  {Pentericci}, {Boutsia}, {Newman}, \& {Puccetti}}]{2019ApJ...884...19G}
{Giallongo}, E., {Grazian}, A., {Fiore}, F., {et~al.} 2019, \apj, 884, 19

\bibitem[{{Goulding} {et~al.}(2023){Goulding}, {Greene}, {Setton}, {Labbe},
  {Bezanson}, {Miller}, {Atek}, {Bogd{\'a}n}, {Brammer}, {Chemerynska},
  {Cutler}, {Dayal}, {Fudamoto}, {Fujimoto}, {Furtak}, {Kokorev}, {Khullar},
  {Leja}, {Marchesini}, {Natarajan}, {Nelson}, {Oesch}, {Pan}, {Papovich},
  {Price}, {van Dokkum}, {Wang}, {Weaver}, {Whitaker}, \&
  {Zitrin}}]{2023ApJ...955L..24G}
{Goulding}, A.~D., {Greene}, J.~E., {Setton}, D.~J., {et~al.} 2023, \apjl, 955,
  L24

\bibitem[{{Green}(2018)}]{2018JOSS....3..695G}
{Green}, G.~M. 2018, The Journal of Open Source Software, 3, 695

\bibitem[{{Greene} {et~al.}(2023){Greene}, {Labbe}, {Goulding}, {Furtak},
  {Chemerynska}, {Kokorev}, {Dayal}, {Williams}, {Wang}, {Setton}, {Burgasser},
  {Bezanson}, {Atek}, {Brammer}, {Cutler}, {Feldmann}, {Fujimoto},
  {Glazebrook}, {de Graaff}, {Leja}, {Marchesini}, {Maseda}, {Matthee},
  {Miller}, {Naidu}, {Nanayakkara}, {Oesch}, {Pan}, {Papovich}, {Price}, {van
  Dokkum}, {Weaver}, {Whitaker}, \& {Zitrin}}]{2023arXiv230905714G}
{Greene}, J.~E., {Labbe}, I., {Goulding}, A.~D., {et~al.} 2023, arXiv e-prints,
  arXiv:2309.05714

\bibitem[{{Grogin} {et~al.}(2011){Grogin}, {Kocevski}, {Faber}, {Ferguson},
  {Koekemoer}, {Riess}, {Acquaviva}, {Alexander}, {Almaini}, {Ashby}, {Barden},
  {Bell}, {Bournaud}, {Brown}, {Caputi}, {Casertano}, {Cassata}, {Castellano},
  {Challis}, {Chary}, {Cheung}, {Cirasuolo}, {Conselice}, {Roshan Cooray},
  {Croton}, {Daddi}, {Dahlen}, {Dav{\'e}}, {de Mello}, {Dekel}, {Dickinson},
  {Dolch}, {Donley}, {Dunlop}, {Dutton}, {Elbaz}, {Fazio}, {Filippenko},
  {Finkelstein}, {Fontana}, {Gardner}, {Garnavich}, {Gawiser}, {Giavalisco},
  {Grazian}, {Guo}, {Hathi}, {H{\"a}ussler}, {Hopkins}, {Huang}, {Huang},
  {Jha}, {Kartaltepe}, {Kirshner}, {Koo}, {Lai}, {Lee}, {Li}, {Lotz}, {Lucas},
  {Madau}, {McCarthy}, {McGrath}, {McIntosh}, {McLure}, {Mobasher},
  {Moustakas}, {Mozena}, {Nandra}, {Newman}, {Niemi}, {Noeske}, {Papovich},
  {Pentericci}, {Pope}, {Primack}, {Rajan}, {Ravindranath}, {Reddy}, {Renzini},
  {Rix}, {Robaina}, {Rodney}, {Rosario}, {Rosati}, {Salimbeni}, {Scarlata},
  {Siana}, {Simard}, {Smidt}, {Somerville}, {Spinrad}, {Straughn}, {Strolger},
  {Telford}, {Teplitz}, {Trump}, {van der Wel}, {Villforth}, {Wechsler},
  {Weiner}, {Wiklind}, {Wild}, {Wilson}, {Wuyts}, {Yan}, \&
  {Yun}}]{2011ApJS..197...35G}
{Grogin}, N.~A., {Kocevski}, D.~D., {Faber}, S.~M., {et~al.} 2011, \apjs, 197,
  35

\bibitem[{{Hainline} {et~al.}(2023){Hainline}, {Johnson}, {Robertson},
  {Tacchella}, {Helton}, {Sun}, {Eisenstein}, {Simmonds}, {Topping}, {Whitler},
  {Willmer}, {Rieke}, {Suess}, {Hviding}, {Cameron}, {Alberts}, {Baker},
  {Bhatawdekar}, {Boyett}, {Bunker}, {Carniani}, {Charlot}, {Chen}, {Curti},
  {Curtis-Lake}, {D'Eugenio}, {Egami}, {Endsley}, {Hausen}, {Ji}, {Looser},
  {Lyu}, {Maiolino}, {Nelson}, {Puskas}, {Rawle}, {Sandles}, {Saxena}, {Smit},
  {Stark}, {Williams}, {Willott}, \& {Witstok}}]{2023arXiv230602468H}
{Hainline}, K.~N., {Johnson}, B.~D., {Robertson}, B., {et~al.} 2023, arXiv
  e-prints, arXiv:2306.02468

\bibitem[{{Harikane} {et~al.}(2023){Harikane}, {Zhang}, {Nakajima}, {Ouchi},
  {Isobe}, {Ono}, {Hatano}, {Xu}, \& {Umeda}}]{2023arXiv230311946H}
{Harikane}, Y., {Zhang}, Y., {Nakajima}, K., {et~al.} 2023, arXiv e-prints,
  arXiv:2303.11946

\bibitem[{{Harris} {et~al.}(2020){Harris}, {Millman}, {van der Walt},
  {Gommers}, {Virtanen}, {Cournapeau}, {Wieser}, {Taylor}, {Berg}, {Smith},
  {Kern}, {Picus}, {Hoyer}, {van Kerkwijk}, {Brett}, {Haldane}, {del R{\'\i}o},
  {Wiebe}, {Peterson}, {G{\'e}rard-Marchant}, {Sheppard}, {Reddy}, {Weckesser},
  {Abbasi}, {Gohlke}, \& {Oliphant}}]{2020Natur.585..357H}
{Harris}, C.~R., {Millman}, K.~J., {van der Walt}, S.~J., {et~al.} 2020, \nat,
  585, 357

\bibitem[{{Heintz} {et~al.}(2023){Heintz}, {Watson}, {Brammer}, {Vejlgaard},
  {Hutter}, {Strait}, {Matthee}, {Oesch}, {Jakobsson}, {Tanvir}, {Laursen},
  {Naidu}, {Mason}, {Killi}, {Jung}, {Hsiao}, {Abdurro'uf}, {Coe}, {Arrabal
  Haro}, {Finkelstein}, \& {Toft}}]{2023arXiv230600647H}
{Heintz}, K.~E., {Watson}, D., {Brammer}, G., {et~al.} 2023, arXiv e-prints,
  arXiv:2306.00647

\bibitem[{{Hosokawa} {et~al.}(2013){Hosokawa}, {Yorke}, {Inayoshi}, {Omukai},
  \& {Yoshida}}]{2013ApJ...778..178H}
{Hosokawa}, T., {Yorke}, H.~W., {Inayoshi}, K., {Omukai}, K., \& {Yoshida}, N.
  2013, \apj, 778, 178

\bibitem[{{Husser} {et~al.}(2013){Husser}, {Wende-von Berg}, {Dreizler},
  {Homeier}, {Reiners}, {Barman}, \& {Hauschildt}}]{2013A&A...553A...6H}
{Husser}, T.~O., {Wende-von Berg}, S., {Dreizler}, S., {et~al.} 2013, \aap,
  553, A6

\bibitem[{{Inayoshi} {et~al.}(2020){Inayoshi}, {Visbal}, \&
  {Haiman}}]{2020ARA&A..58...27I}
{Inayoshi}, K., {Visbal}, E., \& {Haiman}, Z. 2020, \araa, 58, 27

\bibitem[{{Inoue}(2011)}]{2011MNRAS.415.2920I}
{Inoue}, A.~K. 2011, \mnras, 415, 2920

\bibitem[{{Inoue} {et~al.}(2014){Inoue}, {Shimizu}, {Iwata}, \&
  {Tanaka}}]{2014MNRAS.442.1805I}
{Inoue}, A.~K., {Shimizu}, I., {Iwata}, I., \& {Tanaka}, M. 2014, \mnras, 442,
  1805

\bibitem[{{Isobe} {et~al.}(2023){Isobe}, {Ouchi}, {Nakajima}, {Harikane},
  {Ono}, {Xu}, {Zhang}, \& {Umeda}}]{2023ApJ...956..139I}
{Isobe}, Y., {Ouchi}, M., {Nakajima}, K., {et~al.} 2023, \apj, 956, 139

\bibitem[{{Izumi} {et~al.}(2021){Izumi}, {Matsuoka}, {Fujimoto}, {Onoue},
  {Strauss}, {Umehata}, {Imanishi}, {Kohno}, {Kawaguchi}, {Kawamuro}, {Baba},
  {Nagao}, {Toba}, {Inayoshi}, {Silverman}, {Inoue}, {Ikarashi}, {Iwasawa},
  {Kashikawa}, {Hashimoto}, {Nakanishi}, {Ueda}, {Schramm}, {Lee}, \&
  {Suh}}]{2021ApJ...914...36I}
{Izumi}, T., {Matsuoka}, Y., {Fujimoto}, S., {et~al.} 2021, \apj, 914, 36

\bibitem[{{Kaasinen} {et~al.}(2017){Kaasinen}, {Bian}, {Groves}, {Kewley}, \&
  {Gupta}}]{2017MNRAS.465.3220K}
{Kaasinen}, M., {Bian}, F., {Groves}, B., {Kewley}, L.~J., \& {Gupta}, A. 2017,
  \mnras, 465, 3220

\bibitem[{{Kewley} {et~al.}(2019){Kewley}, {Nicholls}, \&
  {Sutherland}}]{2019ARA&A..57..511K}
{Kewley}, L.~J., {Nicholls}, D.~C., \& {Sutherland}, R.~S. 2019, \araa, 57, 511

\bibitem[{{Kocevski} {et~al.}(2023){Kocevski}, {Onoue}, {Inayoshi}, {Trump},
  {Haro}, {Grazian}, {Dickinson}, {Finkelstein}, {Kartaltepe}, {Hirschmann},
  {Aird}, {Holwerda}, {Fujimoto}, {Juneau}, {Amor{\'\i}n}, {Backhaus},
  {Bagley}, {Barro}, {Bell}, {Bisigello}, {Calabr{\`o}}, {Cleri}, {Cooper},
  {Ding}, {Grogin}, {Ho}, {Hutchison}, {Inoue}, {Jiang}, {Jones}, {Koekemoer},
  {Li}, {Li}, {McGrath}, {Molina}, {Papovich}, {P{\'e}rez-Gonz{\'a}lez},
  {Pirzkal}, {Wilkins}, {Yang}, \& {Yung}}]{2023ApJ...954L...4K}
{Kocevski}, D.~D., {Onoue}, M., {Inayoshi}, K., {et~al.} 2023, \apjl, 954, L4

\bibitem[{{Koekemoer} {et~al.}(2007){Koekemoer}, {Aussel}, {Calzetti}, {Capak},
  {Giavalisco}, {Kneib}, {Leauthaud}, {Le F{\`e}vre}, {McCracken}, {Massey},
  {Mobasher}, {Rhodes}, {Scoville}, \& {Shopbell}}]{2007ApJS..172..196K}
{Koekemoer}, A.~M., {Aussel}, H., {Calzetti}, D., {et~al.} 2007, \apjs, 172,
  196

\bibitem[{{Koekemoer} {et~al.}(2011){Koekemoer}, {Faber}, {Ferguson}, {Grogin},
  {Kocevski}, {Koo}, {Lai}, {Lotz}, {Lucas}, {McGrath}, {Ogaz}, {Rajan},
  {Riess}, {Rodney}, {Strolger}, {Casertano}, {Castellano}, {Dahlen},
  {Dickinson}, {Dolch}, {Fontana}, {Giavalisco}, {Grazian}, {Guo}, {Hathi},
  {Huang}, {van der Wel}, {Yan}, {Acquaviva}, {Alexander}, {Almaini}, {Ashby},
  {Barden}, {Bell}, {Bournaud}, {Brown}, {Caputi}, {Cassata}, {Challis},
  {Chary}, {Cheung}, {Cirasuolo}, {Conselice}, {Roshan Cooray}, {Croton},
  {Daddi}, {Dav{\'e}}, {de Mello}, {de Ravel}, {Dekel}, {Donley}, {Dunlop},
  {Dutton}, {Elbaz}, {Fazio}, {Filippenko}, {Finkelstein}, {Frazer}, {Gardner},
  {Garnavich}, {Gawiser}, {Gruetzbauch}, {Hartley}, {H{\"a}ussler},
  {Herrington}, {Hopkins}, {Huang}, {Jha}, {Johnson}, {Kartaltepe},
  {Khostovan}, {Kirshner}, {Lani}, {Lee}, {Li}, {Madau}, {McCarthy},
  {McIntosh}, {McLure}, {McPartland}, {Mobasher}, {Moreira}, {Mortlock},
  {Moustakas}, {Mozena}, {Nandra}, {Newman}, {Nielsen}, {Niemi}, {Noeske},
  {Papovich}, {Pentericci}, {Pope}, {Primack}, {Ravindranath}, {Reddy},
  {Renzini}, {Rix}, {Robaina}, {Rosario}, {Rosati}, {Salimbeni}, {Scarlata},
  {Siana}, {Simard}, {Smidt}, {Snyder}, {Somerville}, {Spinrad}, {Straughn},
  {Telford}, {Teplitz}, {Trump}, {Vargas}, {Villforth}, {Wagner}, {Wandro},
  {Wechsler}, {Weiner}, {Wiklind}, {Wild}, {Wilson}, {Wuyts}, \&
  {Yun}}]{2011ApJS..197...36K}
{Koekemoer}, A.~M., {Faber}, S.~M., {Ferguson}, H.~C., {et~al.} 2011, \apjs,
  197, 36

\bibitem[{{Kokorev} {et~al.}(2022){Kokorev}, {Brammer}, {Fujimoto}, {Kohno},
  {Magdis}, {Valentino}, {Toft}, {Oesch}, {Davidzon}, {Bauer}, {Coe}, {Egami},
  {Oguri}, {Ouchi}, {Postman}, {Richard}, {Jolly}, {Knudsen}, {Sun}, {Weaver},
  {Ao}, {Baker}, {Bradley}, {Caputi}, {Dessauges-Zavadsky}, {Espada},
  {Hatsukade}, {Koekemoer}, {Mu{\~n}oz Arancibia}, {Shimasaku}, {Umehata},
  {Wang}, \& {Wang}}]{2022ApJS..263...38K}
{Kokorev}, V., {Brammer}, G., {Fujimoto}, S., {et~al.} 2022, \apjs, 263, 38

\bibitem[{{Kokorev} {et~al.}(2024){Kokorev}, {Caputi}, {Greene}, {Dayal},
  {Trebitsch}, {Cutler}, {Fujimoto}, {Miller}, {Iani}, {Navarro-Carrera}, \&
  {Rinaldi}}]{2024arXiv240109981K}
{Kokorev}, V., {Caputi}, K.~I., {Greene}, J.~E., {et~al.} 2024, arXiv e-prints,
  arXiv:2401.09981

\bibitem[{{Kokorev} {et~al.}(2023){Kokorev}, {Fujimoto}, {Labbe}, {Greene},
  {Bezanson}, {Dayal}, {Nelson}, {Atek}, {Brammer}, {Caputi}, {Chemerynska},
  {Cutler}, {Feldmann}, {Fudamoto}, {Furtak}, {Goulding}, {de Graaff}, {Leja},
  {Marchesini}, {Miller}, {Nanayakkara}, {Oesch}, {Pan}, {Price}, {Setton},
  {Smit}, {Stefanon}, {Wang}, {Weaver}, {Whitaker}, {Williams}, \&
  {Zitrin}}]{2023arXiv230811610K}
{Kokorev}, V., {Fujimoto}, S., {Labbe}, I., {et~al.} 2023, arXiv e-prints,
  arXiv:2308.11610

\bibitem[{{Kormendy} \& {Ho}(2013)}]{2013ARA&A..51..511K}
{Kormendy}, J. \& {Ho}, L.~C. 2013, \araa, 51, 511

\bibitem[{{Labbe} {et~al.}(2023){Labbe}, {Greene}, {Bezanson}, {Fujimoto},
  {Furtak}, {Goulding}, {Matthee}, {Naidu}, {Oesch}, {Atek}, {Brammer},
  {Chemerynska}, {Coe}, {Cutler}, {Dayal}, {Feldmann}, {Franx}, {Glazebrook},
  {Leja}, {Marchesini}, {Maseda}, {Nanayakkara}, {Nelson}, {Pan}, {Papovich},
  {Price}, {Suess}, {Wang}, {Whitaker}, {Williams}, \&
  {Zitrin}}]{2023arXiv230607320L}
{Labbe}, I., {Greene}, J.~E., {Bezanson}, R., {et~al.} 2023, arXiv e-prints,
  arXiv:2306.07320

\bibitem[{{Larson} {et~al.}(2023){Larson}, {Finkelstein}, {Kocevski},
  {Hutchison}, {Trump}, {Haro}, {Bromm}, {Cleri}, {Dickinson}, {Fujimoto},
  {Kartaltepe}, {Koekemoer}, {Papovich}, {Pirzkal}, {Tacchella}, {Zavala},
  {Bagley}, {Behroozi}, {Champagne}, {Cole}, {Jung}, {Morales}, {Yang},
  {Zhang}, {Zitrin}, {Amor{\'\i}n}, {Burgarella}, {Casey}, {Ch{\'a}vez Ortiz},
  {Cox}, {Chworowsky}, {Fontana}, {Gawiser}, {Grazian}, {Grogin}, {Harish},
  {Hathi}, {Hirschmann}, {Holwerda}, {Juneau}, {Leung}, {Lucas}, {McGrath},
  {P{\'e}rez-Gonz{\'a}lez}, {Rigby}, {Seill{\'e}}, {Simons}, {de La Vega},
  {Weiner}, {Wilkins}, {Yung}, \& {Ceers Team}}]{2023ApJ...953L..29L}
{Larson}, R.~L., {Finkelstein}, S.~L., {Kocevski}, D.~D., {et~al.} 2023, \apjl,
  953, L29

\bibitem[{{Larson} {et~al.}(2022){Larson}, {Hutchison}, {Bagley},
  {Finkelstein}, {Yung}, {Somerville}, {Hirschmann}, {Brammer}, {Holwerda},
  {Papovich}, {Morales}, \& {Wilkins}}]{2022arXiv221110035L}
{Larson}, R.~L., {Hutchison}, T.~A., {Bagley}, M., {et~al.} 2022, arXiv
  e-prints, arXiv:2211.10035

\bibitem[{{Latif} {et~al.}(2021){Latif}, {Khochfar}, {Schleicher}, \&
  {Whalen}}]{2021MNRAS.508.1756L}
{Latif}, M.~A., {Khochfar}, S., {Schleicher}, D., \& {Whalen}, D.~J. 2021,
  \mnras, 508, 1756

\bibitem[{{Lauer} {et~al.}(2007){Lauer}, {Tremaine}, {Richstone}, \&
  {Faber}}]{2007ApJ...670..249L}
{Lauer}, T.~R., {Tremaine}, S., {Richstone}, D., \& {Faber}, S.~M. 2007, \apj,
  670, 249

\bibitem[{{Lawrence} {et~al.}(2007){Lawrence}, {Warren}, {Almaini}, {Edge},
  {Hambly}, {Jameson}, {Lucas}, {Casali}, {Adamson}, {Dye}, {Emerson},
  {Foucaud}, {Hewett}, {Hirst}, {Hodgkin}, {Irwin}, {Lodieu}, {McMahon},
  {Simpson}, {Smail}, {Mortlock}, \& {Folger}}]{2007MNRAS.379.1599L}
{Lawrence}, A., {Warren}, S.~J., {Almaini}, O., {et~al.} 2007, \mnras, 379,
  1599

\bibitem[{{Leitherer} {et~al.}(2002){Leitherer}, {Li}, {Calzetti}, \&
  {Heckman}}]{2002ApJS..140..303L}
{Leitherer}, C., {Li}, I.~H., {Calzetti}, D., \& {Heckman}, T.~M. 2002, \apjs,
  140, 303

\bibitem[{{Li} {et~al.}(2021){Li}, {Silverman}, {Ding}, {Strauss}, {Goulding},
  {Schramm}, {Yesuf}, {Sun}, {Xue}, {Birrer}, {Shi}, {Toba}, {Nagao}, \&
  {Imanishi}}]{2021ApJ...922..142L}
{Li}, J., {Silverman}, J.~D., {Ding}, X., {et~al.} 2021, \apj, 922, 142

\bibitem[{{Li} {et~al.}(2022){Li}, {Silverman}, {Izumi}, {He}, {Akiyama},
  {Inayoshi}, {Matsuoka}, {Onoue}, \& {Toba}}]{2022ApJ...931L..11L}
{Li}, J., {Silverman}, J.~D., {Izumi}, T., {et~al.} 2022, \apjl, 931, L11

\bibitem[{{Liu} {et~al.}(2019){Liu}, {Lang}, {Magnelli}, {Schinnerer},
  {Leslie}, {Fudamoto}, {Bondi}, {Groves}, {Jim{\'e}nez-Andrade}, {Harrington},
  {Karim}, {Oesch}, {Sargent}, {Vardoulaki}, {B{\v{a}}descu}, {Moser},
  {Bertoldi}, {Battisti}, {da Cunha}, {Zavala}, {Vaccari}, {Davidzon},
  {Riechers}, \& {Aravena}}]{2019ApJS..244...40L}
{Liu}, D., {Lang}, P., {Magnelli}, B., {et~al.} 2019, \apjs, 244, 40

\bibitem[{{Lodato} \& {Natarajan}(2006)}]{2006MNRAS.371.1813L}
{Lodato}, G. \& {Natarajan}, P. 2006, \mnras, 371, 1813

\bibitem[{{Lupi} {et~al.}(2016){Lupi}, {Haardt}, {Dotti}, {Fiacconi}, {Mayer},
  \& {Madau}}]{2016MNRAS.456.2993L}
{Lupi}, A., {Haardt}, F., {Dotti}, M., {et~al.} 2016, \mnras, 456, 2993

\bibitem[{{Madau} {et~al.}(2014){Madau}, {Haardt}, \&
  {Dotti}}]{2014ApJ...784L..38M}
{Madau}, P., {Haardt}, F., \& {Dotti}, M. 2014, \apjl, 784, L38

\bibitem[{{Maiolino} {et~al.}(2023{\natexlab{a}}){Maiolino}, {Scholtz},
  {Curtis-Lake}, {Carniani}, {Baker}, {de Graaff}, {Tacchella}, {{\"U}bler},
  {D'Eugenio}, {Witstok}, {Curti}, {Arribas}, {Bunker}, {Charlot},
  {Chevallard}, {Eisenstein}, {Egami}, {Ji}, {Jones}, {Lyu}, {Rawle},
  {Robertson}, {Rujopakarn}, {Perna}, {Sun}, {Venturi}, {Williams}, \&
  {Willott}}]{2023arXiv230801230M}
{Maiolino}, R., {Scholtz}, J., {Curtis-Lake}, E., {et~al.} 2023{\natexlab{a}},
  arXiv e-prints, arXiv:2308.01230

\bibitem[{{Maiolino} {et~al.}(2023{\natexlab{b}}){Maiolino}, {Scholtz},
  {Witstok}, {Carniani}, {D'Eugenio}, {de Graaff}, {Uebler}, {Tacchella},
  {Curtis-Lake}, {Arribas}, {Bunker}, {Charlot}, {Chevallard}, {Curti},
  {Looser}, {Maseda}, {Rawle}, {Rodriguez Del Pino}, {Willott}, {Egami},
  {Eisenstein}, {Hainline}, {Robertson}, {Williams}, {Willmer}, {Baker},
  {Boyett}, {DeCoursey}, {Fabian}, {Helton}, {Ji}, {Jones}, {Kumari},
  {Laporte}, {Nelson}, {Perna}, {Sandles}, {Shivaei}, \&
  {Sun}}]{2023arXiv230512492M}
{Maiolino}, R., {Scholtz}, J., {Witstok}, J., {et~al.} 2023{\natexlab{b}},
  arXiv e-prints, arXiv:2305.12492

\bibitem[{{Marley} {et~al.}(2021){Marley}, {Saumon}, {Visscher}, {Lupu},
  {Freedman}, {Morley}, {Fortney}, {Seay}, {Smith}, {Teal}, \&
  {Wang}}]{2021ApJ...920...85M}
{Marley}, M.~S., {Saumon}, D., {Visscher}, C., {et~al.} 2021, \apj, 920, 85

\bibitem[{{Massonneau} {et~al.}(2023){Massonneau}, {Volonteri}, {Dubois}, \&
  {Beckmann}}]{2023A&A...670A.180M}
{Massonneau}, W., {Volonteri}, M., {Dubois}, Y., \& {Beckmann}, R.~S. 2023,
  \aap, 670, A180

\bibitem[{{Matsuoka} {et~al.}(2019){Matsuoka}, {Iwasawa}, {Onoue}, {Kashikawa},
  {Strauss}, {Lee}, {Imanishi}, {Nagao}, {Akiyama}, {Asami}, {Bosch},
  {Furusawa}, {Goto}, {Gunn}, {Harikane}, {Ikeda}, {Izumi}, {Kawaguchi},
  {Kato}, {Kikuta}, {Kohno}, {Komiyama}, {Koyama}, {Lupton}, {Minezaki},
  {Miyazaki}, {Murayama}, {Niida}, {Nishizawa}, {Noboriguchi}, {Oguri}, {Ono},
  {Ouchi}, {Price}, {Sameshima}, {Schulze}, {Silverman}, {Sugiyama}, {Tait},
  {Takada}, {Takata}, {Tanaka}, {Tang}, {Toba}, {Utsumi}, {Wang}, \&
  {Yamashita}}]{2019ApJ...883..183M}
{Matsuoka}, Y., {Iwasawa}, K., {Onoue}, M., {et~al.} 2019, \apj, 883, 183

\bibitem[{{Matsuoka} {et~al.}(2018){Matsuoka}, {Strauss}, {Kashikawa}, {Onoue},
  {Iwasawa}, {Tang}, {Lee}, {Imanishi}, {Nagao}, {Akiyama}, {Asami}, {Bosch},
  {Furusawa}, {Goto}, {Gunn}, {Harikane}, {Ikeda}, {Izumi}, {Kawaguchi},
  {Kato}, {Kikuta}, {Kohno}, {Komiyama}, {Lupton}, {Minezaki}, {Miyazaki},
  {Murayama}, {Niida}, {Nishizawa}, {Noboriguchi}, {Oguri}, {Ono}, {Ouchi},
  {Price}, {Sameshima}, {Schulze}, {Shirakata}, {Silverman}, {Sugiyama},
  {Tait}, {Takada}, {Takata}, {Tanaka}, {Toba}, {Utsumi}, {Wang}, \&
  {Yamashita}}]{2018ApJ...869..150M}
{Matsuoka}, Y., {Strauss}, M.~A., {Kashikawa}, N., {et~al.} 2018, \apj, 869,
  150

\bibitem[{{Mayer} \& {Bonoli}(2019)}]{2019RPPh...82a6901M}
{Mayer}, L. \& {Bonoli}, S. 2019, Reports on Progress in Physics, 82, 016901

\bibitem[{{McCracken} {et~al.}(2012){McCracken}, {Milvang-Jensen}, {Dunlop},
  {Franx}, {Fynbo}, {Le F{\`e}vre}, {Holt}, {Caputi}, {Goranova}, {Buitrago},
  {Emerson}, {Freudling}, {Hudelot}, {L{\'o}pez-Sanjuan}, {Magnard}, {Mellier},
  {M{\o}ller}, {Nilsson}, {Sutherland}, {Tasca}, \&
  {Zabl}}]{2012A&A...544A.156M}
{McCracken}, H.~J., {Milvang-Jensen}, B., {Dunlop}, J., {et~al.} 2012, \aap,
  544, A156

\bibitem[{{Meiksin}(2006)}]{2006MNRAS.365..807M}
{Meiksin}, A. 2006, \mnras, 365, 807

\bibitem[{{Meyer} {et~al.}(2023){Meyer}, {Barrufet}, {Boogaard}, {Naidu},
  {Oesch}, \& {Walter}}]{2023arXiv231020675M}
{Meyer}, R.~A., {Barrufet}, L., {Boogaard}, L.~A., {et~al.} 2023, arXiv
  e-prints, arXiv:2310.20675

\bibitem[{{Middleton} {et~al.}(2013){Middleton}, {Miller-Jones}, {Markoff},
  {Fender}, {Henze}, {Hurley-Walker}, {Scaife}, {Roberts}, {Walton},
  {Carpenter}, {Macquart}, {Bower}, {Gurwell}, {Pietsch}, {Haberl}, {Harris},
  {Daniel}, {Miah}, {Done}, {Morgan}, {Dickinson}, {Charles}, {Burwitz}, {Della
  Valle}, {Freyberg}, {Greiner}, {Hernanz}, {Hartmann}, {Hatzidimitriou},
  {Riffeser}, {Sala}, {Seitz}, {Reig}, {Rau}, {Orio}, {Titterington}, \&
  {Grainge}}]{2013Natur.493..187M}
{Middleton}, M.~J., {Miller-Jones}, J. C.~A., {Markoff}, S., {et~al.} 2013,
  \nat, 493, 187

\bibitem[{{Mineshige} {et~al.}(2000){Mineshige}, {Kawaguchi}, {Takeuchi}, \&
  {Hayashida}}]{2000PASJ...52..499M}
{Mineshige}, S., {Kawaguchi}, T., {Takeuchi}, M., \& {Hayashida}, K. 2000,
  \pasj, 52, 499

\bibitem[{{Morishita} {et~al.}(2023){Morishita}, {Roberts-Borsani}, {Treu},
  {Brammer}, {Mason}, {Trenti}, {Vulcani}, {Wang}, {Acebron}, {Bah{\'e}},
  {Bergamini}, {Boyett}, {Bradac}, {Calabr{\`o}}, {Castellano}, {Chen}, {De
  Lucia}, {Filippenko}, {Fontana}, {Glazebrook}, {Grillo}, {Henry}, {Jones},
  {Kelly}, {Koekemoer}, {Leethochawalit}, {Lu}, {Marchesini}, {Mascia},
  {Mercurio}, {Merlin}, {Metha}, {Nanayakkara}, {Nonino}, {Paris},
  {Pentericci}, {Rosati}, {Santini}, {Strait}, {Vanzella}, {Windhorst}, \&
  {Xie}}]{2023ApJ...947L..24M}
{Morishita}, T., {Roberts-Borsani}, G., {Treu}, T., {et~al.} 2023, \apjl, 947,
  L24

\bibitem[{{Mortlock} {et~al.}(2011){Mortlock}, {Warren}, {Venemans}, {Patel},
  {Hewett}, {McMahon}, {Simpson}, {Theuns}, {Gonz{\'a}les-Solares}, {Adamson},
  {Dye}, {Hambly}, {Hirst}, {Irwin}, {Kuiper}, {Lawrence}, \&
  {R{\"o}ttgering}}]{2011Natur.474..616M}
{Mortlock}, D.~J., {Warren}, S.~J., {Venemans}, B.~P., {et~al.} 2011, \nat,
  474, 616

\bibitem[{{Nabizadeh} {et~al.}(2023){Nabizadeh}, {Zackrisson}, {Pacucci},
  {Maksym}, {Li}, {Civano}, {Cohen}, {D'Silva}, {Koekemoer}, {Summers},
  {Windhorst}, {Adams}, {Conselice}, {Coe}, {Driver}, {Frye}, {Grogin},
  {Jansen}, {Marshall}, {Nonino}, {Pirzkal}, {Robotham}, {Rutkowski}, {Ryan},
  {Tompkins}, {Willmer}, {Yan}, {Diego}, {Cheng}, {Finkelstein}, {Willner},
  {Zitrin}, {Bhatawdekar}, \& {Gim}}]{2023arXiv230807260N}
{Nabizadeh}, A., {Zackrisson}, E., {Pacucci}, F., {et~al.} 2023, arXiv
  e-prints, arXiv:2308.07260

\bibitem[{{Nakajima} {et~al.}(2023){Nakajima}, {Ouchi}, {Isobe}, {Harikane},
  {Zhang}, {Ono}, {Umeda}, \& {Oguri}}]{2023ApJS..269...33N}
{Nakajima}, K., {Ouchi}, M., {Isobe}, Y., {et~al.} 2023, \apjs, 269, 33

\bibitem[{{Natarajan}(2021)}]{2021MNRAS.501.1413N}
{Natarajan}, P. 2021, \mnras, 501, 1413

\bibitem[{{Natarajan} {et~al.}(2023){Natarajan}, {Pacucci}, {Ricarte},
  {Bogdan}, {Goulding}, \& {Cappelluti}}]{2023arXiv230802654N}
{Natarajan}, P., {Pacucci}, F., {Ricarte}, A., {et~al.} 2023, arXiv e-prints,
  arXiv:2308.02654

\bibitem[{{Nishizawa} {et~al.}(2020){Nishizawa}, {Hsieh}, {Tanaka}, \&
  {Takata}}]{2020arXiv200301511N}
{Nishizawa}, A.~J., {Hsieh}, B.-C., {Tanaka}, M., \& {Takata}, T. 2020, arXiv
  e-prints, arXiv:2003.01511

\bibitem[{{Oesch} {et~al.}(2023){Oesch}, {Brammer}, {Naidu}, {Bouwens},
  {Chisholm}, {Illingworth}, {Matthee}, {Nelson}, {Qin}, {Reddy}, {Shapley},
  {Shivaei}, {van Dokkum}, {Weibel}, {Whitaker}, {Wuyts}, {Covelo-Paz},
  {Endsley}, {Fudamoto}, {Giovinazzo}, {Herard-Demanche}, {Kerutt},
  {Kramarenko}, {Labbe}, {Leonova}, {Lin}, {Magee}, {Marchesini}, {Maseda},
  {Mason}, {Matharu}, {Meyer}, {Neufeld}, {Prieto Lyon}, {Schaerer}, {Sharma},
  {Shuntov}, {Smit}, {Stefanon}, {Wyithe}, \& {Xiao}}]{2023MNRAS.525.2864O}
{Oesch}, P.~A., {Brammer}, G., {Naidu}, R.~P., {et~al.} 2023, \mnras, 525, 2864

\bibitem[{{Pacucci} \& {Loeb}(2020)}]{2020ApJ...895...95P}
{Pacucci}, F. \& {Loeb}, A. 2020, \apj, 895, 95

\bibitem[{{Pacucci} \& {Loeb}(2022)}]{2022MNRAS.509.1885P}
{Pacucci}, F. \& {Loeb}, A. 2022, \mnras, 509, 1885

\bibitem[{{Pacucci} \& {Loeb}(2024)}]{2024arXiv240104159P}
{Pacucci}, F. \& {Loeb}, A. 2024, arXiv e-prints, arXiv:2401.04159

\bibitem[{{Pacucci} {et~al.}(2017){Pacucci}, {Natarajan}, {Volonteri},
  {Cappelluti}, \& {Urry}}]{2017ApJ...850L..42P}
{Pacucci}, F., {Natarajan}, P., {Volonteri}, M., {Cappelluti}, N., \& {Urry},
  C.~M. 2017, \apjl, 850, L42

\bibitem[{{Pacucci} {et~al.}(2023){Pacucci}, {Nguyen}, {Carniani}, {Maiolino},
  \& {Fan}}]{2023ApJ...957L...3P}
{Pacucci}, F., {Nguyen}, B., {Carniani}, S., {Maiolino}, R., \& {Fan}, X. 2023,
  \apjl, 957, L3

\bibitem[{{Pacucci} {et~al.}(2015){Pacucci}, {Volonteri}, \&
  {Ferrara}}]{2015MNRAS.452.1922P}
{Pacucci}, F., {Volonteri}, M., \& {Ferrara}, A. 2015, \mnras, 452, 1922

\bibitem[{{Parsa} {et~al.}(2018){Parsa}, {Dunlop}, \&
  {McLure}}]{2018MNRAS.474.2904P}
{Parsa}, S., {Dunlop}, J.~S., \& {McLure}, R.~J. 2018, \mnras, 474, 2904

\bibitem[{{P{\'e}rez-Gonz{\'a}lez} {et~al.}(2024){P{\'e}rez-Gonz{\'a}lez},
  {Barro}, {Rieke}, {Lyu}, {Rieke}, {Alberts}, {Williams}, {Hainline}, {Sun},
  {Puskas}, {Annunziatella}, {Baker}, {Bunker}, {Egami}, {Ji}, {Johnson},
  {Robertson}, {Rodriguez Del Pino}, {Rujopakarn}, {Shivaei}, {Tacchella},
  {Willmer}, \& {Willott}}]{2024arXiv240108782P}
{P{\'e}rez-Gonz{\'a}lez}, P.~G., {Barro}, G., {Rieke}, G.~H., {et~al.} 2024,
  arXiv e-prints, arXiv:2401.08782

\bibitem[{{Pezzulli} {et~al.}(2016){Pezzulli}, {Valiante}, \&
  {Schneider}}]{2016MNRAS.458.3047P}
{Pezzulli}, E., {Valiante}, R., \& {Schneider}, R. 2016, \mnras, 458, 3047

\bibitem[{{Reback} {et~al.}(2022){Reback}, {jbrockmendel}, {McKinney}, {Van den
  Bossche}, {Augspurger}, {Roeschke}, {Hawkins}, {Cloud}, {gfyoung}, {Sinhrks},
  {Hoefler}, {Klein}, {Petersen}, {Tratner}, {She}, {Ayd}, {Naveh},
  {Darbyshire}, {Garcia}, {Shadrach}, {Schendel}, {Hayden}, {Saxton},
  {Gorelli}, {Li}, {Zeitlin}, {Jancauskas}, {McMaster}, {W{\"o}rtwein}, \&
  {Battiston}}]{2022zndo...3509134R}
{Reback}, J., {jbrockmendel}, {McKinney}, W., {et~al.} 2022,
  {pandas-dev/pandas: Pandas 1.4.2}, Zenodo

\bibitem[{{Reines} \& {Volonteri}(2015)}]{2015ApJ...813...82R}
{Reines}, A.~E. \& {Volonteri}, M. 2015, \apj, 813, 82

\bibitem[{{Richards} {et~al.}(2006){Richards}, {Lacy}, {Storrie-Lombardi},
  {Hall}, {Gallagher}, {Hines}, {Fan}, {Papovich}, {Vanden Berk}, {Trammell},
  {Schneider}, {Vestergaard}, {York}, {Jester}, {Anderson}, {Budav{\'a}ri}, \&
  {Szalay}}]{2006ApJS..166..470R}
{Richards}, G.~T., {Lacy}, M., {Storrie-Lombardi}, L.~J., {et~al.} 2006, \apjs,
  166, 470

\bibitem[{{Roberts-Borsani} {et~al.}(2022){Roberts-Borsani}, {Morishita},
  {Treu}, {Brammer}, {Strait}, {Wang}, {Bradac}, {Acebron}, {Bergamini},
  {Boyett}, {Calabr{\'o}}, {Castellano}, {Fontana}, {Glazebrook}, {Grillo},
  {Henry}, {Jones}, {Malkan}, {Marchesini}, {Mascia}, {Mason}, {Mercurio},
  {Merlin}, {Nanayakkara}, {Pentericci}, {Rosati}, {Santini}, {Scarlata},
  {Trenti}, {Vanzella}, {Vulcani}, \& {Willott}}]{2022ApJ...938L..13R}
{Roberts-Borsani}, G., {Morishita}, T., {Treu}, T., {et~al.} 2022, \apjl, 938,
  L13

\bibitem[{Rocklin(2015)}]{matthew_rocklin-proc-scipy-2015}
Rocklin, M. 2015, in Proceedings of the 14th Python in Science Conference, ed.
  K.~Huff \& J.~Bergstra, 130--136

\bibitem[{{Sanders} {et~al.}(2023){Sanders}, {Shapley}, {Topping}, {Reddy}, \&
  {Brammer}}]{2023arXiv230308149S}
{Sanders}, R.~L., {Shapley}, A.~E., {Topping}, M.~W., {Reddy}, N.~A., \&
  {Brammer}, G.~B. 2023, arXiv e-prints, arXiv:2303.08149

\bibitem[{{Saxena} {et~al.}(2023){Saxena}, {Robertson}, {Bunker}, {Endsley},
  {Cameron}, {Charlot}, {Simmonds}, {Tacchella}, {Witstok}, {Willott},
  {Carniani}, {Curtis-Lake}, {Ferruit}, {Jakobsen}, {Arribas}, {Chevallard},
  {Curti}, {D'Eugenio}, {De Graaff}, {Jones}, {Looser}, {Maseda}, {Rawle},
  {Rix}, {Del Pino}, {Smit}, {{\"U}bler}, {Eisenstein}, {Hainline}, {Hausen},
  {Johnson}, {Rieke}, {Williams}, {Willmer}, {Baker}, {Bhatawdekar}, {Bowler},
  {Boyett}, {Chen}, {Egami}, {Ji}, {Kumari}, {Nelson}, {Perna}, {Sandles},
  {Scholtz}, \& {Shivaei}}]{2023A&A...678A..68S}
{Saxena}, A., {Robertson}, B.~E., {Bunker}, A.~J., {et~al.} 2023, \aap, 678,
  A68

\bibitem[{{Schindler} {et~al.}(2023){Schindler}, {Ba{\~n}ados}, {Connor},
  {Decarli}, {Fan}, {Farina}, {Mazzucchelli}, {Nanni}, {Rix}, {Stern},
  {Venemans}, \& {Walter}}]{2023ApJ...943...67S}
{Schindler}, J.-T., {Ba{\~n}ados}, E., {Connor}, T., {et~al.} 2023, \apj, 943,
  67

\bibitem[{{Schlegel} {et~al.}(1998){Schlegel}, {Finkbeiner}, \&
  {Davis}}]{1998ApJ...500..525S}
{Schlegel}, D.~J., {Finkbeiner}, D.~P., \& {Davis}, M. 1998, \apj, 500, 525

\bibitem[{{Scoville} {et~al.}(2007){Scoville}, {Aussel}, {Brusa}, {Capak},
  {Carollo}, {Elvis}, {Giavalisco}, {Guzzo}, {Hasinger}, {Impey}, {Kneib},
  {LeFevre}, {Lilly}, {Mobasher}, {Renzini}, {Rich}, {Sanders}, {Schinnerer},
  {Schminovich}, {Shopbell}, {Taniguchi}, \& {Tyson}}]{2007ApJS..172....1S}
{Scoville}, N., {Aussel}, H., {Brusa}, M., {et~al.} 2007, \apjs, 172, 1

\bibitem[{{Smith} \& {Bromm}(2019)}]{2019ConPh..60..111S}
{Smith}, A. \& {Bromm}, V. 2019, Contemporary Physics, 60, 111

\bibitem[{{Stalevski} {et~al.}(2012){Stalevski}, {Fritz}, {Baes}, {Nakos}, \&
  {Popovi{\'c}}}]{2012MNRAS.420.2756S}
{Stalevski}, M., {Fritz}, J., {Baes}, M., {Nakos}, T., \& {Popovi{\'c}},
  L.~{\v{C}}. 2012, \mnras, 420, 2756

\bibitem[{{Stalevski} {et~al.}(2016){Stalevski}, {Ricci}, {Ueda}, {Lira},
  {Fritz}, \& {Baes}}]{2016MNRAS.458.2288S}
{Stalevski}, M., {Ricci}, C., {Ueda}, Y., {et~al.} 2016, \mnras, 458, 2288

\bibitem[{{Stone} {et~al.}(2023){Stone}, {Lyu}, {Rieke}, \&
  {Alberts}}]{2023ApJ...953..180S}
{Stone}, M.~A., {Lyu}, J., {Rieke}, G.~H., \& {Alberts}, S. 2023, \apj, 953,
  180

\bibitem[{{Suh} {et~al.}(2020){Suh}, {Civano}, {Trakhtenbrot}, {Shankar},
  {Hasinger}, {Sanders}, \& {Allevato}}]{2020ApJ...889...32S}
{Suh}, H., {Civano}, F., {Trakhtenbrot}, B., {et~al.} 2020, \apj, 889, 32

\bibitem[{{Szalay} {et~al.}(1999){Szalay}, {Connolly}, \&
  {Szokoly}}]{1999AJ....117...68S}
{Szalay}, A.~S., {Connolly}, A.~J., \& {Szokoly}, G.~P. 1999, \aj, 117, 68

\bibitem[{{Tang} {et~al.}(2023){Tang}, {Stark}, {Chen}, {Mason}, {Topping},
  {Endsley}, {Senchyna}, {Plat}, {Lu}, {Whitler}, {Robertson}, \&
  {Charlot}}]{2023MNRAS.526.1657T}
{Tang}, M., {Stark}, D.~P., {Chen}, Z., {et~al.} 2023, \mnras, 526, 1657

\bibitem[{{Taylor}(2005)}]{2005ASPC..347...29T}
{Taylor}, M.~B. 2005, in Astronomical Society of the Pacific Conference Series,
  Vol. 347, Astronomical Data Analysis Software and Systems XIV, ed.
  P.~{Shopbell}, M.~{Britton}, \& R.~{Ebert}, 29

\bibitem[{{Trakhtenbrot} {et~al.}(2017){Trakhtenbrot}, {Volonteri}, \&
  {Natarajan}}]{2017ApJ...836L...1T}
{Trakhtenbrot}, B., {Volonteri}, M., \& {Natarajan}, P. 2017, \apjl, 836, L1

\bibitem[{{Trinca} {et~al.}(2023){Trinca}, {Schneider}, {Maiolino}, {Valiante},
  {Graziani}, \& {Volonteri}}]{2023MNRAS.519.4753T}
{Trinca}, A., {Schneider}, R., {Maiolino}, R., {et~al.} 2023, \mnras, 519, 4753

\bibitem[{{{\"U}bler} {et~al.}(2023){{\"U}bler}, {Maiolino}, {Curtis-Lake},
  {P{\'e}rez-Gonz{\'a}lez}, {Curti}, {Perna}, {Arribas}, {Charlot}, {Marshall},
  {D'Eugenio}, {Scholtz}, {Bunker}, {Carniani}, {Ferruit}, {Jakobsen}, {Rix},
  {Rodr{\'\i}guez Del Pino}, {Willott}, {Boeker}, {Cresci}, {Jones}, {Kumari},
  \& {Rawle}}]{2023A&A...677A.145U}
{{\"U}bler}, H., {Maiolino}, R., {Curtis-Lake}, E., {et~al.} 2023, \aap, 677,
  A145

\bibitem[{{Valentino} {et~al.}(2023){Valentino}, {Brammer}, {Gould}, {Kokorev},
  {Fujimoto}, {Jespersen}, {Vijayan}, {Weaver}, {Ito}, {Tanaka}, {Ilbert},
  {Magdis}, {Whitaker}, {Faisst}, {Gallazzi}, {Gillman}, {Gim{\'e}nez-Arteaga},
  {G{\'o}mez-Guijarro}, {Kubo}, {Heintz}, {Hirschmann}, {Oesch}, {Onodera},
  {Rizzo}, {Lee}, {Strait}, \& {Toft}}]{2023ApJ...947...20V}
{Valentino}, F., {Brammer}, G., {Gould}, K. M.~L., {et~al.} 2023, \apj, 947, 20

\bibitem[{{Valiante} {et~al.}(2016){Valiante}, {Schneider}, {Volonteri}, \&
  {Omukai}}]{2016MNRAS.457.3356V}
{Valiante}, R., {Schneider}, R., {Volonteri}, M., \& {Omukai}, K. 2016, \mnras,
  457, 3356

\bibitem[{{Venemans} {et~al.}(2020){Venemans}, {Walter}, {Neeleman}, {Novak},
  {Otter}, {Decarli}, {Ba{\~n}ados}, {Drake}, {Farina}, {Kaasinen},
  {Mazzucchelli}, {Carilli}, {Fan}, {Rix}, \& {Wang}}]{2020ApJ...904..130V}
{Venemans}, B.~P., {Walter}, F., {Neeleman}, M., {et~al.} 2020, \apj, 904, 130

\bibitem[{{Volonteri}(2010)}]{2010A&ARv..18..279V}
{Volonteri}, M. 2010, \aapr, 18, 279

\bibitem[{{Volonteri}(2012)}]{2012Sci...337..544V}
{Volonteri}, M. 2012, Science, 337, 544

\bibitem[{{Volonteri} {et~al.}(2021){Volonteri}, {Habouzit}, \&
  {Colpi}}]{2021NatRP...3..732V}
{Volonteri}, M., {Habouzit}, M., \& {Colpi}, M. 2021, Nature Reviews Physics,
  3, 732

\bibitem[{{Volonteri} {et~al.}(2023){Volonteri}, {Habouzit}, \&
  {Colpi}}]{2023MNRAS.521..241V}
{Volonteri}, M., {Habouzit}, M., \& {Colpi}, M. 2023, \mnras, 521, 241

\bibitem[{{Volonteri} {et~al.}(2015){Volonteri}, {Silk}, \&
  {Dubus}}]{2015ApJ...804..148V}
{Volonteri}, M., {Silk}, J., \& {Dubus}, G. 2015, \apj, 804, 148

\bibitem[{{Wang} {et~al.}(2021){Wang}, {Yang}, {Fan}, {Hennawi}, {Barth},
  {Banados}, {Bian}, {Boutsia}, {Connor}, {Davies}, {Decarli}, {Eilers},
  {Farina}, {Green}, {Jiang}, {Li}, {Mazzucchelli}, {Nanni}, {Schindler},
  {Venemans}, {Walter}, {Wu}, \& {Yue}}]{2021ApJ...907L...1W}
{Wang}, F., {Yang}, J., {Fan}, X., {et~al.} 2021, \apjl, 907, L1

\bibitem[{{Waskom}(2021)}]{2021JOSS....6.3021W}
{Waskom}, M. 2021, The Journal of Open Source Software, 6, 3021

\bibitem[{{Weaver} {et~al.}(2023{\natexlab{a}}){Weaver}, {Cutler}, {Pan},
  {Whitaker}, {Labbe}, {Price}, {Bezanson}, {Brammer}, {Marchesini}, {Leja},
  {Wang}, {Furtak}, {Zitrin}, {Atek}, {Coe}, {Dayal}, {van Dokkum}, {Feldmann},
  {Forster Schreiber}, {Franx}, {Fujimoto}, {Fudamoto}, {Glazebrook}, {de
  Graaff}, {Greene}, {Juneau}, {Kassin}, {Kriek}, {Khullar}, {Maseda}, {Mowla},
  {Muzzin}, {Nanayakkara}, {Nelson}, {Oesch}, {Pacifici}, {Papovich}, {Setton},
  {Shapley}, {Smit}, {Stefanon}, {Taylor}, {Weibel}, \&
  {Williams}}]{2023arXiv230102671W}
{Weaver}, J.~R., {Cutler}, S.~E., {Pan}, R., {et~al.} 2023{\natexlab{a}}, arXiv
  e-prints, arXiv:2301.02671

\bibitem[{{Weaver} {et~al.}(2022){Weaver}, {Kauffmann}, {Ilbert}, {McCracken},
  {Moneti}, {Toft}, {Brammer}, {Shuntov}, {Davidzon}, {Hsieh}, {Laigle},
  {Anastasiou}, {Jespersen}, {Vinther}, {Capak}, {Casey}, {McPartland},
  {Milvang-Jensen}, {Mobasher}, {Sanders}, {Zalesky}, {Arnouts}, {Aussel},
  {Dunlop}, {Faisst}, {Franx}, {Furtak}, {Fynbo}, {Gould}, {Greve}, {Gwyn},
  {Kartaltepe}, {Kashino}, {Koekemoer}, {Kokorev}, {Le F{\`e}vre}, {Lilly},
  {Masters}, {Magdis}, {Mehta}, {Peng}, {Riechers}, {Salvato}, {Sawicki},
  {Scarlata}, {Scoville}, {Shirley}, {Silverman}, {Sneppen}, {Smolc̆i{\'c}},
  {Steinhardt}, {Stern}, {Tanaka}, {Taniguchi}, {Teplitz}, {Vaccari}, {Wang},
  \& {Zamorani}}]{2022ApJS..258...11W}
{Weaver}, J.~R., {Kauffmann}, O.~B., {Ilbert}, O., {et~al.} 2022, \apjs, 258,
  11

\bibitem[{{Weaver} {et~al.}(2023{\natexlab{b}}){Weaver}, {Zalesky}, {Kokorev},
  {McPartland}, {Chartab}, {Gould}, {Shuntov}, {Davidzon}, {Faisst},
  {Stickley}, {Capak}, {Toft}, {Masters}, {Mobasher}, {Sanders}, {Kauffmann},
  {McCracken}, {Ilbert}, {Brammer}, \& {Moneti}}]{2023arXiv231007757W}
{Weaver}, J.~R., {Zalesky}, L., {Kokorev}, V., {et~al.} 2023{\natexlab{b}},
  arXiv e-prints, arXiv:2310.07757

\bibitem[{{Wenger} {et~al.}(2000){Wenger}, {Ochsenbein}, {Egret}, {Dubois},
  {Bonnarel}, {Borde}, {Genova}, {Jasniewicz}, {Lalo{\"e}}, {Lesteven}, \&
  {Monier}}]{2000A&AS..143....9W}
{Wenger}, M., {Ochsenbein}, F., {Egret}, D., {et~al.} 2000, \aaps, 143, 9

\bibitem[{{Whitaker} {et~al.}(2019){Whitaker}, {Ashas}, {Illingworth}, {Magee},
  {Leja}, {Oesch}, {van Dokkum}, {Mowla}, {Bouwens}, {Franx}, {Holden},
  {Labb{\'e}}, {Rafelski}, {Teplitz}, \& {Gonzalez}}]{2019ApJS..244...16W}
{Whitaker}, K.~E., {Ashas}, M., {Illingworth}, G., {et~al.} 2019, \apjs, 244,
  16

\bibitem[{{Williams} {et~al.}(2023{\natexlab{a}}){Williams}, {Alberts}, {Ji},
  {Hainline}, {Lyu}, {Rieke}, {Endsley}, {Suess}, {Johnson}, {Florian},
  {Shivaei}, {Rujopakarn}, {Baker}, {Bhatawdekar}, {Boyett}, {Bunker},
  {Carniani}, {Charlot}, {Curtis-Lake}, {DeCoursey}, {de Graaff}, {Egami},
  {Eisenstein}, {Gibson}, {Hausen}, {Helton}, {Maiolino}, {Maseda}, {Nelson},
  {Perez-Gonzalez}, {Rieke}, {Robertson}, {Sun}, {Tacchella}, {Willmer}, \&
  {Willott}}]{2023arXiv231107483W}
{Williams}, C.~C., {Alberts}, S., {Ji}, Z., {et~al.} 2023{\natexlab{a}}, arXiv
  e-prints, arXiv:2311.07483

\bibitem[{{Williams} {et~al.}(2023{\natexlab{b}}){Williams}, {Tacchella},
  {Maseda}, {Robertson}, {Johnson}, {Willott}, {Eisenstein}, {Willmer}, {Ji},
  {Hainline}, {Helton}, {Alberts}, {Baum}, {Bhatawdekar}, {Boyett}, {Bunker},
  {Carniani}, {Charlot}, {Chevallard}, {Curtis-Lake}, {de Graaf}, {Egami},
  {Franx}, {Kumari}, {Maiolino}, {Nelson}, {Rieke}, {Sandles}, {Shivaei},
  {Simmonds}, {Smit}, {Suess}, {Sun}, {Ubler}, \&
  {Witstok}}]{2023arXiv230109780W}
{Williams}, C.~C., {Tacchella}, S., {Maseda}, M.~V., {et~al.}
  2023{\natexlab{b}}, arXiv e-prints, arXiv:2301.09780

\bibitem[{{Woods} {et~al.}(2019){Woods}, {Agarwal}, {Bromm}, {Bunker}, {Chen},
  {Chon}, {Ferrara}, {Glover}, {Haemmerl{\'e}}, {Haiman}, {Hartwig}, {Heger},
  {Hirano}, {Hosokawa}, {Inayoshi}, {Klessen}, {Kobayashi}, {Koliopanos},
  {Latif}, {Li}, {Mayer}, {Mezcua}, {Natarajan}, {Pacucci}, {Rees}, {Regan},
  {Sakurai}, {Salvadori}, {Schneider}, {Surace}, {Tanaka}, {Whalen}, \&
  {Yoshida}}]{2019PASA...36...27W}
{Woods}, T.~E., {Agarwal}, B., {Bromm}, V., {et~al.} 2019, \pasa, 36, e027

\bibitem[{{Wu} \& {Shen}(2022)}]{2022ApJS..263...42W}
{Wu}, Q. \& {Shen}, Y. 2022, \apjs, 263, 42

\bibitem[{{Yang} {et~al.}(2022){Yang}, {Boquien}, {Brandt}, {Buat},
  {Burgarella}, {Ciesla}, {Lehmer}, {Ma{\l}ek}, {Mountrichas}, {Papovich},
  {Pons}, {Stalevski}, {Theul{\'e}}, \& {Zhu}}]{2022ApJ...927..192Y}
{Yang}, G., {Boquien}, M., {Brandt}, W.~N., {et~al.} 2022, \apj, 927, 192

\bibitem[{{Yang} {et~al.}(2020){Yang}, {Boquien}, {Buat}, {Burgarella},
  {Ciesla}, {Duras}, {Stalevski}, {Brandt}, \&
  {Papovich}}]{2020MNRAS.491..740Y}
{Yang}, G., {Boquien}, M., {Buat}, V., {et~al.} 2020, \mnras, 491, 740

\bibitem[{{Yang} {et~al.}(2021){Yang}, {Wang}, {Fan}, {Barth}, {Hennawi},
  {Nanni}, {Bian}, {Davies}, {Farina}, {Schindler}, {Ba{\~n}ados}, {Decarli},
  {Eilers}, {Green}, {Guo}, {Jiang}, {Li}, {Venemans}, {Walter}, {Wu}, \&
  {Yue}}]{2021ApJ...923..262Y}
{Yang}, J., {Wang}, F., {Fan}, X., {et~al.} 2021, \apj, 923, 262

\bibitem[{{Yoo} \& {Miralda-Escud{\'e}}(2004)}]{2004ApJ...614L..25Y}
{Yoo}, J. \& {Miralda-Escud{\'e}}, J. 2004, \apjl, 614, L25

\bibitem[{{Yue} {et~al.}(2023){Yue}, {Eilers}, {Simcoe}, {Mackenzie},
  {Matthee}, {Kashino}, {Bordoloi}, {Lilly}, \& {Naidu}}]{2023arXiv230904614Y}
{Yue}, M., {Eilers}, A.-C., {Simcoe}, R.~A., {et~al.} 2023, arXiv e-prints,
  arXiv:2309.04614

\bibitem[{{Zavala} {et~al.}(2023){Zavala}, {Buat}, {Casey}, {Finkelstein},
  {Burgarella}, {Bagley}, {Ciesla}, {Daddi}, {Dickinson}, {Ferguson}, {Franco},
  {Jim{\'e}nez-Andrade}, {Kartaltepe}, {Koekemoer}, {Le Bail}, {Murphy},
  {Papovich}, {Tacchella}, {Wilkins}, {Aretxaga}, {Behroozi}, {Champagne},
  {Fontana}, {Giavalisco}, {Grazian}, {Grogin}, {Kewley}, {Kocevski},
  {Kirkpatrick}, {Lotz}, {Pentericci}, {P{\'e}rez-Gonz{\'a}lez}, {Pirzkal},
  {Ravindranath}, {Somerville}, {Trump}, {Yang}, {Yung}, {Almaini},
  {Amor{\'\i}n}, {Annunziatella}, {Arrabal Haro}, {Backhaus}, {Barro}, {Bell},
  {Bhatawdekar}, {Bisigello}, {Buitrago}, {Calabr{\`o}}, {Castellano},
  {Ch{\'a}vez Ortiz}, {Chworowsky}, {Cleri}, {Cohen}, {Cole}, {Cooke},
  {Cooper}, {Cooray}, {Costantin}, {Cox}, {Croton}, {Dav{\'e}}, {de La Vega},
  {Dekel}, {Elbaz}, {Estrada-Carpenter}, {Fern{\'a}ndez}, {Finkelstein},
  {Freundlich}, {Fujimoto}, {Garc{\'\i}a-Argum{\'a}nez}, {Gardner}, {Gawiser},
  {G{\'o}mez-Guijarro}, {Guo}, {Hamilton}, {Hathi}, {Holwerda}, {Hirschmann},
  {Huertas-Company}, {Hutchison}, {Iyer}, {Jaskot}, {Jha}, {Jogee}, {Juneau},
  {Jung}, {Kassin}, {Kurczynski}, {Larson}, {Leung}, {Long}, {Lucas},
  {Magnelli}, {Mantha}, {Matharu}, {McGrath}, {McIntosh}, {Medrano}, {Merlin},
  {Mobasher}, {Morales}, {Newman}, {Nicholls}, {Pandya}, {Rafelski}, {Ronayne},
  {Rose}, {Ryan}, {Santini}, {Seill{\'e}}, {Shah}, {Shen}, {Simons}, {Snyder},
  {Stanway}, {Straughn}, {Teplitz}, {Vanderhoof}, {Vega-Ferrero}, {Wang},
  {Weiner}, {Willmer}, {Wuyts}, \& {CEERS Team}}]{2023ApJ...943L...9Z}
{Zavala}, J.~A., {Buat}, V., {Casey}, C.~M., {et~al.} 2023, \apjl, 943, L9

\bibitem[{{Zhang} {et~al.}(2023){Zhang}, {Behroozi}, {Volonteri}, {Silk},
  {Fan}, {Hopkins}, {Yang}, \& {Aird}}]{2023MNRAS.518.2123Z}
{Zhang}, H., {Behroozi}, P., {Volonteri}, M., {et~al.} 2023, \mnras, 518, 2123

\end{thebibliography}
%

\onecolumn
\begin{appendix}
\normalsize

\section{Comparative analysis of properties between active and inactive galaxies} \label{sec:agnvgal}

We present the properties of known sources at $4<z<9$ observed with the James Webb Space Telescope (JWST) taken from the literature to investigate how well our selection criteria distinguish between active galactic nuclei (AGNs) and inactive galaxies in real data, as well as to assess the accuracy of the derived physical traits.
To construct the samples for comparison, we first select sources with available spectroscopic redshifts in the DAWN JWST Archive's \citep[DJA;][]{2023ApJ...947...20V} version of CEERS and GOODS-S/N photometric tables as well as the second data release of UNCOVER catalog.
Confirmed AGNs\footnote{
	For completeness, known bright quasars showcased previously in Figure~\ref{fig:mbh_mstar} but not used for analysis in this section since they reside outside the public JWST fields are also listed in Table~\ref{tab:qso} \citep[i.e.,][]{2023arXiv230904614Y,2023Natur.621...51D,2023A&A...677A.145U,2023ApJ...953..180S}.
} reported in previous studies are listed in Table~\ref{tab:qso}, where all of them have been spectroscopically characterized with JWST.
In total, there are 36 AGNs with publicly available JWST/NIRCam data \citep[i.e.,][]{2023arXiv230311946H,2023arXiv230905714G,2023ApJ...953L..29L,2023ApJ...954L...4K,2023arXiv230811610K}, including 6 identified as dual AGN candidates \citep{2023arXiv230801230M}.
We note that these dual AGN candidates are then removed from the samples to avoid complications due to their potential peculiar properties, leaving us with the remaining 30 AGNs.
To construct the inactive galaxy samples, we subsequently chose 93 sources at $z=4$--9 from the DJA's JWST sources repository, characterized and classified as galaxies not containing AGNs, as defined by their originating publication \citep[i.e.,][]{2022ApJ...938L..13R,2023ApJ...947L..24M,2023ApJS..269...33N,2023arXiv230308149S,2023MNRAS.526.1657T,2023ApJ...956..139I,2023ApJ...951L..22A,2023ApJ...949L..25F,2023arXiv230600647H,2023arXiv230602468H,2023arXiv230602467B,2023A&A...678A..68S}.

\begin{figure*}[htb!]
	\centering
	\resizebox{\hsize}{!}{\includegraphics{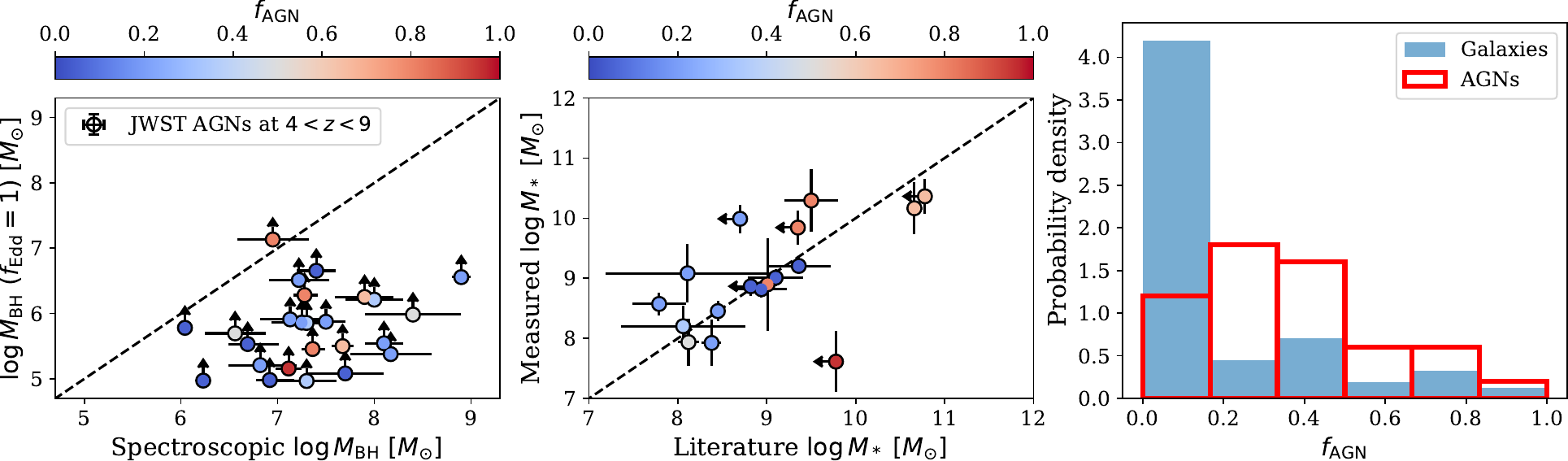}}
	\caption{
		Distribution of black hole masses ($\MBH$), stellar masses ($\MStar$), and the fraction of AGN emission ($f_\mathrm{AGN}$) of known sources (see text).
		The left panel compares the lower limit $\MBH$ assuming an Eddington ratio of $f_\mathrm{Edd}=1$ that we calculated and actual values reported in the literature.
		The data points are color-coded according to the inferred $f_\mathrm{AGN}$ of each source.
		The middle panel shows the $\MStar$ from other studies versus our own measurements.
		The right panel illustrates the distribution of $f_\mathrm{AGN}$ for active and inactive galaxies.
		To compensate for the difference in sample sizes, we normalize the bin heights of the histogram, ensuring that the integral of the distribution equals unity.
	}
	\label{fig:known_prop}
\end{figure*}

Spectral energy distribution (SED) fitting with \texttt{CIGALE} is then performed to the sources compiled above to calculate their stellar mass ($\MStar$) and AGN fraction of the total emission ($f_\mathrm{AGN}$) within the rest-frame wavelengths of 0.1--0.7~$\mu$m.
The right panel of Figure~\ref{fig:known_prop} depicts the distribution of $f_\mathrm{AGN}$ for two distinct populations: AGNs versus inactive galaxies.
Correspondingly, it is observed that 80\% of the known AGNs (24 out of 30) have $f_\mathrm{AGN}\geq0.2$, whereas approximately 30\% of the galaxies (28 out of 93) display $f_\mathrm{AGN}\geq0.2$.
This might indicate that if we adopt $f_\mathrm{AGN}\geq0.2$ as a limit for the quasar candidates selection, we would expect a contamination from the high-$z$ inactive galaxies as high as 30\%.
Of course, we could increase the $f_\mathrm{AGN}$ cutoff to a higher value to get a more pure quasar samples.
However, given the scarcity of quasar number density in the sky, we prefer to adopt 0.2 to aim for a completeness level of up to 80\%.
Accordingly, of the discussed parent samples, only 11 AGNs reside at $z\gtrsim6$, for which we successfully recover 9 of them using the selection method explained in Section~\ref{sec:sedfit}.
Two sources are missed because their $f_\mathrm{AGN}$ is less than our selection threshold of 0.2.

We then proceed to estimate the lower limit black hole masses of the 30 JWST-confirmed AGNs at $4<z<9$ as explained in Section \ref{sec:bh_dist} by adopting Equation~\ref{eq:ledd} and assuming Eddington ratio of $f_\mathrm{Edd}=1$.
Accordingly, we compare these limits with the actual $\MBH$ reported in the literature, determined based on the broad emission line analysis.
In this case, only 25 of 30 AGNs have available spectroscopic $\MBH$ (see Table~\ref{tab:qso}).
As shown by the left panel of Figure~\ref{fig:known_prop}, our $\MBH$ estimations are, on average, systematically lower by $\approx$1.6~dex than the spectroscopic $\MBH$ reported in other studies, consistent with the notion that the $\MBH$ values that we inferred are really lower limits.
This offset is expected since $f_\mathrm{Edd}$ strongly affects our $\MBH$ estimation.
If we change the assumed $f_\mathrm{Edd}$ to be much lower, like 0.1, our data points will become 1~dex higher, closer to those spectroscopic $\MBH$.

The comparison between stellar masses ($\MStar$) we computed via SED modeling and values from the literature is depicted in the middle panel of Figure~\ref{fig:known_prop}.
We note that this further excludes samples from \cite{2023arXiv230905714G} since they do not provide $\MStar$ measurements, leaving us with the remaining 17 AGNs.
Our measurements and other studies are reasonably consistent within the expected uncertainties as the data points are positively correlated with a scatter around the one-to-one relation of $\approx$0.5~dex.
This scatter is expected since we calculated $\MStar$ without performing decomposition of the quasar and host galaxy lights via 2D image modeling.
Instead, we performed the SED decomposition directly using \texttt{CIGALE} to the photometric catalog, as explained in Section~\ref{sec:cigalefit}.

\begin{figure*}[htb!]
	\centering
	\resizebox{\hsize}{!}{\includegraphics{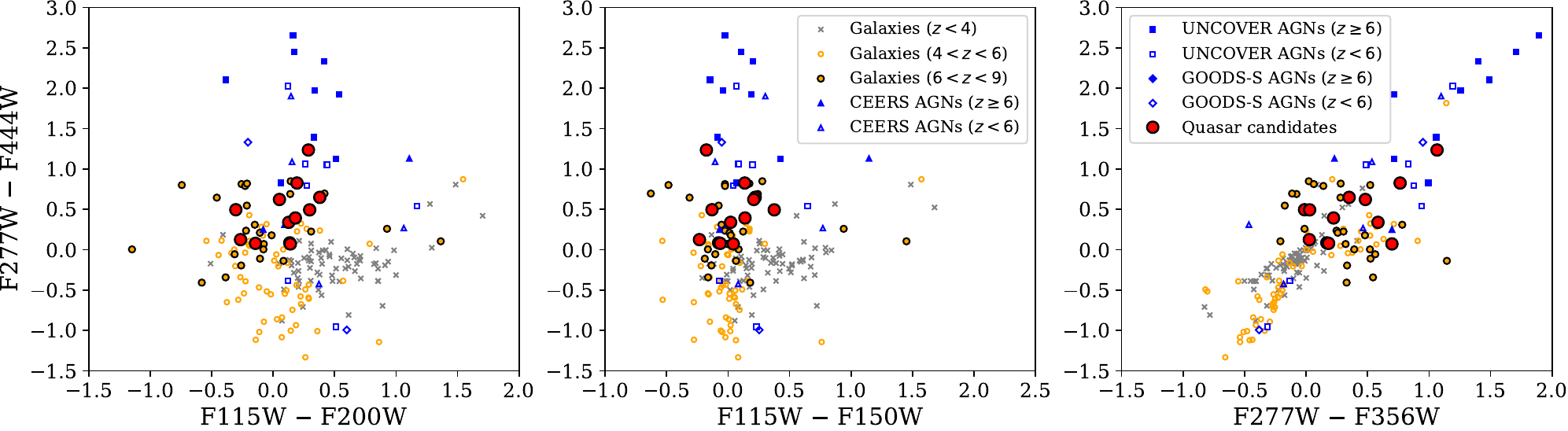}}
	\caption{
		JWST/NIRCam color diagram of spectroscopically confirmed sources residing in the CEERS, UNCOVER, and GOODS-S datasets.
		Galaxy samples from the DJA's JWST sources repository at low, medium, and high redshifts are marked with gray crosses, orange empty circles, and orange-filled circles, respectively.
		Broad-line AGNs are indicated with blue colors, where filled symbols denote objects at $6 < z <9$, while empty symbols show those at $4 < z < 6$ (see the figure legend).
		On the other hand, samples of our quasar candidates existing in the same extragalactic fields are portrayed with red circles.
		Substantial overlap between the colors of unobscured AGNs -- that is, those with blue rest-frame UV continuum -- and galaxies make it challenging to separate them using simple color cuts, indicating that full SED fitting is a better way to recover those blue quasars.
	}
	\label{fig:known_color}
\end{figure*}

As additional information, we find that distinguishing unobscured AGNs having blue rest-frame UV continuum from galaxies is challenging due to their color similarity in specific filter pairs.
To illustrate this better, we present the color distribution of our quasar candidates compared to confirmed AGNs and inactive galaxies in Figure~\ref{fig:known_color}.
Since a substantial overlap between the colors of unobscured AGNs and galaxies is observed, employing a more advanced technique, such as full SED fitting as done here, is more effective than using simple color cuts for accurately identifying these blue quasars.
In the future, complementing our current datasets with more mid-infrared (MIR) measurements will be instrumental in differentiating AGNs from star-forming galaxies.
This distinction arises from the fact that the presence of hot dust emission in MIR bands is a unique feature not easily attributable to stellar light or cold dust within the interstellar medium.

\begin{center}
	\centering
	\tiny
	\begin{longtable}{ccccHHcccc}
		\caption{Compilation of AGN samples that have been spectroscopically characterized with JWST from the literature.} \\
		\label{tab:qso} \\
		\hline\hline
		Source & RA & Dec &  $z_\mathrm{spec}$  &  $\log M_\mathrm{BH, ref}$  & $\log M_\mathrm{*, ref}$ & $f_\mathrm{AGN}$ & $\log M_\mathrm{BH}$ & $\log M_\mathrm{*}$ & Reference \\
		& [J2000] & [J2000] & & [$M_\odot$] & [$M_\odot$] & & [$M_\odot$] & [$M_\odot$] & \\
		\hline
		\endfirsthead
		\caption{continued.} \\
		\hline\hline
		Source & RA & Dec &  $z_\mathrm{spec}$  &  $\log M_\mathrm{BH, ref}$  & $\log M_\mathrm{*, ref}$ & $f_\mathrm{AGN}$ & $\log M_\mathrm{BH, calc}$ & $\log M_\mathrm{*, calc}$ & Reference \\
		& [J2000] & [J2000] & & [$M_\odot$] & [$M_\odot$] & & [$M_\odot$] & [$M_\odot$] & \\
		\hline
		\endhead
		\hline
		\endfoot
J0148+0600 & 177.06933 & 52.86397 & 5.98 & $9.89^{+0.05}_{-0.06}$ & $10.74^{+0.31}_{-0.30}$ & \nodata & \nodata & \nodata & \cite{2023arXiv230904614Y} \\
J159$-$02 & 159.22579 & $-$2.54387 & 6.38 & $9.10^{+0.01}_{-0.01}$ & $10.14^{+0.34}_{-0.36}$ & \nodata & \nodata & \nodata & \cite{2023arXiv230904614Y} \\
J1120+0641 & 170.00617 & 6.69008 & 7.09 & $9.08^{+0.03}_{-0.03}$ & $9.81^{+0.23}_{-0.31}$ & \nodata & \nodata & \nodata & \cite{2023arXiv230904614Y} \\
J0100+2802 & 15.05425 & 28.04050 & 6.33 & $10.06^{+0.01}_{-0.01}$ & $<11.58$ & \nodata & \nodata & \nodata & \cite{2023arXiv230904614Y} \\
J1030+0524 & 157.61296 & 5.41529 & 6.30 & $9.19^{+0.01}_{-0.01}$ & $<10.65$ & \nodata & \nodata & \nodata & \cite{2023arXiv230904614Y} \\
J1148+5251 & 27.15683 & 6.00556 & 5.98 & $9.64^{+0.01}_{-0.01}$ & $<10.93$ & \nodata & \nodata & \nodata & \cite{2023arXiv230904614Y} \\
J2236+0032 & 339.18575 & 0.54914 & 6.40 & $9.19^{+0.08}_{-0.08}$ & $11.12^{+0.40}_{-0.27}$ & \nodata & \nodata & \nodata & \cite{2023Natur.621...51D} \\
J2255+0251 & 343.90850 & 2.85739 & 6.34 & $8.31^{+0.04}_{-0.04}$ & $10.53^{+0.51}_{-0.37}$ & \nodata & \nodata & \nodata & \cite{2023Natur.621...51D} \\
CEERS01244 & 215.24067 & 53.03606 & 4.48 & $7.18^{+0.03}_{-0.03}$ & $8.63^{+0.63}_{-1.03}$ & \nodata & \nodata & \nodata & \cite{2023arXiv230311946H} \\
GLASS160133 & 3.58029 & $-$30.42439 & 4.02 & $6.04^{+0.04}_{-0.04}$ & $<8.82$ & $0.05\pm0.20$ & $>5.78$ & $8.87\pm0.16$ & \cite{2023arXiv230311946H} \\
GLASS150029 & 3.57717 & $-30.42258$ & 4.58 & $6.23^{+0.03}_{-0.05}$ & $9.10^{+0.31}_{-0.37}$ & $0.05\pm0.20$ & $>4.97$ & $9.01\pm0.14$ & \cite{2023arXiv230311946H} \\
CEERS00746 & 214.80913 & 52.86847 & 5.62 & $7.43^{+0.11}_{-0.10}$ & $<9.11$ & \nodata & \nodata & \nodata & \cite{2023arXiv230311946H} \\
CEERS01665 & 215.17821 & 53.05936 & 4.48 & $6.95^{+0.20}_{-0.11}$ & $9.92^{+0.51}_{-0.68}$ & \nodata & \nodata & \nodata & \cite{2023arXiv230311946H} \\
CEERS00672 & 214.88967 & 52.83297 & 5.67 & $7.36^{+0.13}_{-0.11}$ & $<9.01$ & $0.80\pm0.17$ & $>5.46$ & $8.90\pm0.77$ & \cite{2023arXiv230311946H} \\
CEERS02782 & 214.82346 & 52.83028 & 5.24 & $7.28^{+0.14}_{-0.11}$ & $<9.35$ & $0.80\pm0.28$ & $>6.28$ & $9.84\pm0.28$ & \cite{2023arXiv230311946H} \\
CEERS00397 & 214.83621 & 52.88269 & 6.00 & $6.69^{+0.32}_{-0.19}$ & $9.36^{+0.36}_{-0.45}$ & $0.05\pm0.17$ & $>5.53$ & $9.21\pm0.14$ & \cite{2023arXiv230311946H} \\
CEERS00717 & 215.08142 & 52.97219 & 6.94 & $7.66^{+0.19}_{-0.14}$ & $9.61^{+0.77}_{-1.18}$ & \nodata & \nodata & \nodata & \cite{2023arXiv230311946H} \\
CEERS01236 & 215.14529 & 52.96728 & 4.48 & $6.92^{+0.26}_{-0.14}$ & $8.94^{+0.29}_{-0.54}$ & $0.05\pm0.24$ & $>4.98$ & $8.82\pm0.14$ & \cite{2023arXiv230311946H} \\
MSAID2008 & 3.59242 & $-$30.43283 & 6.74 & \nodata & \nodata & $0.20\pm0.25$ & $>5.12$ & $9.21\pm0.41$ & \cite{2023arXiv230905714G} \\
MSAID4286 & 3.61920 & $-$30.42327 & 5.84 & $8.00^{+0.30}_{-0.30}$ & \nodata & $0.35\pm0.08$ & $>6.21$ & $11.25\pm0.21$ & \cite{2023arXiv230905714G} \\
MSAID10686 & 3.55084 & $-$30.40660 & 5.05 & \nodata & \nodata & $0.50\pm0.08$ & $>6.07$ & $10.61\pm0.18$ & \cite{2023arXiv230905714G} \\
MSAID13123a & 3.57983 & $-$30.40157 & 7.04 & $7.30^{+0.20}_{-0.20}$ & \nodata & $0.35\pm0.06$ & $>5.86$ & $10.50\pm0.10$ & \cite{2023arXiv230905714G} \\
MSAID13821 & 3.62061 & $-$30.39995 & 6.34 & $8.10^{+0.20}_{-0.20}$ & \nodata & $0.20\pm0.11$ & $>5.54$ & $10.12\pm0.17$ & \cite{2023arXiv230905714G} \\
MSAID15383a & 3.58353 & $-$30.39668 & 7.04 & \nodata & \nodata & $0.65\pm0.01$ & $>5.95$ & $10.43\pm0.11$ & \cite{2023arXiv230905714G} \\
MSAID16594a & 3.59720 & $-$30.39433 & 7.04 & \nodata & \nodata & $0.50\pm0.06$ & $>5.61$ & $10.24\pm0.14$ & \cite{2023arXiv230905714G} \\
MSAID23608 & 3.54282 & $-$30.38065 & 5.80 & $7.50^{+0.20}_{-0.20}$ & \nodata & $0.20\pm0.16$ & $>5.87$ & $9.34\pm0.18$ & \cite{2023arXiv230905714G} \\
MSAID28876 & 3.56960 & $-$30.37322 & 7.04 & \nodata & \nodata & $0.50\pm0.11$ & $>5.54$ & $10.1\pm0.20$ & \cite{2023arXiv230905714G} \\
MSAID32265 & 3.53753 & $-$30.37017 & \nodata & \nodata & \nodata & \nodata & \nodata & \nodata & \cite{2023arXiv230905714G} \\
MSAID33437 & 3.54642 & $-$30.36625 & \nodata & \nodata & \nodata & \nodata & \nodata & \nodata & \cite{2023arXiv230905714G} \\
MSAID35488 & 3.57898 & $-$30.36260 & 6.26 & $7.40^{+0.20}_{-0.20}$ & \nodata & $0.05\pm0.17$ & $>6.66$ & $9.51\pm0.12$ & \cite{2023arXiv230905714G} \\
MSAID38108 & 3.53001 & $-$30.35801 & 4.96 & $8.40^{+0.50}_{-0.50}$ & \nodata & $0.50\pm0.07$ & $>5.98$ & $10.67\pm0.12$ & \cite{2023arXiv230905714G} \\
MSAID39243 & 3.51389 & $-$30.35602 & \nodata & \nodata & \nodata & \nodata & \nodata & \nodata & \cite{2023arXiv230905714G} \\
MSAID41225 & 3.53399 & $-$30.35331 & 6.76 & $7.70^{+0.40}_{-0.40}$ & \nodata & $0.05\pm0.18$ & $>5.08$ & $9.69\pm0.21$ & \cite{2023arXiv230905714G} \\
MSAID45924 & 3.58476 & $-$30.34363 & 4.46 & $8.90^{+0.10}_{-0.10}$ & \nodata & $0.20\pm0.01$ & $>6.56$ & $11.22\pm0.02$ & \cite{2023arXiv230905714G} \\
CEERS1019 & 215.03539 & 52.89066 & 8.68 & $6.95^{+0.37}_{-0.37}$ & $9.50^{+0.30}_{-0.30}$ & $0.80\pm0.26$ & $>7.13$ & $10.29\pm0.53$ & \cite{2023ApJ...953L..29L} \\
GS3073 & 53.07888 & $-$27.88416 & 5.55 & $8.20^{+0.40}_{-0.40}$ & $9.40^{+0.7}_{-0.20}$ & \nodata & \nodata & \nodata & \protect{\cite{2023A&A...677A.145U}} \\
1670 & 214.82345 & 52.83028 & 5.24 & $7.11^{+0.13}_{-0.13}$ & $<9.78$ & $0.95\pm0.30$ & $>5.16$ & $7.61\pm0.51$ & \cite{2023ApJ...954L...4K} \\
3210AV4 & 214.80914 & 52.86848 & 5.62 & $7.67^{+0.11}_{-0.11}$ & $<10.78$ & $0.65\pm0.11$ & $>5.50$ & $10.36\pm0.29$ & \cite{2023ApJ...954L...4K} \\
J2239+0207 & 339.94779 & 2.12986 & 6.25 & $8.78^{+0.4}_{-0.4}$ & $10.00^{+0.30}_{-0.50}$ & \nodata & \nodata & \nodata & \cite{2023ApJ...953..180S} \\
10013704a & 53.12654 & $-$27.81809 & 5.92 & $5.65^{+0.31}_{-0.31}$ & $8.88^{+0.66}_{-0.66}$ & \nodata & \nodata & \nodata & \cite{2023arXiv230801230M} \\
10013704b & 53.12654 & $-$27.81809 & 5.92 & $7.50^{+0.31}_{-0.31}$ & $8.88^{+0.66}_{-0.03}$ & \nodata & \nodata & \nodata & \cite{2023arXiv230801230M} \\
8083 & 53.13284 & $-$27.80186 & 4.65 & $7.25^{+0.31}_{-0.31}$ & $8.45^{+0.03}_{-0.03}$ & $0.20\pm0.24$ & $>5.86$ & $8.45\pm0.17$ & \cite{2023arXiv230801230M} \\
1093 & 189.17974 & 62.22463 & 5.60 & $7.36^{+0.31}_{-0.31}$ & $8.34^{+0.20}_{-0.20}$ & \nodata & \nodata & \nodata & \cite{2023arXiv230801230M} \\
3608 & 189.11794 & 62.23552 & 5.27 & $6.82^{+0.38}_{-0.33}$ & $8.38^{+0.11}_{-0.15}$ & $0.20\pm0.31$ & $>5.21$ & $7.92\pm0.39$ & \cite{2023arXiv230801230M} \\
11836 & 189.22059 & 62.26368 & 4.41 & $7.13^{+0.31}_{-0.31}$ & $7.79^{+0.30}_{-0.30}$ & $0.20\pm0.22$ & $>5.91$ & $8.57\pm0.19$ & \cite{2023arXiv230801230M} \\
20621 & 189.12252 & 62.29285 & 4.68 & $7.30^{+0.31}_{-0.31}$ & $8.06^{+0.7}_{-0.7}$ & $0.35\pm0.33$ & $>4.97$ & $8.20\pm0.34$ & \cite{2023arXiv230801230M} \\
73488a & 189.19740 & 62.17723 & 4.13 & $6.18^{+0.30}_{-0.30}$ & $9.78^{+0.20}_{-0.20}$ & \nodata & \nodata & \nodata & \cite{2023arXiv230801230M} \\
73488b & 189.19740 & 62.17723 & 4.13 & $7.71^{+0.30}_{-0.30}$ & $9.78^{+0.20}_{-0.20}$ & \nodata & \nodata & \nodata & \cite{2023arXiv230801230M} \\
77652 & 189.29323 & 62.19900 & 5.23 & $6.86^{+0.35}_{-0.34}$ & $7.87^{+0.16}_{-0.28}$ & \nodata & \nodata & \nodata & \cite{2023arXiv230801230M} \\
61888 & 189.16802 & 62.21701 & 5.88 & $7.22^{+0.31}_{-0.31}$ & $8.11^{+0.92}_{-0.92}$ & $0.20\pm0.27$ & $>6.51$ & $9.08\pm0.48$ & \cite{2023arXiv230801230M} \\
62309 & 189.24898 & 62.21835 & 5.17 & $6.56^{+0.32}_{-0.31}$ & $8.12^{+0.12}_{-0.13}$ & $0.50\pm0.3$ & $>5.7$ & $7.94\pm0.39$ & \cite{2023arXiv230801230M} \\
53757a & 189.26978 & 62.19421 & 4.45 & $6.29^{+0.33}_{-0.32}$ & $10.18^{+0.13}_{-0.12}$ & \nodata & \nodata & \nodata & \cite{2023arXiv230801230M} \\
53757b & 189.26978 & 62.19421 & 4.45 & $7.69^{+0.32}_{-0.31}$ & $10.18^{+0.13}_{-0.12}$ & \nodata & \nodata & \nodata & \cite{2023arXiv230801230M} \\
954 & 189.15197 & 62.25964 & 6.76 & $7.90^{+0.30}_{-0.31}$ & $10.66^{+0.09}_{-0.1}$ & $0.65\pm0.2$ & $>6.25$ & $10.17\pm0.43$ & \cite{2023arXiv230801230M} \\
MSAID20466 & 3.64041 & $-$30.38644 & 8.50 & $8.17^{+0.42}_{-0.42}$ & $<8.70$ & $0.20\pm0.12$ & $>5.38$ & $9.99\pm0.24$ & \cite{2023arXiv230811610K} \\
	\end{longtable}
	\tablefoot{
		Column (1): name or identifier of each source.
		Column (2)--(3): right ascension (RA) and declination (Dec) in decimal degrees. 
		Column (4): spectroscopic redshift.
		Column (5): fraction of AGN component to the total emission within the rest-wavelengths of 0.1--0.7~$\mu$m.
		Column (6): lower limit black hole mass assuming an accretion at Eddington limit.
		Column (7): total stellar mass calculated following the method presented in this work.
		Column (8): original literature describing the object.
		Empty columns indicate that the data is unavailable in the public JWST datasets, and the corresponding source is not used to benchmark our quasar selection method.
	}
\end{center}

\section{Complete list of the quasar candidates} \label{sec:qcand}
The complete list of our high-$z$ quasar candidates chosen following the method explained in the main text is reported here.
Table~\ref{tab:qcand} summarizes the subset of photometric properties for these sources, and the complete catalog will be available in electronic form at the CDS\footnote{
Accessible through anonymous ftp to \href{http://cdsarc.u-strasbg.fr/}{cdsarc.u-strasbg.fr} (130.79.128.5) or via \href{http://cdsweb.u-strasbg.fr/cgi-bin/qcat?J/A+A/}{http://cdsweb.u-strasbg.fr/cgi-bin/qcat?J/A+A/}.
}.
Additionally, figures containing the SED fitting results for each quasar candidate can be provided upon reasonable request.

\begin{center}
	\centering
	\tiny
	\begin{longtable}{ccccccccccc}
		\caption{
			List of high-$z$ quasar candidates selected in this work.
			Here, we only show the first ten rows of the catalog as an example, while the entire catalog can be accessed at the CDS (see text).
		} \\
		\label{tab:qcand} \\
		\hline\hline
Source & RA & Dec & F444W & $z_\mathrm{phot}$ & $\log L_\mathrm{bol}$ & $\log M_*$ & $\log$ SFR & $\log M_\mathrm{BH}$ & $f_\mathrm{AGN}$ & Grade \\
& [J2000] & [J2000] & [nJy] & & [erg~s$^{-1}$] & [$M_\odot$] & [$M_\odot$ yr$^{-1}$] & [$M_\odot$] & & \\
		\hline
		\endfirsthead
		\caption{continued.} \\
		\hline\hline
Source & RA & Dec & F444W & $z_\mathrm{phot}$ & $\log L_\mathrm{bol}$ & $\log M_*$ & $\log$ SFR & $\log M_\mathrm{BH}$ & $f_\mathrm{AGN}$ & Grade \\
& [J2000] & [J2000] & [nJy] & & [erg~s$^{-1}$] & [$M_\odot$] & [$M_\odot$ yr$^{-1}$] & [$M_\odot$] & &\\
		\hline
		\endhead
		\hline
		\endfoot
CWB-663 & 149.75099 & $2.15091$ & $351.86\pm10.87$ & $6.05\pm0.12$ & $43.66\pm0.90$ & $9.13\pm0.31$ & $0.83\pm0.35$ & > 5.55 & $0.05\pm0.35$ & B \\
CWB-8286 & 149.76383 & $2.19782$ & $203.96\pm3.92$ & $7.81\pm0.30$ & $43.58\pm1.54$ & $9.35\pm0.25$ & $1.03\pm0.22$ & > 5.46 & $0.05\pm0.32$ & B \\
CWB-19858 & 149.84033 & $2.24807$ & $395.54\pm3.92$ & $6.70\pm0.13$ & $43.66\pm1.31$ & $9.38\pm0.16$ & $1.08\pm0.15$ & > 5.54 & $0.05\pm0.19$ & B \\
CWB-24983 & 149.85942 & $2.27450$ & $234.01\pm3.98$ & $6.14\pm0.17$ & $44.01\pm0.67$ & $9.12\pm0.26$ & $0.53\pm0.91$ & > 5.89 & $0.20\pm0.27$ & A \\
CWB-26445 & 149.80063 & $2.30474$ & $183.36\pm3.92$ & $6.93\pm0.27$ & $44.81\pm0.82$ & $9.19\pm0.44$ & $0.87\pm0.45$ & > 6.70 & $0.95\pm0.33$ & A \\
CWB-35877 & 149.95743 & $2.11303$ & $129.15\pm3.92$ & $6.88\pm0.22$ & $44.76\pm0.49$ & $8.72\pm0.46$ & $0.44\pm0.47$ & > 6.65 & $0.95\pm0.28$ & A \\
CWB-40773 & 149.92404 & $2.15948$ & $253.99\pm6.37$ & $7.30\pm0.20$ & $44.93\pm0.32$ & $8.88\pm0.39$ & $0.62\pm0.42$ & > 6.82 & $0.80\pm0.29$ & A \\
CWB-41512 & 149.93536 & $2.16019$ & $836.82\pm5.79$ & $6.02\pm0.12$ & $45.22\pm0.29$ & $9.75\pm0.15$ & $1.46\pm0.16$ & > 7.11 & $0.20\pm0.23$ & A \\
CWB-42214 & 149.93324 & $2.16687$ & $1760.68\pm5.20$ & $6.09\pm0.12$ & $44.88\pm0.79$ & $10.21\pm0.38$ & $1.67\pm0.34$ & > 6.77 & $0.50\pm0.17$ & A \\
CWB-43536 & 150.02278 & $2.14372$ & $545.53\pm11.98$ & $8.10\pm0.19$ & $45.29\pm0.59$ & $9.83\pm0.25$ & $1.52\pm0.25$ & > 7.17 & $0.20\pm0.27$ & A \\
	\end{longtable}
	\tablefoot{
		Column (1): name of each candidate with specific prefixes indicating the originating dataset, that is, CWB (COSMOS-Web), JDS/GDS (JADES/GOODS-S), GDN (GOODS-N), UCV (UNCOVER), CRS (CEERS), PMC (PRIMER-COSMOS), and PMU (PRIMER-UDS).
		Column (2)--(3): right ascension (RA) and declination (Dec) in decimal degrees. 
		Column (4): fluxes measured using the JWST/NIRCam images.
		Column (5): calculated photometric redshift of the target derived from the best-fitted SED template.
		Column (6): bolometric luminosity of the AGN SED component.
		Column (7): total stellar mass of the presumed host galaxy.
		Column (8): SFR averaged over 100 Myr.
		Column (9): lower limit of the black hole mass, assuming an accretion at the Eddington limit.
		Column (10): fraction of AGN component to the total spectral emission within the rest-frame wavelengths of 0.1--0.7~$\mu$m.
		Column (11): grade after employing the visual inspection, black hole mass limit, and AGN fraction threshold criteria.
	}
\end{center}

\end{appendix}

\end{document}